\documentclass[%
reprint,
showpacs,
amsmath,amssymb, aps,
prb,
]{revtex4-1}

\usepackage{graphicx}
\usepackage{dcolumn}
\usepackage{bm}
\usepackage{hyperref}
\hypersetup{colorlinks,linkcolor={red},citecolor={blue},urlcolor={blue}}
\def\be{\begin{equation}}
\def\ee{\end{equation}}
\def\bea{\begin{eqnarray}}
\def\eea{\end{eqnarray}}

\def\ra{\rangle}
\def\la{\langle}
\def\s{\sigma}
\def\t{\tau}

\usepackage{color}

\begin{document}

\title{Frustrated mixed-spin ladders:  Evidence for a bond order wave phase between rung-singlet and Haldane phases }

\author{N. Ahmadi}
\affiliation{Department of Physics, Institute for Advanced Studies
	in Basic Sciences (IASBS), Zanjan 45137-66731, Iran}

\author{J. Abouie}
\email[]{jahan@iasbs.ac.ir}
\affiliation{Department of Physics,
	Institute for Advanced Studies in Basic Sciences (IASBS), Zanjan
	45137-66731, Iran}

\author{R. Haghshenas}
\affiliation{Division of Chemistry and Chemical Engineering, California Institute of Technology, Pasadena, California 91125, USA}

\author{D. Baeriswyl}
\affiliation{Department of Physics, University of Fribourg, CH-1700 Fribourg, Switzerland}

\begin{abstract}
In frustrated spin ladders the interplay of frustration and correlations leads to the familiar Haldane (H) and rung-singlet (RS) phases. The nature of the transition between these two phases is still under debate. In this paper we tackle this issue using tools of quantum information theory. We consider frustrated mixed-spin-(1, 1/2) ladders with antiferromagnetic leg, rung and diagonal couplings, and  calculate various quantities, such as the entanglement entropy (EE), the Schmidt gap, and the level degeneracy of the entanglement spectrum (ES). We use two numerical techniques, the infinite time-evolving block decimation (iTEBD) and the density matrix renormalization group (DMRG). We demonstrate that there exists an intermediate phase in which the ES levels do not exhibit the characteristic degeneracies of the H and RS phases. To understand the underlying physics in this phase, we investigate short-range spin correlations along legs, rungs and diagonals and show that in this intermediate phase long-wavelength modulations occur, akin to bond order waves.

\end{abstract}

\date{\today}


\maketitle

\section{Introduction}\label{sec:intro}
Low-dimensional frustrated spin systems have attracted great interest due to their importance for the understanding of emergent phenomena, such as reentrant phase transitions\cite{doi:10.1142/5697}, flat-band physics\cite{Derzhko}, anomalous robustness of topological order\cite{Schmidt-toric,Schulz-toric,Zarei}, and spin liquid phases.\cite{RevModPhys.89.025003,RevModPhys.85.1473,RevModPhys.91.041003,RevModPhys.86.563,RevModPhys.86.1453}
Frustrated spin ladders have received special attention for several reasons; i) due to their low dimensionality, the interplay of frustration and quantum correlations leads to a variety of topological and nontopological phases such as the Haldane phase, dimer order and various spin liquids\cite{PhysRevB.63.134403, PhysRevB.62.14965, PhysRevB.72.014449, PhysRevB.91.115130, PhysRevB.89.094424, PhysRevB.82.174410,doi:10.1080/00018730701265383,PhysRevB.81.064432,PhysRevA.91.043614,PhysRevB.68.094405,PhysRevB.97.144424,PhysRevB.101.144407,PhysRevB.73.224433,PhysRevA.97.013634,PhysRevB.73.214427,PhysRevB.77.214418},
ii) they are quasi-one dimensional, and show characteristics of both one- and two-dimensional systems\cite{Dagotto618}, and iii) powerful numerical and analytical methods are available for studying their low-energy properties.\cite{PhysRevB.61.8871,PhysRevB.73.214405}
\begin{figure}[h!]
	\centering
	\includegraphics[scale=0.9]{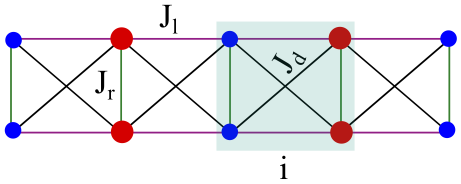}
	\caption{(Color online) Pictorial representation of a mixed-spin ladder with different intra- and inter-leg exchange couplings. The blue and red dots represent the $\s$ and $\t$ spins, respectively. The index $i$ numbers the unit cells.}
	\label{Fig:ladder}
\end{figure}

Most studies of spin ladders have considered a single type of spin, in particular $\sigma=\frac{1}{2}$. Ladders with two types of spin ($\t>\s$), so-called mixed-spin ladders, have received less attention, although this heterogeneity produces qualitatively new effects.\cite{PhysRevB.70.184416,PhysRevB.73.014411,PhysRevB.96.104406,PhysRevB.81.060401,Heydarinasab_2020,Langari_2011}
In this paper we study the ground state phase diagram of a mixed-spin (1, 1/2) ladder with coupling constants $J_l, J_r, J_d$ (illustrated in Fig. \ref{Fig:ladder}), focussing on the effects of diagonal interactions ($J_d$). We limit ourselves to $J_l\ge 0, J_d\ge 0$ but admit both ferromagnetic and antiferromagnetic rung couplings $J_r$.

\begin{figure}[h!]
	\centering
	\includegraphics[width=6cm]{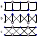}
	\caption{Sublattices $\mathcal{A}$ (full symbols) and $\mathcal{B}$ (empty symbols) for three special cases: a) $J_d=0, J_l>0, J_r>0$, b) $J_l=0, J_r>0, J_d>0$,
	c) $J_r=0, J_l>0, J_d>0$.}
	\label{Fig:sublattices}
\end{figure}

Some knowledge about the ground state can be gained thanks to the Lieb-Mattis theorem\cite{Lieb_62}, which fixes the total spin $S$ for cases where the lattice can be subdivided into sublattices $\mathcal{A}$ and $\mathcal{B}$ in such a way that $J_{ij}=0$ if $i,j\in\mathcal{A}$ or $i,j\in\mathcal{B}$ and $J_{ij}\ge 0$ if $i\in\mathcal{A}$ and $j\in\mathcal{B}$. Then $S=\vert S_{\mathcal{A}}-S_{\mathcal{B}}\vert$, where $S_{\mathcal{A}}$ and $S_{\mathcal{B}}$ are the largest possible values of the spin on sublattices
${\mathcal{A}}$ and ${\mathcal{B}}$, respectively. Fig. \ref{Fig:sublattices} shows three cases where this separation is particularly simple.
In the special case of vanishing diagonal coupling, $J_d=0$ (but $J_l>0, J_r>0$), the ladder can be viewed as two zigzag chains, whereas for vanishing coupling on the legs, $J_l=0$ (but $J_r>0, J_d>0$), the two legs can be taken as subsystems $\mathcal{A}$ and $\mathcal{B}$. In both cases $S_{\mathcal{A}}=S_{\mathcal{B}}=\frac{3}{2}N$, where $N$ is the number of unit cells, and therefore the Lieb-Mattis theorem predicts a spin-singlet ground state. For vanishing rung coupling, $J_r=0$ (but $J_l>0,J_d>0$), the sites with spin 1 may be taken as subsystem $\mathcal{A}$ and those with spin $\frac{1}{2}$ as subsystem $\mathcal{B}$. In this case $S_{\mathcal{A}}=2N$ and $S_{\mathcal{B}}=N$, and therefore the ground state is a ferrimagnet with total spin $S=N$.

A previous study \cite{PhysRevB.81.024409} of our ladder model in the range $J_l=J_r=1$, $0\le J_d\le1.5$  has revealed four different phases, a rung-singlet phase (RS)
for $0\le J_d\le 0.710$, a Haldane phase (H) for $0.710\le J_d\le 0.875$ and two ferrimagnetic phases for $J_d\ge0.875$. The authors of this study also suggested that for
$0\le J_d\lesssim 0.875$ the mixed-spin ladder can be mapped onto the homogeneous spin $\frac{1}{2}$ ladder with modified couplings $J_l',J_r',J_d'$.
The (frustrated) spin $\frac{1}{2}$ ladder has been intensively studied and is essentially understood. Its ground state exhibits the RS and H phases and the location of the RS-H transition is in good agreement with that found for the mixed-spin ladder, thus lending further support to the mapping proposed in Ref. \onlinecite{PhysRevB.81.024409}.

The detailed nature of the RS-H transition in the frustrated spin $\frac{1}{2}$ ladder has been a matter of debate. A first-order transition is well established in a wide region of parameter space, but for weak interchain coupling an intermediate phase, a ``columnar dimer phase'', has been conjectured \cite{PhysRevLett.93.127202}. Numerous studies have since tried to find evidence for this phase, with little success\cite{PhysRevB.77.205121,PhysRevB.73.224433,PhysRevB.57.11439,PhysRevB.77.214418,PhysRevB.73.214427}. More recent calculations fully agree with a single transition, of first order for not too small couplings\cite{PhysRevB.86.075133,Chen_16,SciPostPhys.3.1.005}.
Similarly the RS-H transition found for the mixed-spin model has been interpreted as a single first-order transition due to a level crossing of the RS and H singlet ground states, at the same time  ``some peculiarities of the RS-H transition in the mixed-spin system, as compared to the uniform-spin case'' were noticed\cite{PhysRevB.81.024409}.

Our aim is to shed light on the RS-H transition in the mixed-spin ladder, using tools of quantum information theory, which are better suited for investigating subtle details of ladders\cite{PhysRevLett.105.077202} than, e.g., the dependence of bond energies on coupling parameters, as used in Ref. \onlinecite{PhysRevB.81.024409}.
The remarkable concept of entanglement combined with new algorithms has improved our understanding of many-body systems in general and quantum phase transitions in particular
\cite{PhysRevLett.69.2863,PhysRevLett.91.147902,RevModPhys.77.259,PhysRevLett.91.147902,PhysRevLett.93.040502,PhysRevLett.101.250602,PhysRevB.97.174408, Haghshenas:2019, PhysRevB.100.054404} The best-known measure of entanglement is the von Neumann entanglement entropy (EE) which is widely used to detect quantum phase transitions\cite{osterlo} as well as topological properties of many-body states.\cite{PhysRevLett.96.110405,PhysRevLett.96.110404} Moreover, the entanglement spectrum (ES), the eigenvalues of reduced density matrices, is a remarkable tool in the characterization of topological phases of matter.\cite{PhysRevLett.101.010504, PhysRevB.81.064439, Haghshenas:2014,Ahmadi2020} Actually, the study of the low-lying part of the ES allows us to detect the topological properties of a state or gives direct access to the excitation spectrum of edges.
Using two numerical techniques, iTEBD and DMRG, we obtain the EE and ES of the ground state of mixed-spin (1, 1/2) ladders, and demonstrate that the RS and H phases are separated by an intermediate phase with a different ES level degeneracy.

To identify the intermediate phase, we also calculate short-range spin correlations along rungs, legs and diagonals. The bond pattern differs markedly from that  of a columnar dimer phase, and we attribute it to a long-wavelength incommensurate bond-order wave (BOW), which breaks translational symmetry. Bond order, a well-known concept of quantum chemistry, measures the strength of chemical bonds. Bond alternation in conjugated polymers, a sequence of ``single'' and ``double'' bonds, is in reality a sequence of weakened and strengthened bond orders, a commensurate BOW. While this phenomenon is usually attributed to the bond length dependence of overlap integrals, it was realized that it can also be produced by electron-electron interactions, where {\it a priori} one would only expect a competition between spin-density waves (SDW) and charge-density waves (CDW). In fact, Nakamura
\cite{Nakamura-JPSJ, Nakamura-PRB} and, shortly after, Sengupta and collaborators\cite{Sangupta} realized that in the one-dimensional extended Hubbard model a BOW phase exists in a narrow strip between CDW and SDW phases for small to intermediate coupling strengths. The notion BOW can be extended to spin systems, where it represents again a modulation of bond energies.
A BOW phase was found in the familiar zigzag spin-1/2 Heisenberg chain with frustrated antiferromagnetic exchange.\cite{White-Affleck, J1J2-chigak,Kumar1, Kumar2}. Its nature is
particularly transparent at the Majumdar-Ghosh point where the ground state is an exact dimer state, a product of singlet-paired spins.

The paper is organized as follows.
In Section \ref{sec:Model}, we introduce the Hamiltonian of our mixed-spin ladder and define various spin correlation functions. Section \ref{sec:plaquette} discusses the exact solution for the elementary plaquette. Section \ref{sec:PT} treats the limits of weak and strong rung couplings using perturbation theory.
 In Section \ref{sec:TEBD} the numerical iTEBD technique and its generalization to the mixed-spin  ladder are explained. Section \ref{sec:ladderA} presents a comprehensive study of the ground state phase diagram of the ladder in the absence of diagonal interactions, in terms of the EE and the ES level degeneracies. In Section \ref{sec:Intermediate} the focus is on the intermediate phase, using DMRG.
 A brief summary and suggestions for further studies are presented in Section \ref{sec:summary}. The model is explicitly diagonalized on the plaquette in Appendix \ref{sec:diagonalization} and
some details on the perturbative approach are provided in Appendix \ref{app:expansion}.

\section{Model}\label{sec:Model}
We consider a frustrated mixed-spin ($\t=1, \s=1/2$) ladder, embodied by the Hamiltonian
\begin{equation}
H = H_l +H_r + H_d,
\label{Eq:hamiltonian}
\end{equation}
with
\begin{eqnarray}\label{Eq:Hamiltonian1}	
\nonumber	H_l &=& J_l \sum_{n=1,2}\sum_{i}\left({\pmb\s}_i^{(n)}  \cdot {\pmb\t}_i^{(n)} + {\pmb\t}_i^{(n)} \cdot {\pmb\s}_{i+1}^{(n)} \right) ,\\
\nonumber  H_r &=& J_r \sum_{i}\left({\pmb\s}_i^{(1)} \cdot {\pmb\s}_i^{(2)} +{\pmb\t}_i^{(1)}  \cdot  {\pmb\t}_i^{(2)}\right), \\
\nonumber H_d &=& J_d\sum_{n\neq n'} \sum_{i}\left({\pmb\s}_i^{(n)}\cdot {\pmb\t}_i^{(n')}+{\pmb\t}_i^{(n)}\cdot {\pmb\s}_{i+1}^{(n')}\right),
\end{eqnarray}
where $n$ and $n'$ label the legs, and the summations $\sum_i$ run over unit cells (see Fig. \ref{Fig:ladder}). Here, $J_l$ is the intra-leg exchange interaction between spins $\s$ and $\t$, and the other two couplings refer to the inter-leg interactions, $J_r$ on the rungs and $J_d$ across the diagonals.

The Hamiltonian (\ref{Eq:hamiltonian}) has several symmetries, including SU(2), time-reversal, ``leg-swap'', inversion and discrete translations. Some of them may be explicitly broken by
boundary conditions. Nevertheless, for long enough ladders these symmetries may be partially restored (well inside the ladder if a correlation length exists which is much smaller than the number of unit cells).

If some symmetry is spontaneously broken we can define appropriate order parameters. Important additional informations about the ground state can be gained from spin correlation functions. Those across the rungs are defined as

\begin{align}\label{eq:corr1}
S_{\sigma}(i):=\langle {\pmb\sigma}_i^{(1)}\cdot  {\pmb\sigma}_i^{(2)}\rangle\, ,\nonumber\\
S_{\tau}(i):=\langle {\pmb\tau}_i^{(1)}\cdot  {\pmb\tau}_i^{(2)}\rangle\, ,
\end{align}
where $\langle...\rangle$ denotes the expectation value with respect to the ground state. The ``columnar'' correlation functions (those along legs) are conveniently labeled by rung numbers $\ell$ (instead of cell numbers $i$),
\begin{align}\label{eq:corr2}
S_{ln}(\ell):=\left\{\begin{array}{ll}\langle {\pmb\sigma}_i^{(n)}\cdot  {\pmb\tau}_i^{(n)}\rangle\,
&\ell=2i-1\\
\langle {\pmb\tau}_i^{(n)}\cdot  {\pmb\sigma}_{i+1}^{(n)}\rangle\, ,&\ell=2i\,
\end{array}\, ,\right.
\end{align}
where $n=1,2$ numbers the legs and $\ell$ runs from 1 to $L=2N$. Similarly, the ``diagonal'' correlation functions are
\begin{align}\label{eq:corr3}
S_{d1}(\ell):=\left\{\begin{array}{ll}\langle {\pmb\sigma}_i^{(1)}\cdot  {\pmb\tau}_i^{(2)}\rangle\, &\ell=2i-1\\
\langle {\pmb\tau}_i^{(1)}\cdot  {\pmb\sigma}_{i+1}^{(2)}\rangle\, ,&\ell=2i\,
\end{array}\, ,\right.
\end{align}
and
\begin{align}
S_{d2}(\ell):=\left\{\begin{array}{ll}\langle {\pmb\sigma}_i^{(2)}\cdot  {\pmb\tau}_i^{(1)}\rangle\, &\ell=2i-1\\
\langle {\pmb\tau}_i^{(2)}\cdot  {\pmb\sigma}_{i+1}^{(1)}\rangle\, ,&\ell=2i\,
\end{array}\, .\right.
\end{align}

These correlation functions satisfy the inequalities
\begin{align}\label{eq:bounds}
-\frac{3}{4}\le S_{\sigma}(i)\le \frac{1}{4}\, ,&-2\le S_{\tau}(i)\le 1\, ,\nonumber\\
-1\le S_{an}(\ell) \le \frac{1}{2}\, ,&\quad a=l,d,\,\, n=1,2.
\end{align}
If the spatial symmetries are preserved in the ground state, the rung correlations are independent of the cell number $i$, and both columnar and diagonal correlations do not depend on the rung number $\ell$ (nor on $n$).
\section{Frustrated mixed-spin plaquette}\label{sec:plaquette}
It is instructive to consider first the building block of the mixed-spin ladder, consisting of four spins on the corners of a square, two with spin 1 and two spin $\frac{1}{2}$. Eigenstates of both the total spin and the Hamiltonian can be deduced analytically.
The results may serve as a starting point for the construction of effective low-energy Hamiltonians.

We consider the general case where the rung couplings can be different ($J_r\rightarrow J_\sigma,J_\tau$). The Hamiltonian
\begin{align}\label{eq:plaquette}
H&=J_\sigma\boldsymbol{\sigma}_1\cdot \boldsymbol{\sigma}_2
+J_\tau{\boldsymbol\tau}_1\cdot \boldsymbol{\tau}_2
+J_l\big({\boldsymbol\tau}_1\cdot \boldsymbol{\sigma}_1
+\boldsymbol{\tau}_2\cdot \boldsymbol{\sigma}_2\big)\nonumber\\
&+J_d\big({\boldsymbol\tau}_1\cdot \boldsymbol{\sigma}_2
+\boldsymbol{\tau}_2\cdot \boldsymbol{\sigma}_1\big)
\end{align}
commutes with the total spin
\begin{align}
{\bf S}=\boldsymbol{\tau}_1+\boldsymbol{\tau}_2+\boldsymbol{\sigma}_1+\boldsymbol{\sigma}_2
\end{align}
and is invariant with respect to an interchange of spin operators, $\boldsymbol{\tau}_1\leftrightarrow\boldsymbol{\tau}_2$, $\boldsymbol{\sigma}_1\leftrightarrow\boldsymbol{\sigma}_2$
(``leg-swap symmetry'').

We classify the states of the 36-dimensional Hilbert space according to the eigenstates of ${\bf S}^2$, $S_z$ and the leg-swap operation. This basis is constructed explicitly in Appendix \ref{app:plaquette}. For $S_z=0$ we obtain 5 states which are even under the leg-swap operation (one with $S=3$, one with $S=2$, three with $S=1$) and 5 states which are odd (two with $S=2$, one with $S=1$, two with $S=0$). The Hamiltonian is block-diagonal in this basis, with a $3\times 3$ matrix as largest block.

The ground state symmetry depends on the coupling constants. If all couplings are  equal, $J_l=J_d=J_\sigma=J_\tau=J$, the Hamiltonian is directly related to the total spin,
\begin{align}\label{eq:symmetric}
H=\frac{J}{2}\Big({\bf S}^2-\frac{11}{2}\Big)\, ,
\end{align}
and the ground state has $S=3$ for $J<0$ and $S=0$ for $J>0$.

\begin{figure}
\centering
\includegraphics[width=8cm]{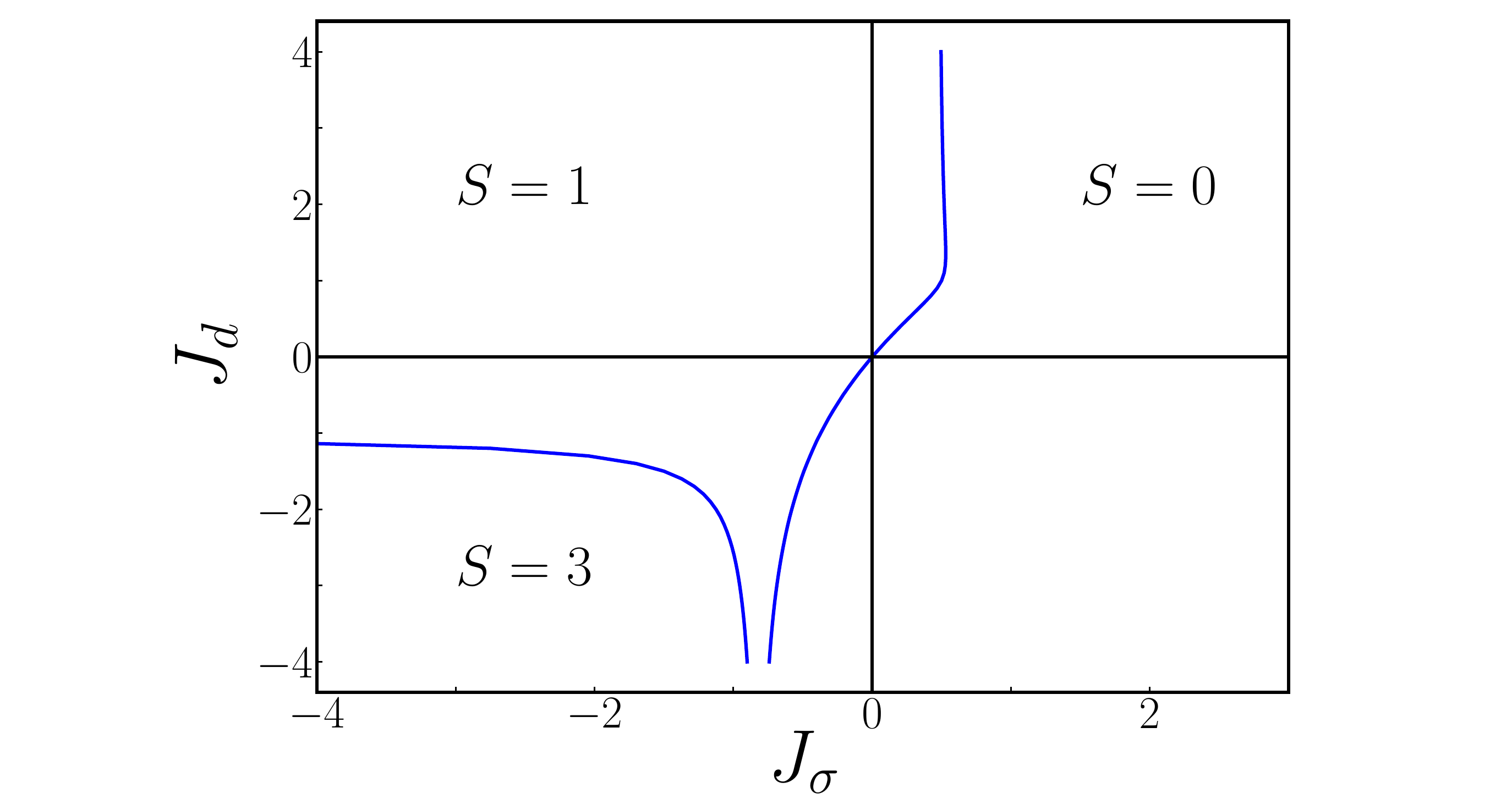}
\caption{Ground state phase diagram of the mixed-spin plaquette for $J_l=1$, $J_\sigma=J_ \tau$ in the
$J_\sigma-J_d$ plain.}
\label{fig:PD}
\end{figure}

\begin{figure}
\centering
\includegraphics[width=8cm]{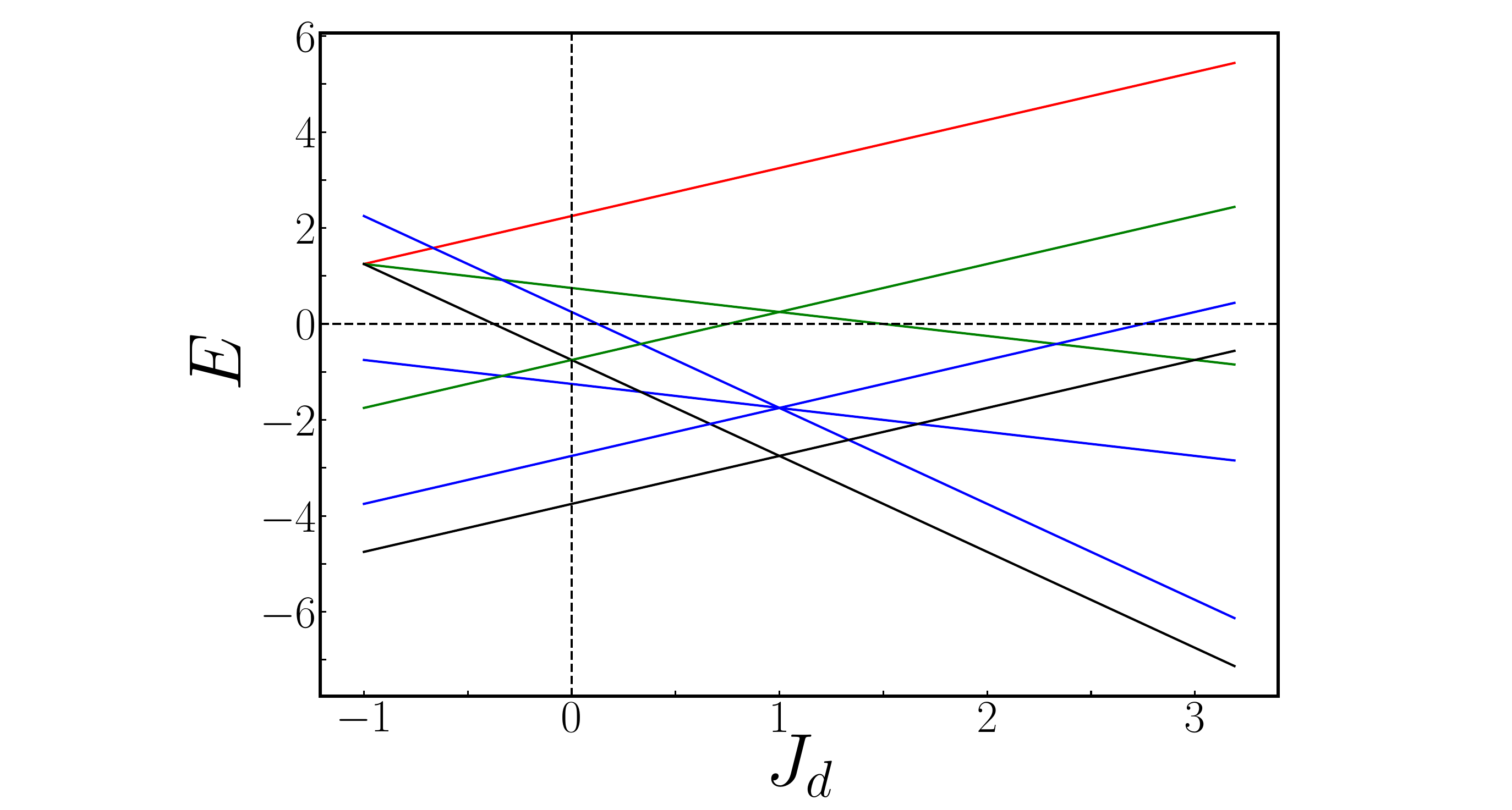}
\caption{Energy spectrum for the plaquette for coupling parameters $J_l=J_\sigma=J_\tau=1$. Colours represent different spin values, $S=3$ (red), $S=2$ (green), $S=1$ (blue), $S=0$ (black). Doubly degenerate eigenvalues occur for $S=1$ (middle line) and $S=2$ (downgoing line).}
\label{fig:EV}
\end{figure}

The diagonalization of the Hamiltonian (Appendix \ref{app:plaquette}) yields the phase diagram of Fig. \ref{fig:PD}. In this paper we concentrate ourselves mostly on the parameter region $J_l=J_\sigma=J_\tau=1$. Fig. \ref{fig:EV} shows the energy spectrum for this case. Remarkably, all eigenvalues are linear functions of $J_d$. At the fully symmetric point the lines collapse to the values predicted by Eq. (\ref{eq:symmetric}). The ground state is always a spin singlet, but of different character above and below $J_d=1$, where the singlet levels cross. It is worthwhile to add that for generic parameter sets the two singlet levels repel each other and are separated by a gap.

The spin correlation functions defined by Eqs. (\ref{eq:corr1}) to (\ref{eq:corr3}) provide valuable insight into the character of the eigenstates. They are readily evaluated using the Hellman-Feynman theorem. As a simple example we first consider the eigenstate with $S=3$, which has an energy $J_l+J_d+\frac{1}{4}J_\sigma+J_\tau$ and therefore correlation functions $S_l=S_d=\frac{1}{2}$, $S_\sigma=\frac{1}{4}$, $S_\tau=1$. These are just the upper bounds of Eq. (\ref{eq:bounds}). We now turn to the ground state. The correlation functions are calculated using the relations
\begin{align}
S_a=\frac{1}{2}\frac{\partial E_-}{\partial J_a},\, a=l,d,\qquad
S_b=\frac{\partial E_-}{\partial J_b},\, b=\sigma,\tau,
\end{align}
where $E_-$ is given by Eq. (\ref{eq:EV_singlet}).
The result for the parameter region of Fig. \ref{fig:EV} is presented in Table \ref{tab:correlations}. Remarkably, the correlation functions do not depend on $J_d$, except at the crossing point where they exhibit steps. Moreover, the ``rung spins'' have the same ``lengths'',
\begin{align}
\langle( \boldsymbol{\sigma}_1+ \boldsymbol{\sigma}_2)^2\rangle
=\langle( \boldsymbol{\tau}_1+ \boldsymbol{\tau}_2)^2\rangle=\left\{
\begin{array}{ll}\frac{2}{3},&J_d<1\\ \frac{4}{3},&J_d>1
\end{array}
\right.
\end{align}
A glance at Eqs. (\ref{eq:EV_singlet}) reveals that this equality is generally valid for the two singlet states. It has also been found for the ladder in Ref. \onlinecite{PhysRevB.81.024409}  in the RS phase and interpreted as a signature of ``zero weight of rung quintet states''.

\begin{table}
\centering
$
\begin{array}{c|c|c}
&J_d<1&J_d>1\\
\hline
S_l&-5/6&1/3\\
S_d&1/2&-1\\
S_\sigma&-5/12&-1/12\\
S_\tau&-5/3&-4/3
\end{array}
$
\caption{Spin correlations as functions of $J_d$ for $J_l=J_\sigma=J_\tau=1$.}
\label{tab:correlations}
\end{table}

It is worthwhile to mention the homogeneous case where both $\sigma$- and $\tau$-operators have spin $\frac{1}{2}$. There are six eigenstates of ${\bf S}^2$ for $S_z=0$,
four of which are even (one with $S=2$, one with $S=1$, two with $S=0$) and two are odd (with $S=1$). For $J_l=J_\sigma=J_\tau=1$ the energy eigenvalues are again linear functions of $J_d$. The ground state is a singlet with different spin correlations below and above the symmetric point $J_d=1$, as in the mixed-spin case.

\section{Weak and strong rung couplings}\label{sec:PT}
We return now to the mixed-spin ladder. In this Section we investigate the limits of weak and strong rung couplings, which can be understood analytically. We limit ourselves to the unfrustrated ladder, $J_d=0$, and consider three special cases, weak antiferromagnetic rung coupling, ferromagnetic rung coupling and strong antiferromagnetic rung coupling. Our mixed-spin ladder is invariant under the exchange of leg and diagonal couplings, \cite{PhysRevB.57.11439} i.e., $H\left(J_l,J_r,J_d \right)=H\left(J_d, J_r,J_l\right)$, therefore the entire discussion below can be applied to the case of $J_l=0$ and $J_d>0$, by replacing $J_l$ by $J_d$.

\subsection{Weak antiferromagnetic rung coupling}
In the limit of vanishing rung coupling, the ladder is decoupled into two equivalent mixed-spin (1,1/2) chains. According to the Lieb-Mattis theorem\cite{Lieb_62}, each chain has a total spin $S_{tot}=N/2$, where $N$ is the number of unit cells, and thus exhibits ferrimagnetic long-range order.
Since the elementary cell (of a chain) consists of two spins, linear spin wave theory yields two types of magnons, a gapless ``acoustic'' branch with dispersion $\omega_k^-/J_l=-\frac 12+(\frac 14+2\sin^2k)^{1/2}$ ($\sim k^2$ for small $k$), and a gapped ``optical'' branch with dispersion $\omega_k^+/J_l=\frac 12+(\frac 14+2\sin^2k)^{1/2}$.\cite{Brehmer_1997,PhysRevB.57.13610,Yamamoto2000}
When an antiferromagnetic rung coupling is switched on, spin wave theory predicts a linear dispersion of the gapless mode, reflecting the
antiferromagnetic character of the ladder system, whereas
the optical mode moves upward.\cite{PhysRevB.64.134408} However, DMRG calculations show that a spin gap opens, which first increases quadratically as a function of $J_r$ up to $J_r\sim 0.3$, and then grows linearly.\cite{PhysRevB.64.134408}

\subsection{Ferromagnetic rung coupling}
For $J_r<0$ and $J_l=0$, the $\s$ spins form rung triplets and the $\t$ spins form rung quintets, therefore the ground state is a product of rung-triplet and rung-quintet states. Low-energy excited states are rung singlets for $\s$ spins and rung triplets for $\t$ spins, separated from the ground state by energy gaps of $J_r$ and $2J_r$, respectively.  As soon as leg couplings are switched on, when $J_l\ll|J_r|$, the ladder behaves like a ferrimagnetic spin (1, 2) chain, with long-range order, a total spin $S_{tot}=N$, an acoustic mode and an optical mode (at $\omega_0^+=2J_l$). In the opposite limit of weak ferromagnetic rung coupling ($\vert J_r\vert\ll J_l$), the magnetic moments of the two chains are aligned, giving again $S_{tot}=N$, while the low-energy excitation spectrum remains essentially that of two independent chains, with acoustic and optical modes as described above. Hence we expect the two limits of weak and strong ferromagnetic rung couplings to be smoothly connected.

\subsection{Strong antiferromagnetic rung coupling}
For $J_r>0$ and $J_l=J_d=0$, the ground state is a product of local rung singlets, with an energy (per unit cell) of
$-(11/4)J_r$. The first excited states are triplets, separated from the ground state by a finite energy gap of size $J_r$. With increasing leg coupling $J_l$ the energy gap decreases monotonically. Appendix \ref{app:expansion} shows that first-order perturbation theory in $J_l$ does not give any contribution from quintets ($S=2$) on the rungs hosting $\t$-spins. This implies that quintets are not involved in the ground-state energy up to second order. We have also checked that the expectation values of $\s_i$ and $\t_i$ vanish, in agreement with the Lieb-Mattis theorem. Quintet states appear in second-order perturbation theory for the ground state and are thus expected to play a role for $J_l$ of the order of $J_r$.

\section{Numerical methods}\label{sec:TEBD}
\subsection{Matrix product states}
Matrix product states (MPSs)\cite{PhysRevLett.75.3537,RevModPhys.93.045003} provide an efficient representation of the ground state of one-dimensional systems obeying the area law, for which the entanglement entropy grows with the boundary of a specific area rather than its volume.\cite{RevModPhys.82.277}
To use MPSs for our model, we map the ladder onto a chain and consider each rung as a supersite with a larger Hilbert space. In Vidal's representation\cite{PhysRevLett.91.147902} a generic state of a one-dimensional system is described in terms of two sets of matrices.
\begin{figure}[h!]
	\centering
	\includegraphics[scale=0.8]{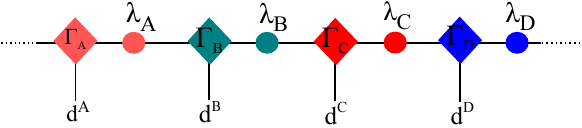}
	\caption{(Color online) MPSs representation for the mixed-spin (1,1/2) ladder. The rungs with spins $\s=1/2$ ($\t=1$) are indexed by A and C (B and D). A spin-$1/2$ (1) rung has a local Hilbert space with dimension $d^A=d^C=4$ $(d^B=d^D=9)$.}
	\label{fig:ladderMPS}
\end{figure}
We choose four pairs of matrices
$(\Gamma_A, \Gamma_B,\Gamma_C, \Gamma_D)$ and $(\lambda_A , \lambda_B, \lambda_C, \lambda_D)$, which allow us to include phases showing a doubling of the unit cell, such as a dimer solid or the ferrimagnetic state $[2,1,1,1,...,2,1,1,1]$\cite{PhysRevB.81.024409}.
In this representation, illustrated in Fig. \ref{fig:ladderMPS}, an arbitrary state of our ladder is
\begin{eqnarray}\label{Eq:MPS}
\nonumber\left| \psi\right\rangle  &=& \sum_{i_1,\cdots,i_N} [\cdots \Gamma_A^{i_m}\lambda_A\Gamma_B^{i_{m+1}}\lambda_B\\
&&\Gamma_C^{i_{m+2}}\lambda_C\Gamma_D^{i_{m+3}}\lambda_D \cdots ]\, \vert i_1\rangle\otimes...\otimes\vert i_N\rangle,
\end{eqnarray}
where $i_n$ numbers the states of rung $n$ ($i_n=1,...,4$ for $\s$-rungs and $i_n=1,...,9$ for $\t$-rungs).
The $\lambda$'s are diagonal matrices with the non-negative ``Schmidt coefficients'' $\lambda_i$ on the diagonal, and $\lambda_i^2$ are the eigenvalues of the reduced density matrix ($\rho_{red}=Tr_{\mathcal{B}(\mathcal{A})}\left| \psi\right\rangle \left\langle \psi\right|$, where $\mathcal{A}$ and $\mathcal{B}$ are two halves of the ladder). The EE is directly connected to these eigenvalues thorough
\begin{equation}
S= -\sum_{i}\lambda_i^ 2 \log \lambda_i^2.
\end{equation}
The matrices $\Gamma$ in (\ref{Eq:MPS}) correspond to transformations between different Schmidt bases. The dimension $\chi$ of the matrices is a key parameter in tensor network states, called bond dimension, and the accuracy of the state (\ref{Eq:MPS}) is controlled by this quantity.
For weakly entangled states a fairly small bond dimension is sufficient to obtain sensible results. But there are situations where very large bond dimensions are required, such as gapless critical systems\cite{Rubio_22}.
\subsection{Time-evolving block decimation}
One of the efficient MPS-based algorithms for simulating one-dimensional quantum many-body systems is the iTEBD technique.\cite{PhysRevLett.91.147902,PhysRevLett.93.040502,PhysRevLett.93.076401,1742-5468-2004-04-P04005,PhysRevB.78.155117} In iTEBD, using the imaginary time evolution of a quantum state we can find the ground state of the Hamiltonian $H$ through the relation
\begin{equation}\label{eq:evolution}
\left|GS\right\rangle = \lim _{\beta\rightarrow  \infty } \exp(-\beta H)  \left| \psi_0\right\rangle,
\end{equation}
where $\left| \psi_0\right\rangle$ is an initial ``guess state'', chosen in the form of Eq. (\ref{Eq:MPS}). If $\vert\psi_0\rangle$ is orthogonal to $\vert GS\rangle$, for instance because it has a different symmetry, the ground state cannot be reached by this method. Practical difficulties may appear in the vicinity of a continuous phase transition, where a judicious choice of the guess state is of crucial importance. However, for a first order transition between two gapped phases, such as that between RS and H phases, the method is expected to work well.

To proceed, we first rewrite the Hamiltonian (\ref{Eq:hamiltonian}) as
\begin{align}
H=\sum_i\left(h_i^{(\s)}+h_i^{(\t)}\right),
\end{align}
where
\begin{align}
h_i^{(\s)}&=J_r{\pmb\s}_i^{(1)}\cdot{\pmb\s}_i^{(2)}+J_l\sum_n{\pmb\s}_i^{(n)}\cdot {\pmb\t}_i^{(n)}\nonumber\\
&+J_d\sum_{n\neq n'}{\pmb\s}_i^{(n)}\cdot {\pmb\t}_i^{(n')},\nonumber\\
h_i^{(\t)}&=J_r{\pmb\t}_i^{(1)}\cdot {\pmb\t}_i^{(2)}+J_l\sum_n {\pmb\t}_i^{(n)}\cdot {\pmb\s}_{i+1}^{(n)}\nonumber\\
&+J_d\sum_{n\neq n'}{\pmb\t}_i^{(n)}\cdot {\pmb\s}_{i+1}^{(n')}\, .
\end{align}
If the ``time'' $\beta$ is divided into a large number of intervals of width $\delta$ one can use the (first-order) Suzuki-Trotter decomposition
\begin{align}
e^{-\delta \, H}\approx\prod_ie^{-\delta \, h_i^{(\sigma)}}e^{-\delta \, h_i^{(\tau)}}\, .
\end{align}
Starting with an initial guess state of the form (\ref{Eq:MPS}), we apply the operator $e^{-\delta \, H}$ iteratively to update the matrix product representation,
until the ground state energy or the entropy converges.
In our ladder system, we use the second-order Suzuki-Trotter decomposition. The simulations are started with a time step $\delta  = 0.5$, which gradually is decreased to
$\delta  =10^{-5}$.
A key feature of the iTEBD algorithm is that it directly treats the infinite system by exploiting translational invariance, therefore it is free of finite-size effects.
\subsection{DMRG}
We also use the DMRG technique, especially to study the ground state in the vicinity of transition points.
Unlike iTEBD, DMRG is a variational method. In other respects, the two methods have many steps in common. We iteratively optimize the MPSs of two
neighboring sites to minimize the ground state energy, and then project the Hamiltonian onto a variational space. We use an iterative algorithm such as Lanczos to lower the energy.
The two-site update is repeated for each pair of neighboring sites until the wave function converges to the ground state.
\section{Unfrustrated ladder}\label{sec:ladderA}
In the absence of diagonal interactions, our mixed-spin ladder (\ref{Eq:hamiltonian}) is unfrustrated and, as discussed in Section \ref{sec:PT}, its ground state is ferrimagnetically long-range ordered in the limit of strong ferromagnetic rung couplings and magnetically disordered in the limit of strong antiferromagnetic rung couplings. In order to obtain the complete ground state phase diagram, we introduce the dimensionless parameter $R=J_r/(J_l+|J_r|)$. The limits $R=\pm1$  correspond to the strong rung coupling regimes, whereas $R=0$ is the limiting case of two decoupled mixed-spin chains.
\subsection{iTEBD}
\begin{figure}[t]
	\centering
	\includegraphics[scale=0.2]{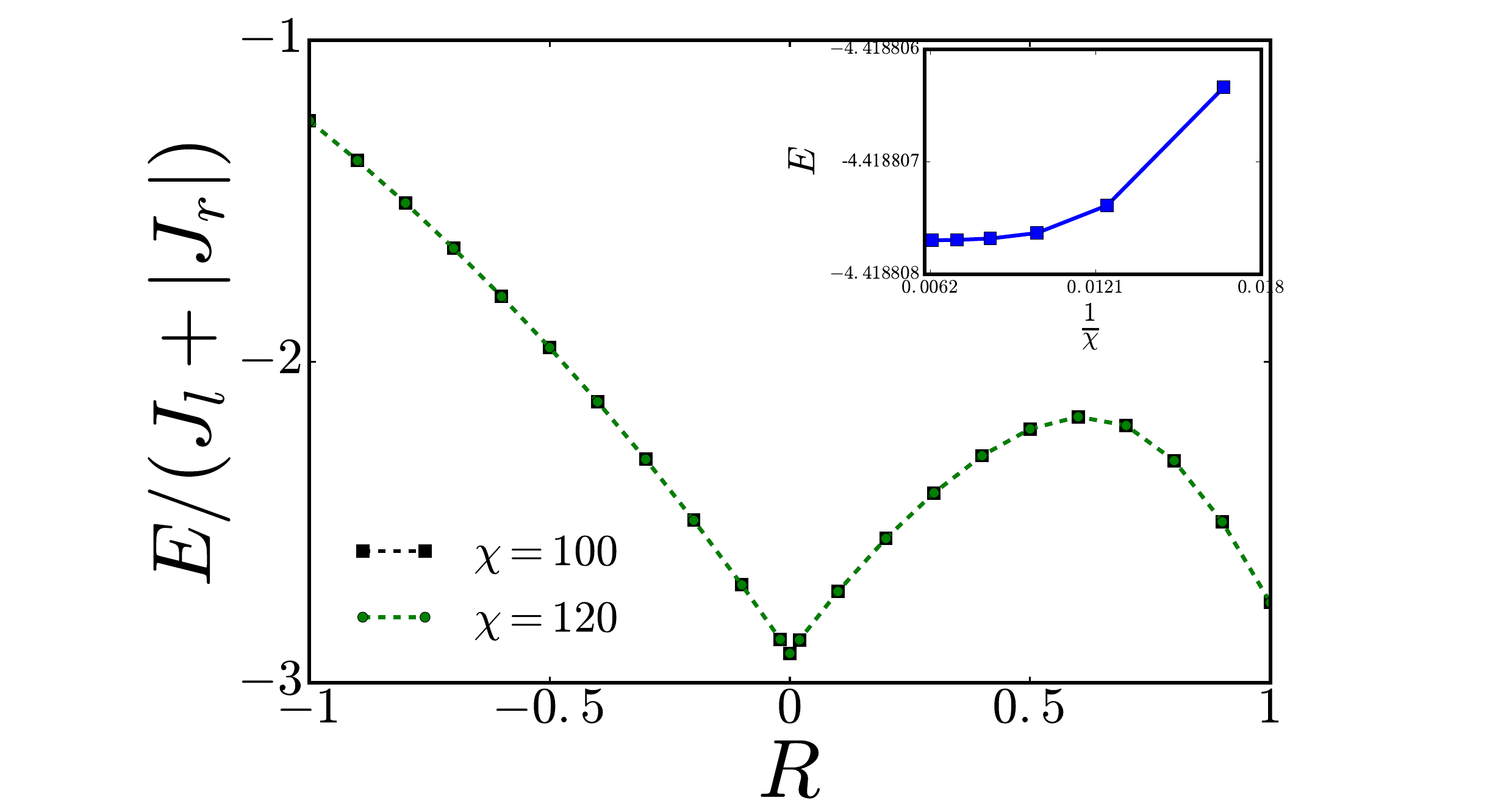}
	\caption{(Color online) Scaled ground state energy of the unfrustrated ladder ($J_d=0$) per unit cell as a function of $R$, obtained by iTEBD with $\chi=100$ and $120$. The curves tend to the exactly known limits for $R\rightarrow\pm1$, namely to $-5/4$ for $R\rightarrow -1$, the total energy of a spin-1/2 rung-triplet and a spin-1 rung-quintet, and to $-11/4$ for $R\rightarrow +1$, the total energy of a spin-1/2 and  a spin-1 rung-singlet, whereas for $R=0$ the energy of a mixed-spin (1,1/2) chain is reproduced.	
	The inset depicts the energy as a function of bond dimension for $R=0.5$.  }
	\label{fig:Ferri_Energy}
\end{figure}
We first discuss results obtained by the iTEBD method. The ground state energy is shown in Fig. \ref{fig:Ferri_Energy} for two different bond dimensions, $\chi=100$ and 120. The results for the two cases are almost the same. The inset of Fig. \ref{fig:Ferri_Energy} confirms that the energy has well converged for $\chi=120$. The cusp at $R=0$ points to a phase transition of first order. Some insight on the two sides of the transition can be gained by calculating the magnetic moments on the rungs.
We find in the entire range $-1\le R<0$ a non-vanishing magnetization and a ground state in the sector $[1,2,...,1,2]$, where numbers stand for spin 1 on the $\s$-rungs and spin 2 on the $\t$-rungs. We refer to this phase as F$_1$, in agreement with Ref. \onlinecite{PhysRevB.81.024409}.

For $R=0$ the ladder is decoupled into two equivalent mixed-spin-(1,1/2) chains with ferrimagnetically ordered ground states in the sector $[1/2,1,\dots, 1/2,1]$. For $R>0$ the ground state must be a spin singlet, because of the Lieb-Mattis theorem, applied to the case of Fig. \ref{Fig:sublattices}a. For $R\gtrsim 0.1$ we do find a non-magnetic ground state, but for very small positive $R$ we routinely obtain finite local moments for a randomly chosen initial state. This is understandable because for $R=0$ the ground state is infinitely degenerate (for infinite chains), which implies a huge density of states for low-energy excitations at very small $R$. Large values of both $\beta$ and $\chi$ would therefore be required to reach a faithful ground state. We have verified that the contentious region indeed shrinks if the bond dimension is increased. However, for very small positive values of $R$ a bond dimension $\chi\gg 120$ is required to obtain satisfactory results, even if a non-magnetic initial state is chosen.

\subsection{DMRG}
\begin{figure}[t]
	\centering
	\includegraphics[scale=0.2]{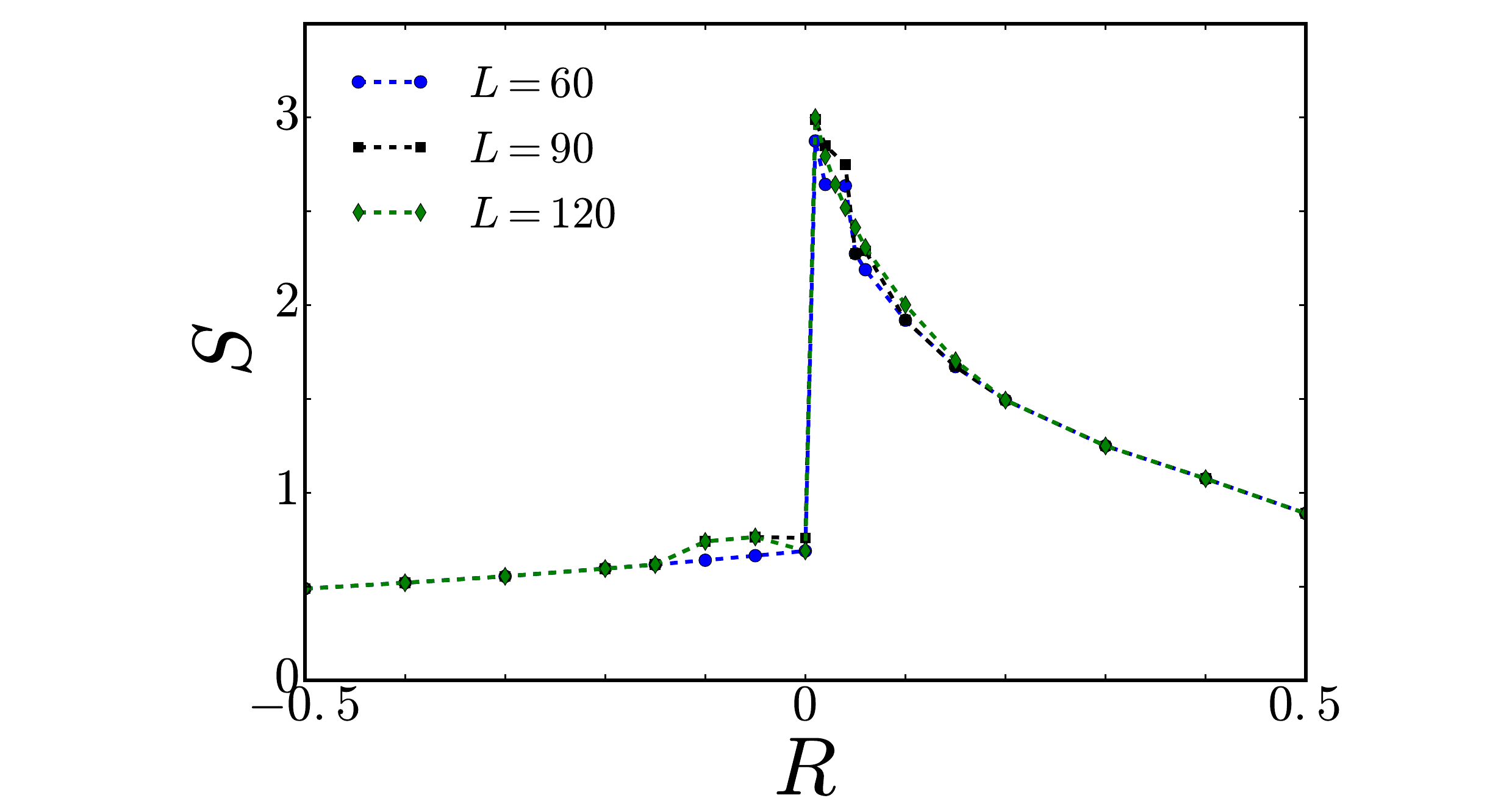}
	\caption{(Color online ) EE for the unfrustrated ladder ($J_d=0$) versus $R$, obtained by DMRG for 60, 90, and 120 rungs, with $\chi$ increasing from 300 to 1000. The discontinuity at $R=0$ indicates a phase transition of first order. The EE vanishes at the two end points, $R=\pm 1$.}
	\label{Fig:EE1}
\end{figure}
\begin{figure}[t]
	\centering
	\includegraphics[scale=0.2]{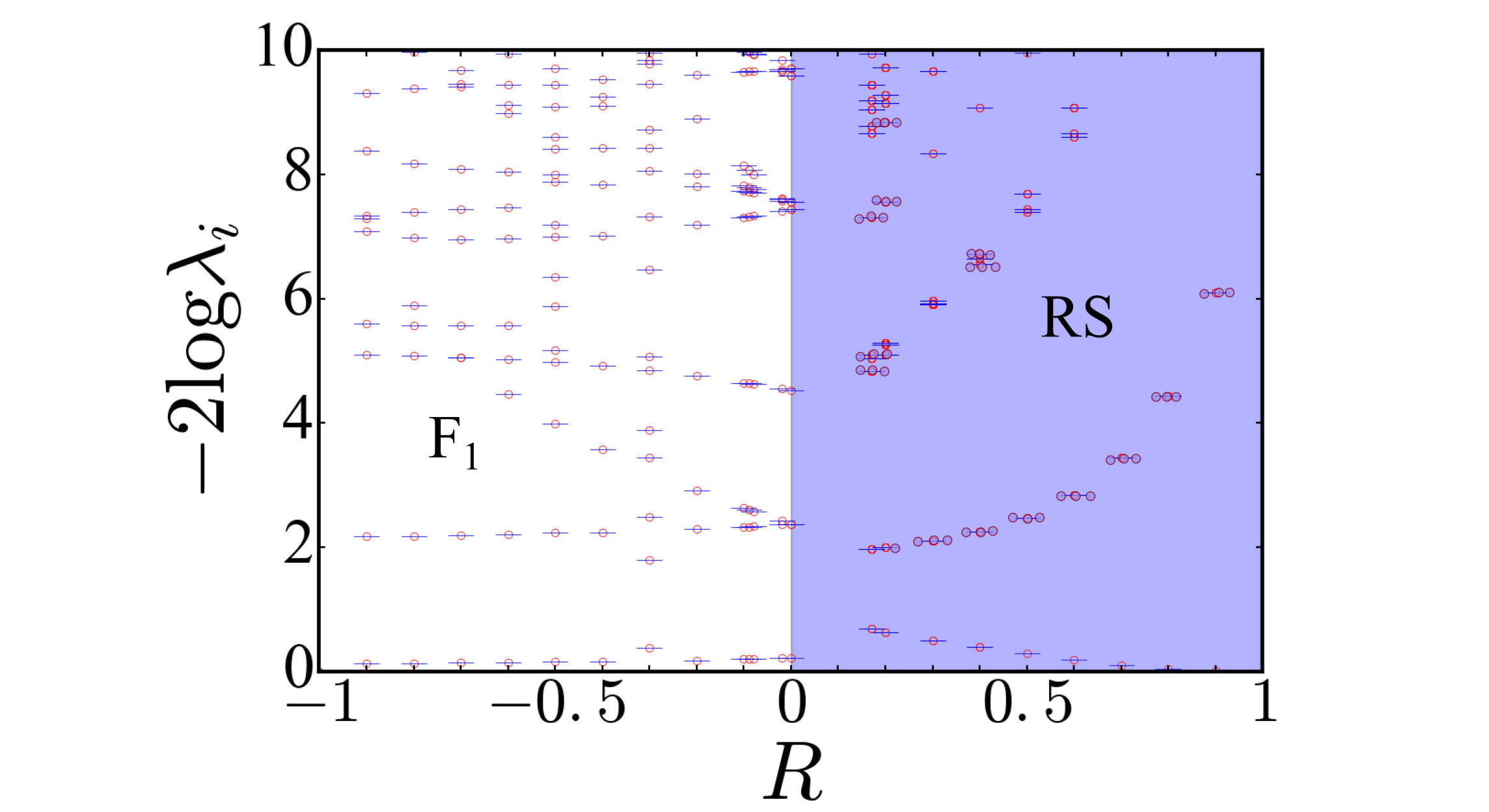}
	\caption{(Color online) Low-lying ES levels and their degeneracies versus $R$, obtained by DMRG with $\chi=800$ for an unfrustrated ladder ($J_d=0$) of 120 rungs. Circles indicate the degeneracy. The RS phase is characterized by odd degeneracies, and the F$_1$ phase has no degenerate ES. }
	\label{Fig:DMRGES1}
\end{figure}

We now discuss results obtained with DMRG, which is less sensitive to initial guess states than iTEBD.
Computations were performed for ladders of different sizes with open boundary conditions. For greater efficiency we increased the bond dimension to 1000.
The ground state energies obtained with DMRG (after extrapolation) match perfectly those calculated with iTEBD, except in the tiny region of $R$ discussed above, where the DMRG values are in general smaller.

The EE is obtained by cutting the ladder into two halves, and tracing out the degrees of freedom of one of the halves.
The results presented in Fig. \ref{Fig:EE1} exhibit a discontinuity at $R=0$ where a phase transition of first order occurs from the ferrimagnetically ordered F$_1$ phase to the RS phase (where all magnetic order parameters are zero).

The degeneracy of the ES levels offers a versatile tool both for identifying different phases and for locating quantum phase transitions
\cite{PhysRevB.81.064439,PhysRevB.85.075125,PhysRevB.84.125110,PhysRevB.94.165167, Haghshenas:2014}.
As seen in Fig. \ref{Fig:DMRGES1}, the level degeneracy of the ES is odd (1 or 3) in the RS phase ($0<R\leq 1$), while in the F$_1$ phase ($-1< R \leq 0$)  the ES levels are nondegenerate.
\section{Frustrated ladder}\label{sec:ladderB}

We now consider the case where all three exchange couplings are finite. Some regions in parameter space can be understood without detailed calculations, especially those with no or small frustration. Thus for $J_d=J_l$ our ladder model can be mapped onto a generalized mixed-spin Heisenberg chain. Its ground state can be guessed in a simple way both for $J_r<0$ and for $0<J_r\ll J_l$. Regions with strong frustration, such as those where the three coupling parameters are all positive and of a similar size, are of course difficult to handle using qualitative arguments. We therefore resort to numerical calculations, limiting ourselves to the region $J_l=J_r=1$, $J_d>0$.
\subsection{Generalized Heisenberg chain}
For $J_d=J_l$ it is convenient to introduce rung spins
\begin{align}
{\bf S}_i:={\pmb\s}_i^{(1)}+{\pmb\s}_i^{(2)},\qquad {\bf T}_i:={\pmb\t}_i^{(1)}+{\pmb\t}_i^{(2)},
\end{align}
because in this case the Hamiltonian can be written in terms of these operators,
\begin{equation}
{\cal H} =\frac{J_r}{2}\sum_i({\bf S}_i^2+{\bf T}_i^2)+J_l\sum_i(S_i\cdot T_i+T_i\cdot S_{i+1}),
 \end{equation}
where we have neglected an additive constant. The rung spin operators do not have a fixed ``length'', but can assume the values $S_i=0,1$, $T_i=0,1,2$. For $J_r<0$ the first term has the lowest eigenvalue if both $S_i$ and $T_i$ are as large as possible, i.e., $S_i=1$ and $T_i=2$. But this is also true for the second term (it is obvious for classical spins or for the N\'eel state). Therefore the ground state is expected to be ferrimagnetic (F$_1$) if $J_l<0$ (and $J_l=J_d>0$). For positive $J_r$ the first term in the Hamiltonian favors singlet rung spins, in contrast to the second term which is lowest for maximal rung spins. The system is frustrated. However, for very small positive values of $J_r$ the first term can be neglected and we obtain again a ferrimagnetic phase of type F$_1$.

Our numerical analysis of nearest-neighbor-spin correlations on the rungs imply that $\langle{\bf T}_i^2\rangle\approx\langle{\bf S}_i^2\rangle$ for $0 \leq J_d \lesssim0.86$,
which implies that the rung-quintet states on the spin-1 rungs have negligible weight and can be projected out. This lends support to the mapping of the frustrated mixed-spin (1, 1/2) ladder onto a frustrated uniform spin-1/2 ladder, as proposed in Ref. \onlinecite{PhysRevB.81.024409} .

\subsection{iTEBD}
 \begin{figure}[t]
	\centering
	\includegraphics[scale=0.2]{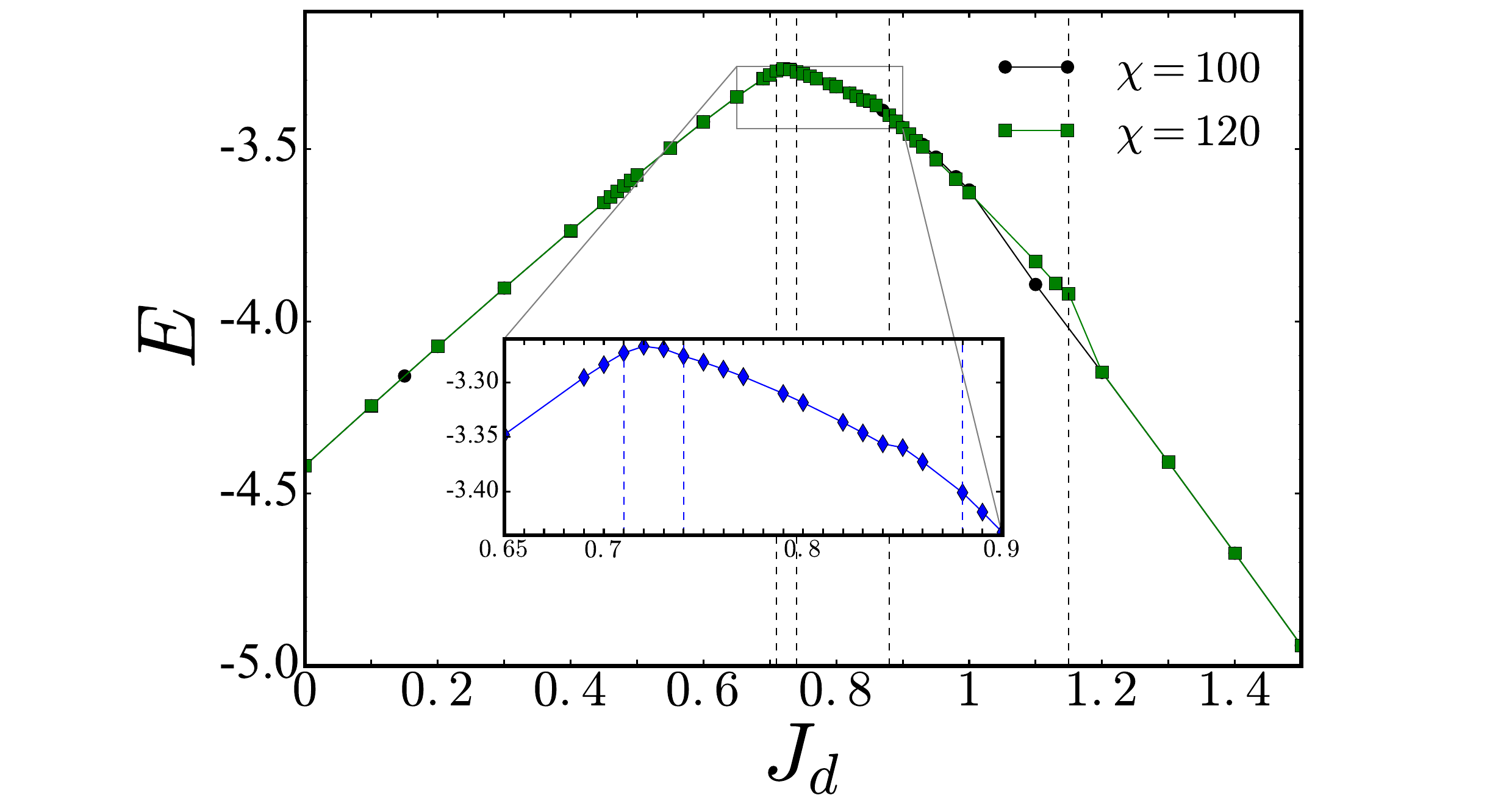}
	\caption{(Color online) Ground state energy per unit cell versus $J_d$ for $J_l=J_r=1$, obtained with iTEBD for $\chi=100$ and 120. Inset: Enlarged view for $0.65<J_d<0.9$.
}
	\label{Fig:diagonal_energy}
\end{figure}

We have studied numerically the ground state energy, both the EE and the Schmidt gap (the difference between the two largest eigenvalues of the reduced density matrix\cite{PhysRevLett.109.237208}), as well as the level degeneracy of the ES. We first present the overall picture obtained with iTEBD. More details will be given in Section \ref{sec:Intermediate} for DMRG results, in particular regarding the intermediate phase.

Fig. \ref{Fig:diagonal_energy} shows the ground-state energy as a function of $J_d$. Cusps at certain values of
$J_d$ (indicated by the dashed vertical lines), are indicative for phase transitions. Corresponding jumps in the EE
are more pronounced, and found to occur at $J_d\simeq 0.71, 0.74, 0.86$ and $1.17$.  The gross features of the ground state energy -- an almost linear increase for $J_d\lesssim 0.7$ and a linear decrease for $J_d\gtrsim 0.7$ -- match the behavior found for the plaquette (Fig. \ref{fig:EV}). The different values of the ``critical points'' can be attributed to the fact that in the ladder a site is connected to two neighbors by $J_d$ and only to one in the plaquette. The situation is reminiscent of that encountered for the antiferromagnetic XX chain in a transverse field $h$, where the transition to the ferromagnetic state occurs at $J=2h$ for two sites, but already at $J=h$ for the chain. An additional similarity between the plaquette and the ladder is the vanishing of the excitation gap at criticality, observed as a level crossing in the case of the plaquette and by numerical evidence in the case of the ladder (for the same specific coupling parameters as used here)\cite{PhysRevB.81.024409}.

We have also studied both the ES and local order parameters in the different phases. No magnetic order was found both for $0\le J_d<0.71$ and for $0.74<J_d\le 0.86$. The low-lying levels of the ES are odd-degenerate for $0\le J_d<0.71$ (as in the RS phase \cite{PhysRevB.81.064439,PhysRevB.85.075125}), and even-degenerate for $0.74<J_d\le 0.86$ (as in the H phase).
For $J_d>0.86$ the ES is non-degenerate and magnetic order does exist. A sharp first-order transition at $J_d\approx 1.17$ separates two different magnetic phases, the F$_1$ phase corresponding to the sector $\left[ 1,2,\cdots,1,2\right]$ for $J_d\gtrsim 1.17$, and the F$_2$ phase corresponding to the sector $\left[ 1,1,1,2,\cdots,1,1,1,2\right]$ for $0.86<J_d\lesssim 1.17$. The F$_2$ phase breaks the translational symmetry and can be considered an intermediate phase between the $H$ and F$_1$ phases. This picture agrees with the phase diagram of Ref. \onlinecite{PhysRevB.81.024409}, although the transition point between F$_1$ and F$_2$ phases is somewhat different.

In the narrow interval $0.71 \lesssim J_d < 0.74$, the ES levels are not found to exhibit any characteristic degeneracy, neither of the $RS$-type nor of the $H$-type, and we often detect magnetic order. We attribute the erratic data to limitations of the method. In fact, the spin gap seems to be very small in this region \cite{PhysRevB.81.024409}, therefore one would need both a large parameter
$\beta$ and a large bond dimension $\chi$ to obtain consistent results, very much like in the unfrustrated case for
$J_d=0,\, \vert J_r\vert\ll J_l$. Moreover, in our ansatz we included a possible period doubling, which allowed us to reproduce the F$_2$ phase, but excluded ground states with longer periods, for which we find good evidence on the basis of DMRG data (to be discussed below).

\section{Intermediate singlet phase}\label{sec:Intermediate}

As discussed above, for $0<J_d\lesssim 0.86$ the ground state of our model appears to be well represented by that of a frustrated spin-1/2 ladder (with couplings $J'_l,J'_r,J'_d$)\cite{PhysRevB.81.024409}, which has been intensively investigated using both analytical and numerical methods\cite{PhysRevB.73.214427, PhysRevB.77.214418,PhysRevB.81.064432}. An intermediate columnar-dimer phase has been reported only in a narrow neighborhood of $J'_{r}=0.38$ (for $J'_l=1$ and $J'_d=0.2$)\cite{PhysRevB.77.214418}, i.e., in a ``weak coupling'' region in parameter space which does not correspond to the ``strong coupling'' region of our intermediate phase. This apparent discrepancy is less serious if one keeps in mind that the low-energy excitations are quite different in the two models (the spin gap remains finite at the RS-H transition in the spin-1/2 ladder\cite{PhysRevB.57.11439}, in contrast to the softening observed in the mixed-spin ladder\cite{PhysRevB.81.024409}). Therefore the (approximate) mapping between the ground states of the two models cannot be used to rule out a strong-coupling intermediate phase in the mixed-spin ladder.

We now discuss DMRG data for the frustrated mixed-spin ladder, focussing on the parameter region $J_l=J_r=1$, $0.4<J_d<1$. All magnetic order parameters  $\la{\pmb\t}_i^{(n)}\ra$
and $\la{\pmb\s}_i^{(n)}\ra$ vanish for $J_d\lesssim 0.86$, and the existence of an intermediate phase is confirmed on the basis of spatially modulated spin correlations.
\subsection{EE, Schmidt gap and ES}

\begin{figure}[t]
	\centering
	\includegraphics[scale=0.2]{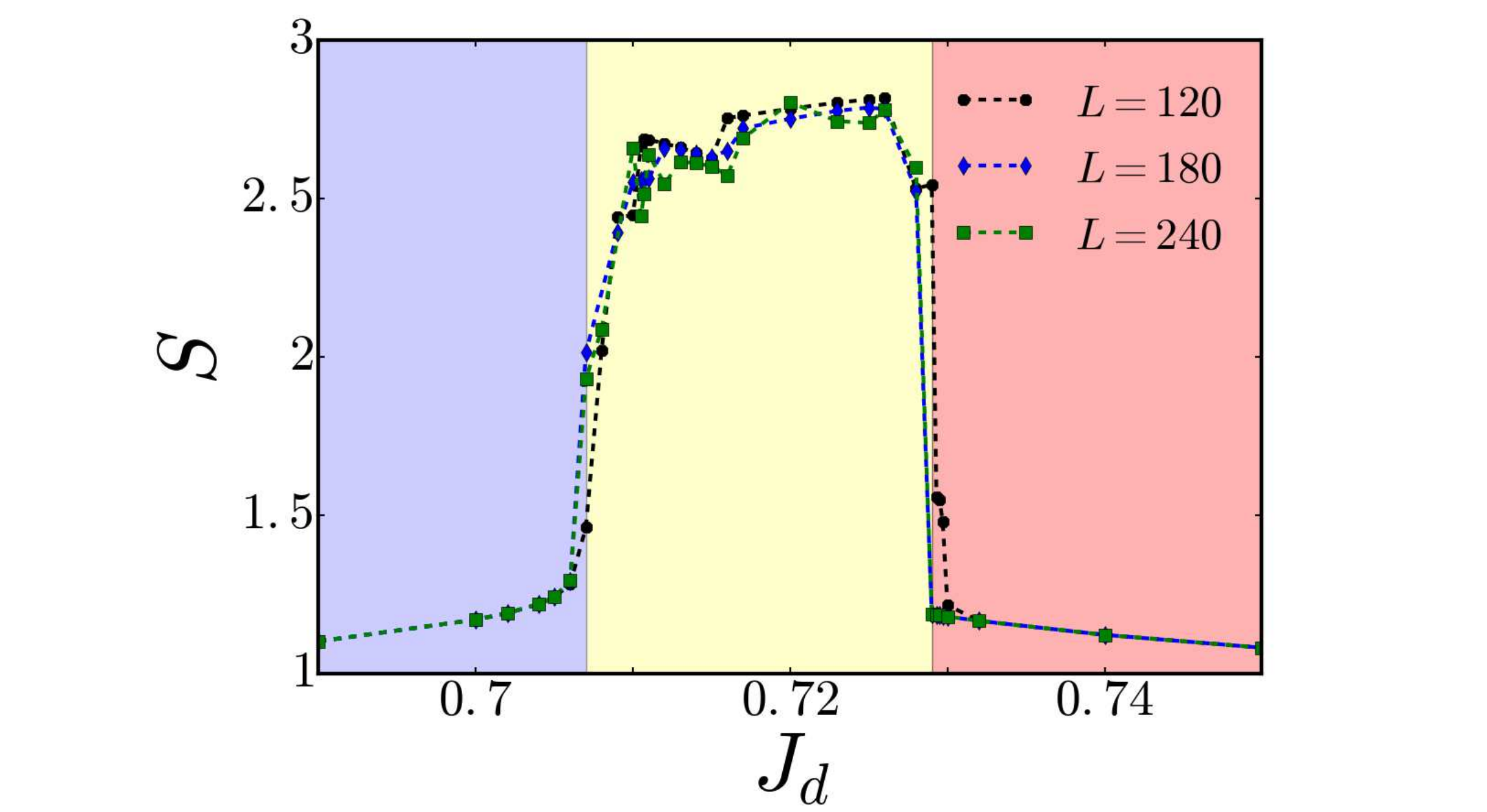}
	\includegraphics[scale=0.2]{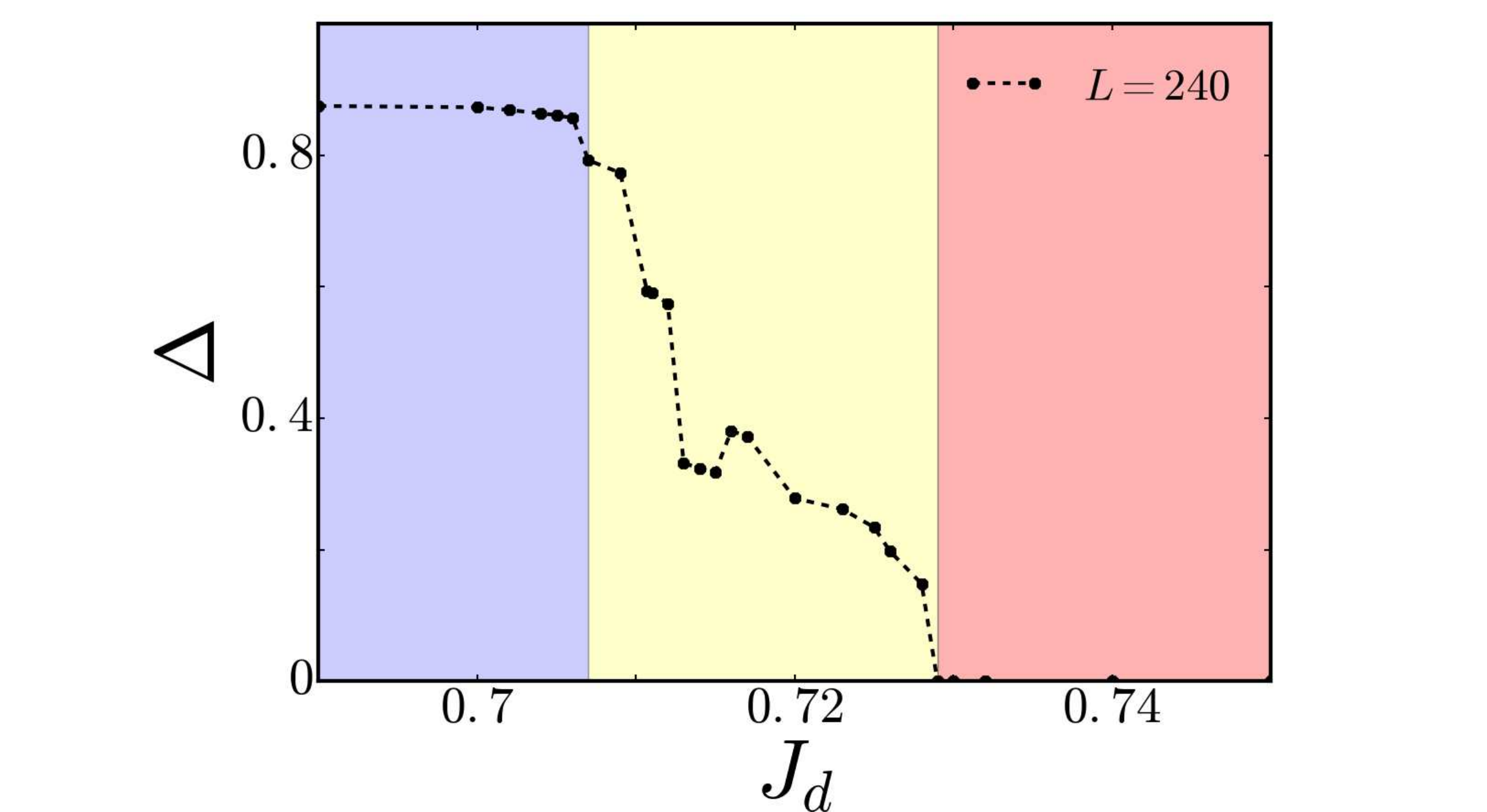}
	\caption{(Color online) EE (top) and Schmidt gap (bottom) versus $J_d$, for different sizes, $L=120,180$ and 240, obtained by DMRG with $\chi$ gradually increased from 300 to 1000.}
	
	\label{Fig:DMRGDiagEE}
\end{figure}
\begin{figure}[t]
	\centering
	\includegraphics[scale=0.22]{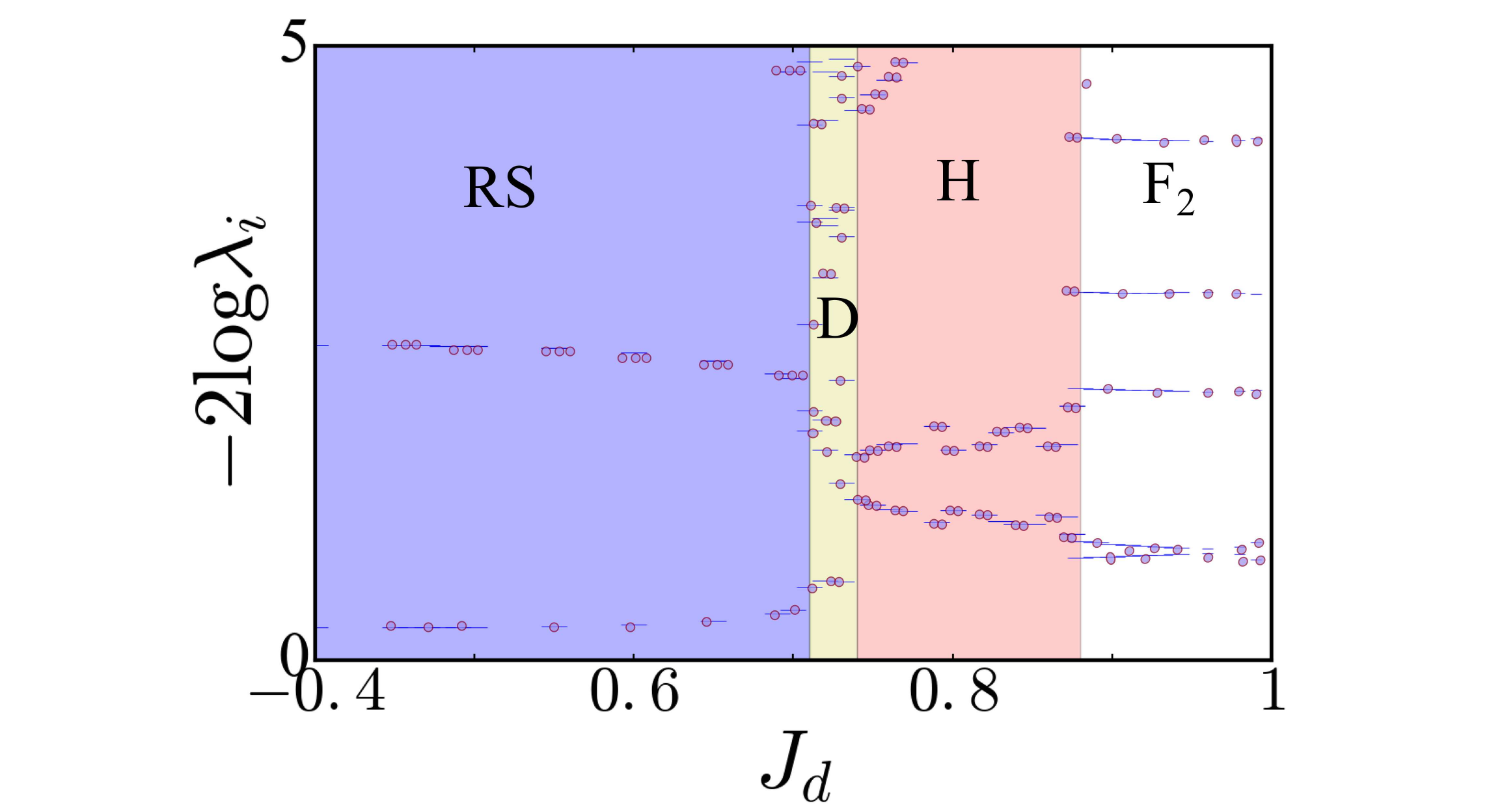}
	\caption{(Color online) DMRG results for the low-lying ES levels and their degeneracies (marked by circles).}
	\label{Fig:DMRGESDiag}
\end{figure}

We first use tools of quantum information theory for pinpointing the intermediate singlet phase. Results for the EE as well as the Schmidt gap are presented in Fig. \ref{Fig:DMRGDiagEE}. The sharp changes of the EE at $J_d\simeq0.707$ and $J_d\simeq0.73$ indeed are clear signatures for a distinct phase, squeezed in between RS and H phases. It will be shown later that dimerization plays a role, and therefore we refer to this new ground state as the D phase. At the first boundary, $J_d=0.707$, the EE changes continuously, which points to a second-order transition from the RS to the D phase. At the second boundary, $J_d=0.73$, a discontinuous jump indicates a first-order transition from the D to the H phase.
The Schmidt gap is almost constant in the RS phase, decreases gradually in the D phase, and tends to 0 in the H phase.

The ES is illustrated in Fig. \ref{Fig:DMRGESDiag}. The low-lying ES levels have odd and even degeneracies in the RS and H phases, respectively, and are non-degenerate in the ferrimagnetic phases. In the D phase, the ES levels have a mixed even-odd degeneracy.

\subsection{Short-range spin correlations}

\begin{figure}[h]
	\centering
	\includegraphics[scale=0.2]{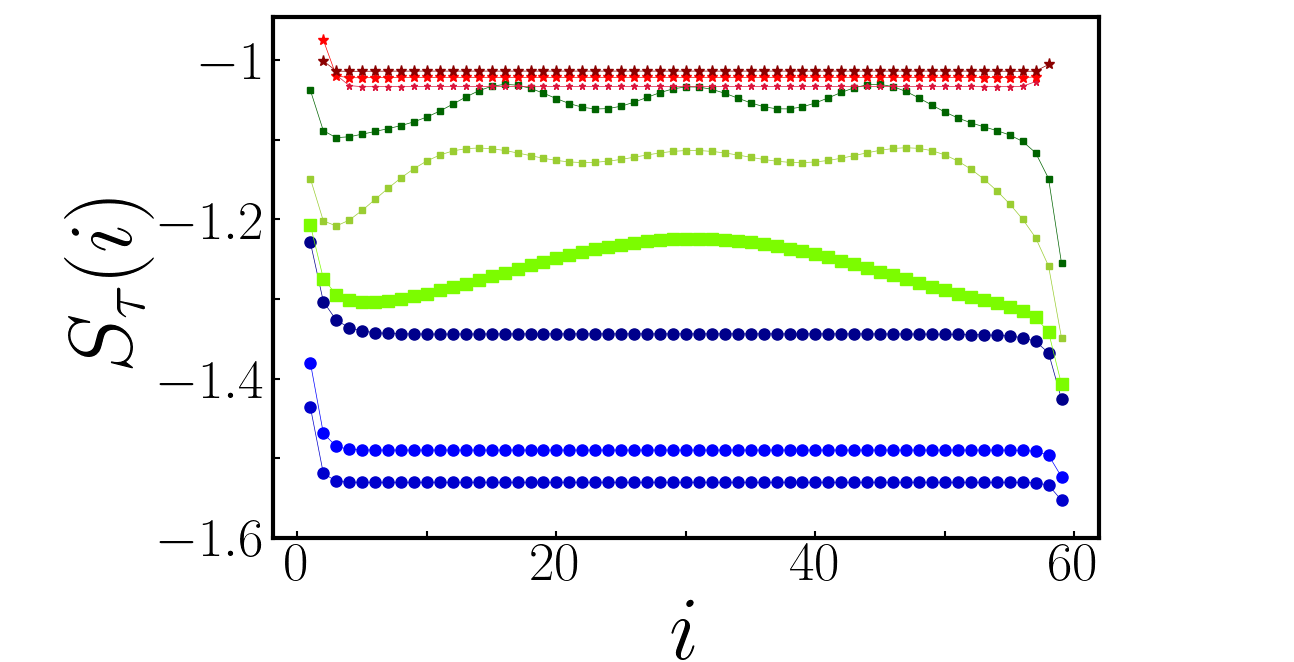}
	\includegraphics[scale=0.2]{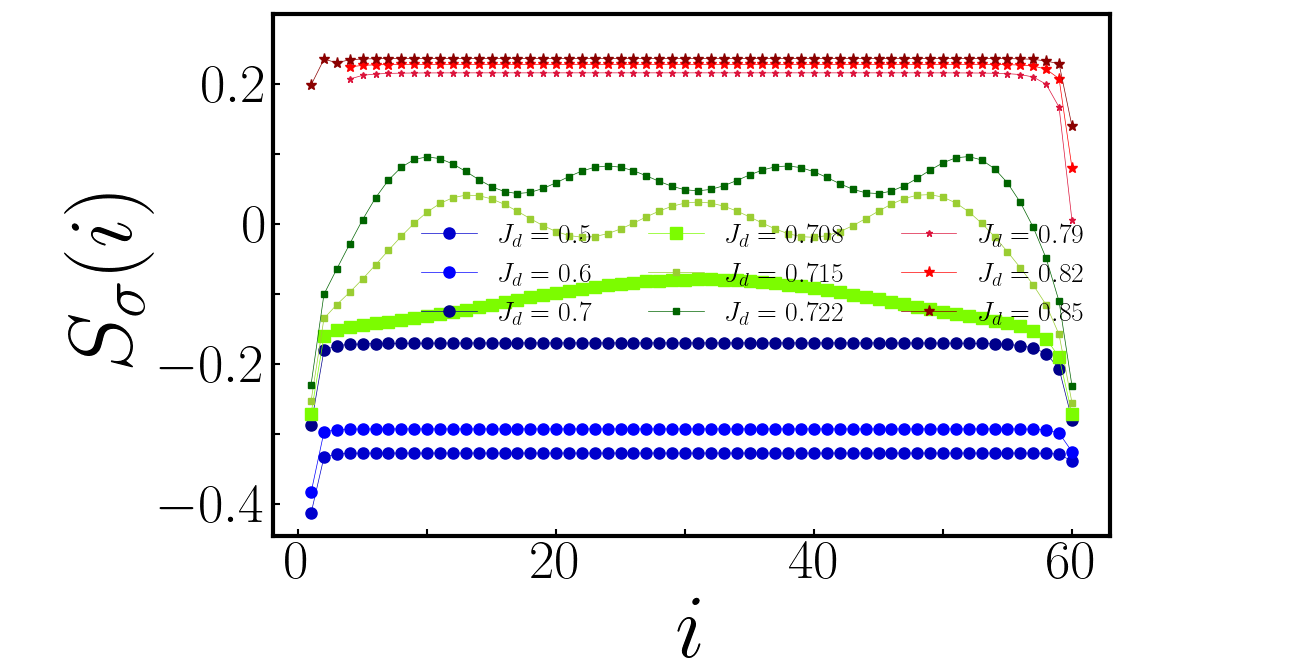}
		\caption{(Color online) Spin correlations across $\tau$-rungs (upper panel) and across $\sigma$-rungs (lower panel), as functions of the cell number $i$, obtained by DMRG.}
	\label{Fig:Rung-correlation}
\end{figure}

Fig. \ref{Fig:Rung-correlation} displays the rung correlations as functions of the cell number $i$ for different values of $J_d$. Clearly, boundary effects are limited to a few cells. Both for $J_d\leq 0.707$ and for $J_d>0.73$ the data do not depend on $i$; for small $J_d$ they are close to their lower (singlet) bounds while for $J_d\gtrsim 0.8$ they approach values corresponding to spin 1. In the D phase, $0.707<J_d<0.73$, the rung correlations are no longer constant but oscillate with a $J_d$-dependent wave vector.

\begin{figure}[h]
	\centering
		\includegraphics[scale=0.2]{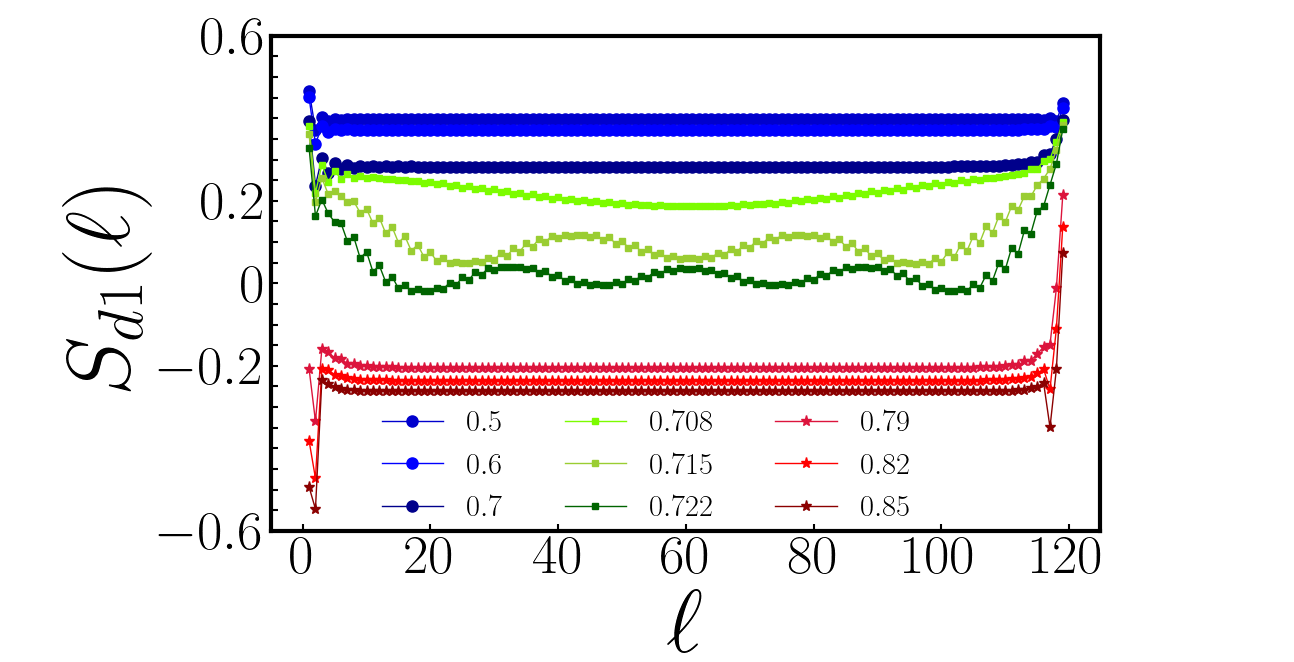}
	\includegraphics[scale=0.2]{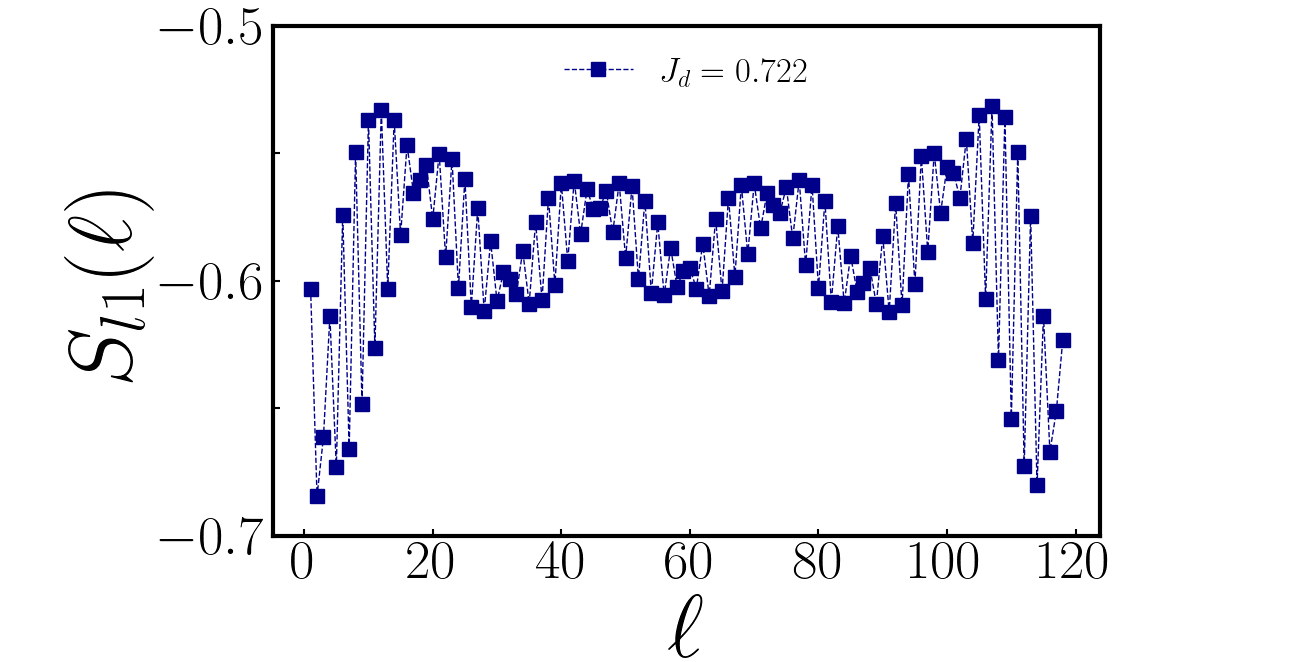}
	\caption{(Color online) Spin correlations along diagonals (upper panel) and legs (lower panel), as functions of the rung number $\ell$, calculated with DMRG for $n=1$ (the results for $n=2$ are identical).}
	\label{Fig:LegDiag-correlation}
\end{figure}

Both columnar and diagonal correlation functions, depicted in Fig. \ref{Fig:LegDiag-correlation}, are $\ell$-independent outside the D phase and show incommensurate modulations inside. However, there is an additional rapid oscillation -- a weak dimerization -- which breaks the inversion symmetry. Both correlation functions are found to be independent of $n$, therefore the leg-swap symmetry is not broken. The translational symmetry is of course broken both by the slow incommensurate modulations and by the dimerization.

We have repeated our computations for ladders of larger sizes (for example $L=240$) and also for ladders with odd $L$. The main features, in particular the oscillations in the bulk, remain the same.

Similar incommensurate oscillations have been seen in DMRG results for an SU(3) spin ladder, where they appear in an intermediate phase between a valence-bond crystal for small rung couplings and a critical Luttinger liquid for large rung couplings\cite{SUN-ladder}.

\subsection{Interpretation}
\begin{figure}[t]
	\centering
	\includegraphics[scale=0.2]{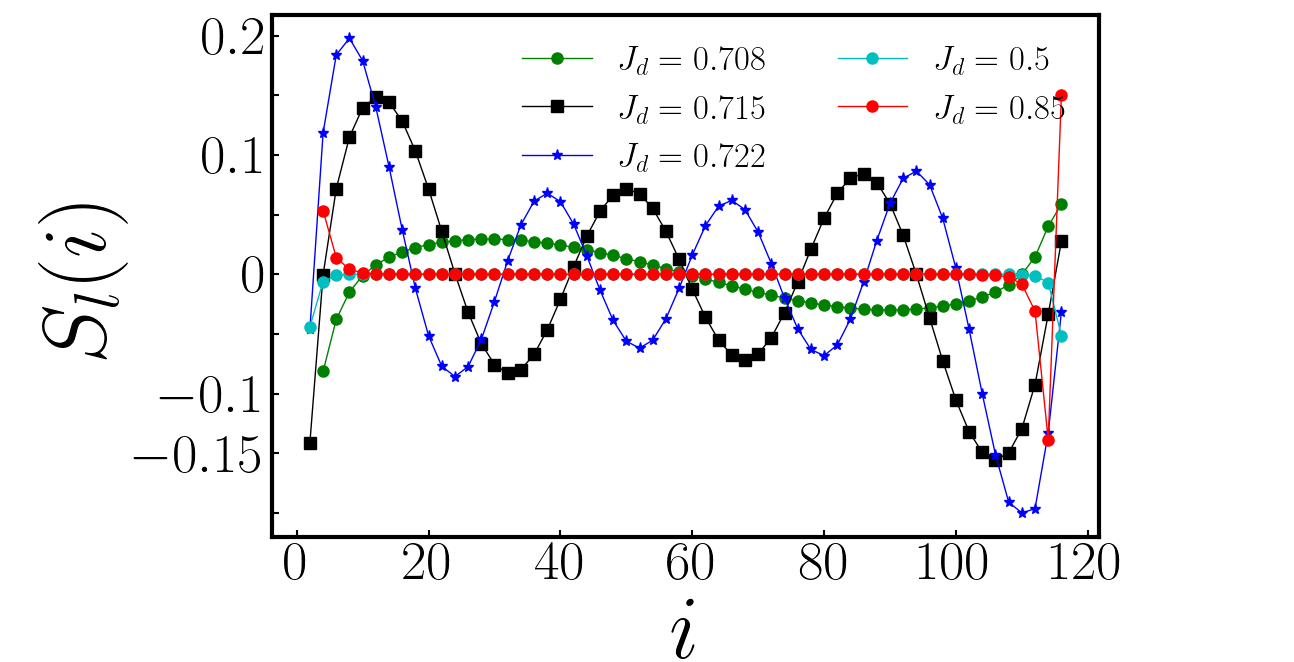}
	\includegraphics[scale=0.2]{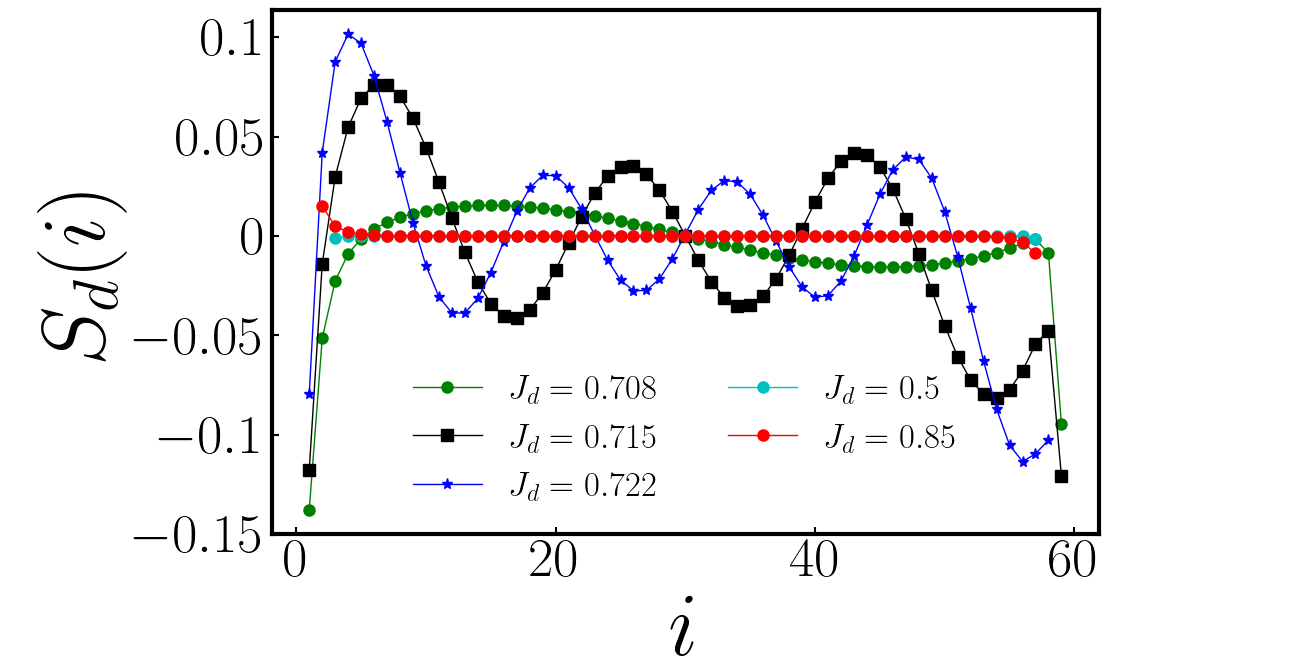}
	\caption{(Color online) Staggered bond orders along legs (upper panel) and diagonals (lower panel).}
	\label{Fig:CD-D}
\end{figure}

To quantify the rapid oscillations of columnar and diagonal correlation functions, we introduce staggered order parameters
\begin{align}\label{Eq:OCD}
S_{a}(i)&:= \sum_n\big[S_{an}(2i-1)-S_{an}(2i)\big],\, a=l,d.\nonumber\\
\end{align}
Both $S_l(i)$ and $S_d(i)$ vanish in the RS and H phases, but are finite in the D phase. Fig. \ref{Fig:CD-D} shows that these functions oscillate, actually with the same wave vectors $q$ as the rung correlations (Fig. \ref{Fig:Rung-correlation}). The oscillations are sinusoidal and therefore we could also state that the original correlation functions $S_{an}(\ell)$ have an oscillatory component with wave vector $\pi-2q$. This reminds us of the BOW in nearly half-filled Peierls systems, where a commensurate-incommensurate transition from bond alternation at half filling to an incommensurate harmonic oscillation away from half filling occurs. Very close to half filling the bond order wave is not simply sinusoidal but has the form of a ``soliton lattice'', consisting of relatively wide regions with constant order parameter and narrow domain walls in which the order parameter changes rapidly.
We do not see any evidence for domain walls, maybe simply because when $q$ approaches 0 the amplitude also tends to 0.

\begin{figure}[t]
	\centering
	\includegraphics[scale=0.19]{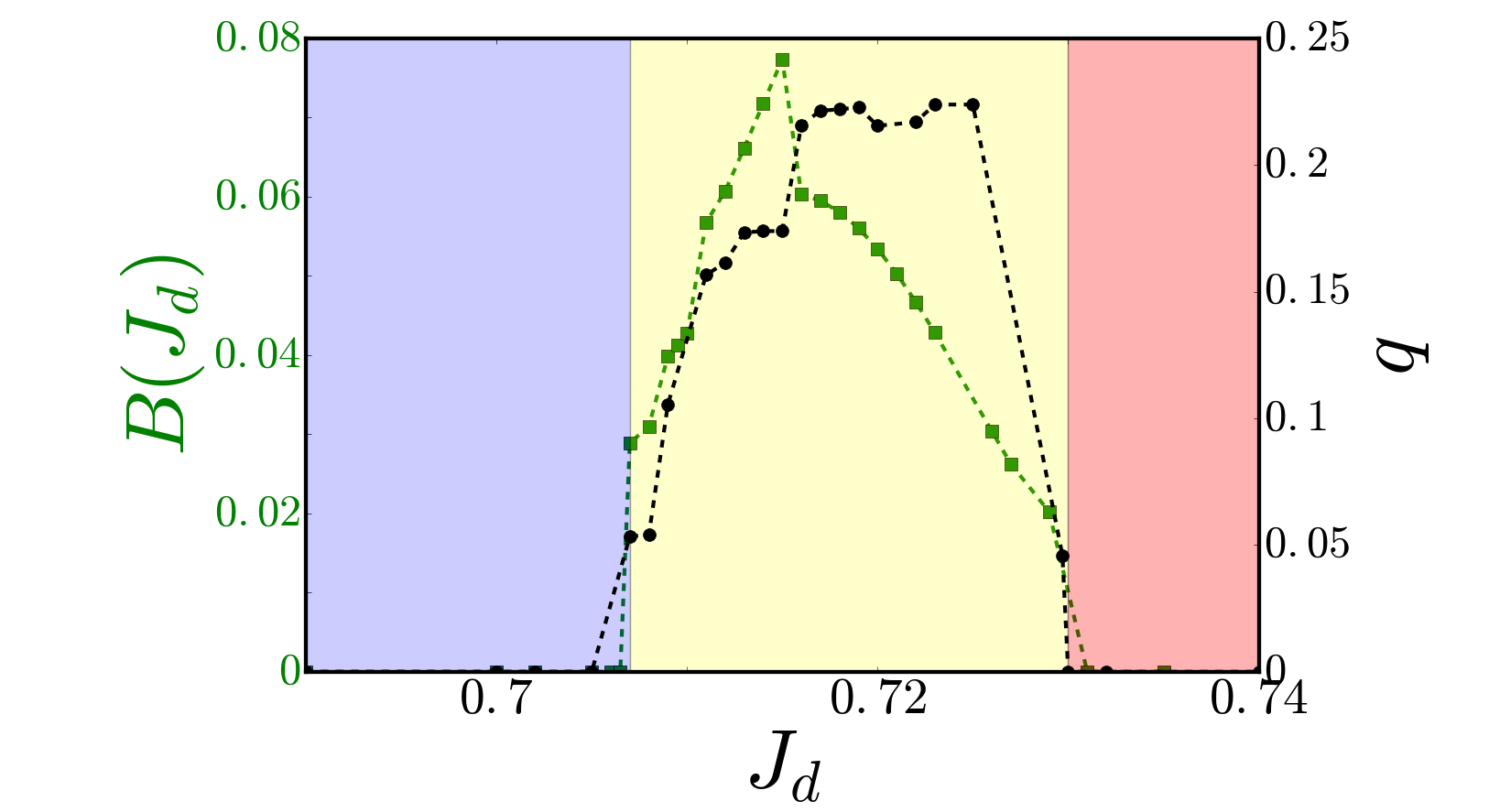}
	\includegraphics[scale=0.19]{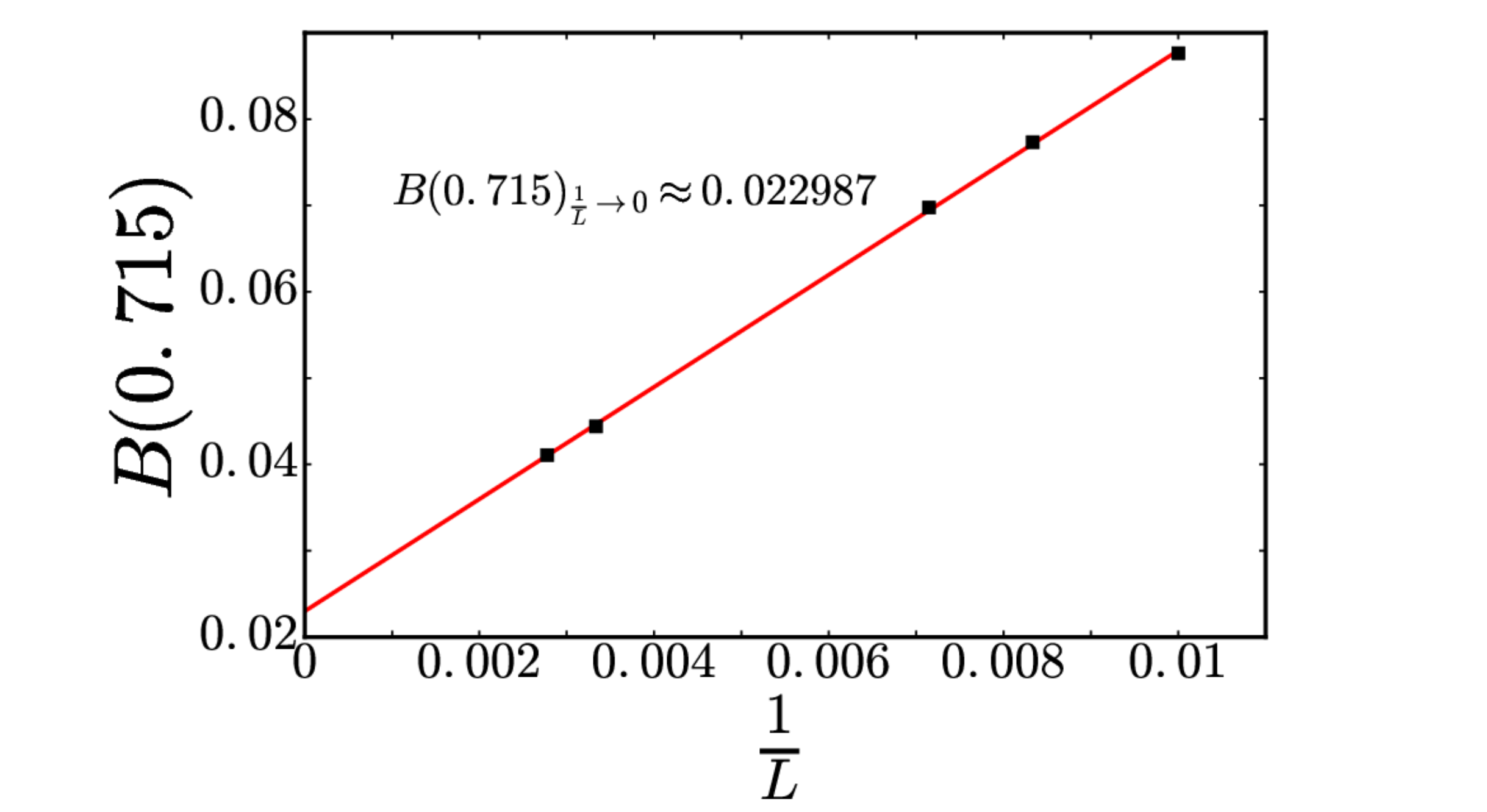}
	\caption{(Color online) Top: BOW parameter $B$ and wave vector $q$ versus $J_d$. The smooth increase of $B$ from zero above the RS-D boundary indicates a continuous phase transition, whereas the jump at the D-H boundary points to a first order transition. Bottom: Dependence of the BOW parameter on the number of rungs and extrapolation to the thermodynamic limit for $J_d=0.715$.
	}
	\label{Fig:q-b}
\end{figure}

We have deduced the wave vector $q$ for the staggered order parameters by fitting sinusoidal functions to the data points far from the edges of the ladder. The result shown in Fig. \ref{Fig:q-b} is consistent with a slow increase from zero at the lower boundary of the intermediate phase and a rapid fall to zero at the upper boundary, similarly to the behavior reported in Ref. \onlinecite{SUN-ladder}.
Fig. \ref{Fig:q-b} also shows the ``BOW parameter'' $B$, the oscillation amplitude at the center of the ladder. Clearly $B$ is only non-zero in the D phase.

An important question is whether finite size effects are responsible for the intermediate phase. This has been an issue in the context of the Hubbard model on the honeycomb lattice, for which Meng and coworkers reported an intermediate spin liquid phase for relatively small system sizes\cite{Meng_2010}, while subsequent work by Sorella and others showed that this phase disappears in larger
systems\cite{Sorella_2012}. Therefore we have investigated the size dependence of the oscillation amplitudes. The results for the BOW parameter $B$, shown in Fig. \ref{Fig:q-b}, indicate that the intermediate phase survives in the thermodynamic limit. However, the limiting value of $B$ is rather small and therefore additional studies for larger system sizes would be very useful for strengthening the case for these incommensurate spin patterns.

\section{Summary and outlook}\label{sec:summary}
\begin{figure}
\centering
\includegraphics[width=2cm]{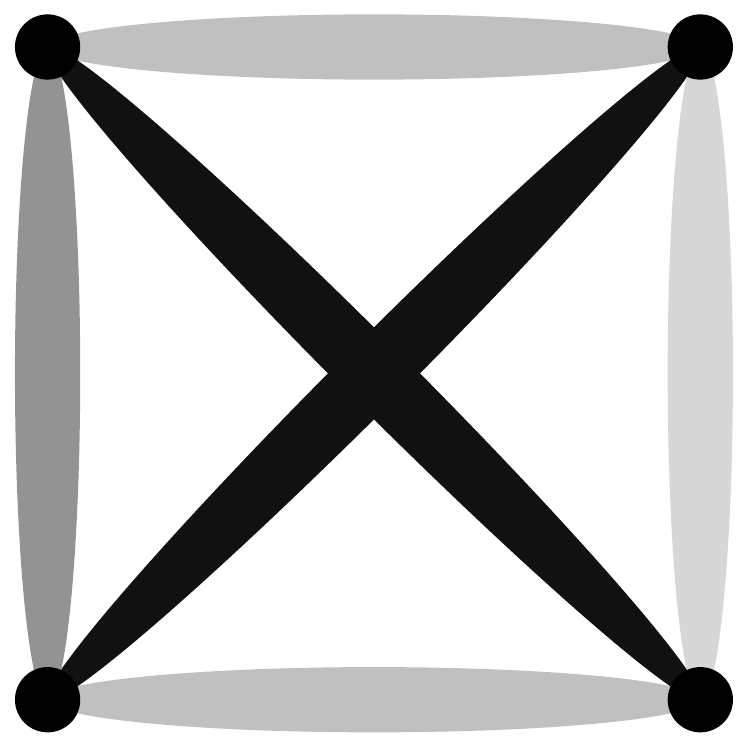}
\hspace{0.7cm}
\includegraphics[width=2cm]{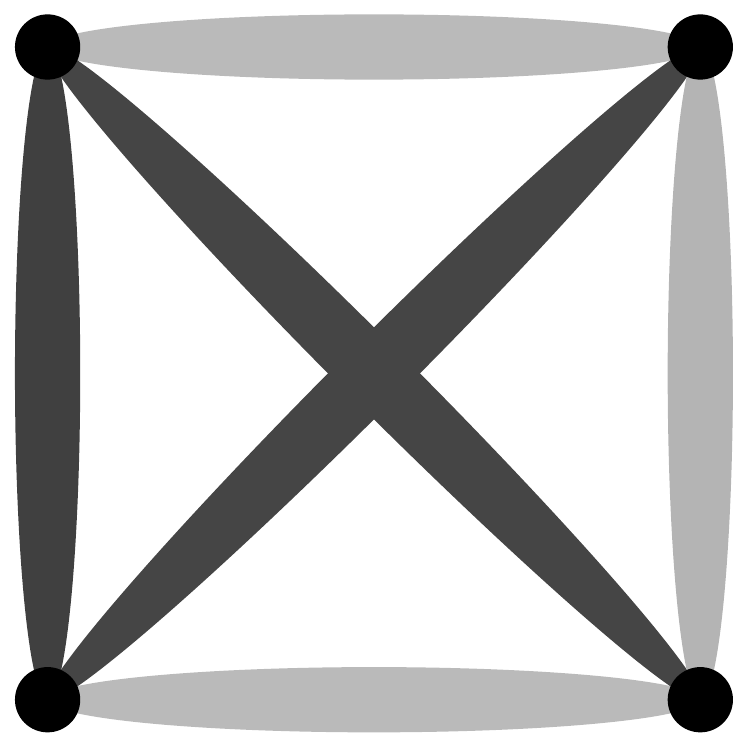}
\hspace{0.7cm}
\includegraphics[width=2cm]{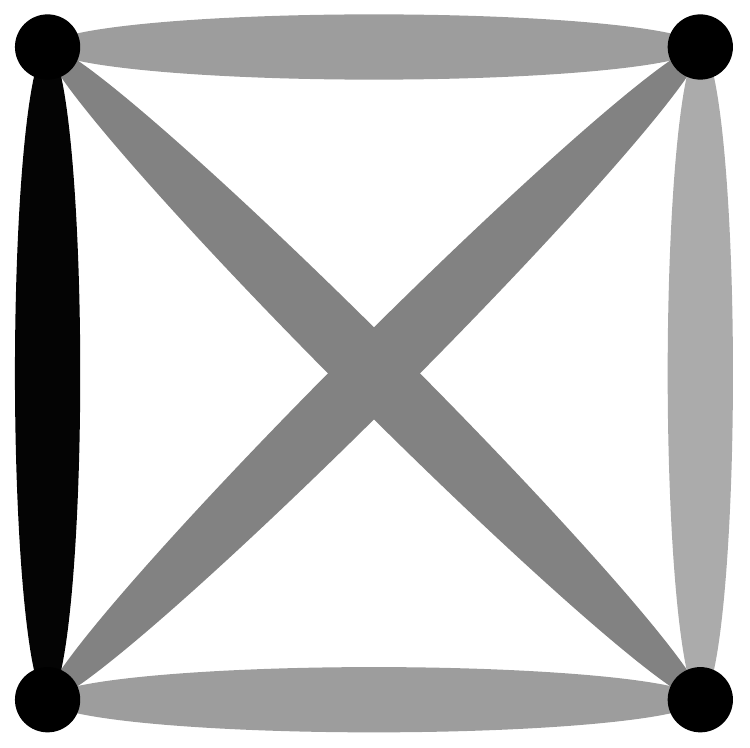}
\caption{Pictorial representation of spin correlations in the unit cell (defined in Fig. \ref{Fig:ladder}) for coupling constants $J_l=J_r=1$ and $J_d=0.5$ (left figure, RS phase), $J_d=0.715$ (middle figure, D phase), $J_d=0.85$ (right figure, H phase). The grayscale reaches from the white shade (lower bound of the correlations -- singlet bonds) to the black shade (upper bound of the correlations -- magnetic bonds).}
\label{fig:bonds}
\end{figure}

In this paper we have described our study of a frustrated mixed-spin ladder, consisting alternatively of spin $\frac{1}{2}$ and spin 1 rungs. We have used tools from quantum information, in particular the entanglement entropy (EE), the Schmidt gap and the entanglement spectrum (ES), as well as correlation functions to characterize the various phases. Three types of interactions were taken into account, one along legs (coupling $J_l$), one on rungs ($J_r$) and one on diagonals ($J_d$). We limited ourselves mostly on the region $J_l=J_r=1$, $J_d\ge 0$. Three distinct nonmagnetic phases were identified, a gapped RS phase with odd degeneracy of the ES levels (for $J_d\lesssim 0.7$), a gapped H phase  with even degeneracy of the ES levels (for $J_d\gtrsim 0.7$), and an intermediate D phase with  mixed even-odd degeneracies of the ES levels (for $J_d\approx 0.7$).

The overall behavior of short-range correlation functions is depicted in Fig. \ref{fig:bonds}. Three bonds are quite prominent, the singlet bond on the $\tau$ rung and the ``magnetic'' bonds on the diagonals, both in the RS phase, as well as the triplet bond on the $\sigma$ rung in the H phase. Therefore, with one exception, the bonds are far from being singlets, the favorite state of two antiferromagnetically coupled spins. This is a clear signature for frustration.

The most striking features of our study are the incommensurate spatial oscillations of spin correlation functions, observed only in the D phase. Their wave vector depends on the coupling strength. Their amplitude can serve as an order parameter; its behavior close to the phase boundaries suggests that the transition is continuous at the RS-D boundary and of first order at the D-H boundary.

Our choice of coupling parameters is quite special, as became apparent in our calculations for the elementary plaquette (Section \ref{sec:plaquette}). In fact, for this choice the energy spectrum of the plaquette is ``integrable'' (level crossing but no level repulsion), for other couplings it is generically non-integrable (level repulsion but no level crossing). Therefore it would be desirable to study the ladder
in a wider region of parameter space.

We have used relatively small system sizes. It is true that the extrapolation to the thermodynamic limit did preserve a finite order parameter in the D phase, at the same time its small value is worrisome. Clearly, additional calculations for larger systems would be very useful.
\begin{acknowledgments}
RH acknowledges support for this work from
the US Department of Energy, Office of Science, via award no DE-SC0019374. We have used Uni10 \cite{Kao2015Uni10AO} and ITensor \cite{itensor} as  middleware libraries to build up the ansatz.
\end{acknowledgments}
\appendix
\section{Diagonalization of the plaquette Hamiltonian}\label{sec:diagonalization}
\label{app:plaquette}
The 10 eigenstates of the Hamiltonian (\ref{eq:plaquette}) with $S_z=0$ can be grouped into even and odd states (with respect to the leg-swap operation), namely
\begin{align}\label{eq:basis}
\vert a\rangle_\pm&=\frac{1}{\sqrt{2}}\big(\vert \!+\!0\!\downarrow\downarrow\rangle\pm \vert 0\!+\!\downarrow\downarrow\rangle\big)\nonumber\\
\vert b\rangle_\pm&=\frac{1}{\sqrt{2}}\big(\vert \!+\!-\!\uparrow\downarrow\rangle\pm \vert \!-\!+\!\downarrow\uparrow\rangle\big)\nonumber\\
\vert c\rangle_\pm&=\frac{1}{\sqrt{2}}\big(\vert \!+\!-\!\downarrow\uparrow\rangle\pm \vert \!-\!+\!\uparrow\downarrow\rangle\big)\nonumber\\
\vert d\rangle_\pm&=\frac{1}{\sqrt{2}}\big(\vert 0\!-\!\uparrow\uparrow\rangle\pm \vert\! -\!0\!\uparrow\uparrow\rangle\big)\nonumber\\
\vert e\rangle_\pm&=\frac{1}{\sqrt{2}}\big(\vert 00\!\uparrow\downarrow\rangle\pm \vert 00\!\downarrow\uparrow\rangle\big)
\end{align}
Once the eigenstates in this subspace are determined, those for $S_z\neq 0$ are easily obtained by applying $S_+$ and $S_-$. These new states are uninteresting as long as we are just searching for the eigenvalues of ${\bf S}^2$ and $H$, because if $\vert\Psi\rangle$ is an eigenstate of ${\bf S}^2$ and $H$, the same holds for $S_\pm\vert\Psi\rangle$, with the same eigenvalues.

It is straightforward to calculate eigenstates and eigenvalues of ${\bf S}^2$ in this basis. We find
one state with $S=3$, one with $S=2$ and three with $S=1$ from the even basis states,
\begin{align}\label{eq:even}
\vert 3,0\rangle&=\frac{1}{\sqrt{5}}\Big[\vert a\rangle_++\vert d\rangle_+
+\frac{1}{\sqrt{2}}\big(\vert b\rangle_++\vert c\rangle_+\big)+\sqrt{2}\vert e\rangle_+\Big]\nonumber\\
\vert 2,0\rangle_+&=\frac{1}{\sqrt{2}}\big(\vert a\rangle_+-\vert d\rangle_+\big)\nonumber\\
\vert 1,0\rangle_{+}^{(1)}&=\frac{1}{\sqrt{2}}\big(\vert b\rangle_+-\vert c\rangle_+\big)\nonumber\\
\vert 1,0\rangle_{+}^{(2)}&=\frac{1}{2}\big(\vert a\rangle_++\vert d\rangle_+-\sqrt{2}\vert e\rangle_+\big)\nonumber\\
\vert 1,0\rangle_{+}^{(3)}&=\frac{1}{\sqrt{5}}\Big[\frac{1}{2}\big(\vert a\rangle_++\vert d\rangle_+\big)
-\sqrt{2}\big(\vert b\rangle_++\vert c\rangle_+\big)\nonumber\\
&\qquad+\frac{1}{\sqrt{2}}\vert e\rangle_+\Big]\, .
\end{align}
The odd basis states yield two eigenstates of ${\bf S}^2$ with $S=2$, one with $S=1$ and two singlet states,
 \begin{align}\label{eq:odd}
 \vert 2,0\rangle_-^{(1)}&=\frac{1}{\sqrt{6}}\, \Big[\vert a\rangle_-+\vert d\rangle_-
 +\sqrt{2}\big(\vert b\rangle_-+\vert c\rangle_-\big)\Big]\nonumber\\
  \vert 2,0\rangle_-^{(2)}&=\frac{1}{\sqrt{6}}\, \big(\vert b\rangle_--\vert c\rangle_-+2\vert e\rangle_-\big)\nonumber\\
  \vert 1,0\rangle_-&=\frac{1}{\sqrt{2}}\, \big(\vert a\rangle_--\vert d\rangle_-\big)\nonumber\\
 \vert 0,0\rangle^{(1)}&=\frac{1}{\sqrt{3}}\, \Big[\vert a\rangle_-+\vert d\rangle_-
 -\frac{1}{\sqrt{2}}\big(\vert b\rangle_-+\vert c\rangle_-\big)\Big]\nonumber\\
  \vert 0,0\rangle^{(2)}&=\frac{1}{\sqrt{3}}\, \big(\vert b\rangle_--\vert c\rangle_--\vert e\rangle_-\big)\, .
 \end{align}

We now determine the eigenstates and eigenvalues of the Hamiltonian using these basis states. We find for the ``non-degenerate'' states
\begin{align}
H\vert 3,0\rangle&=\big(J_l+J_d+\frac{1}{4}J_\sigma+J_\tau\big)\vert 3,0\rangle\nonumber\\
H\vert 2,0\rangle_+&=\big[-\frac{1}{2}(J_l+J_d)+\frac{1}{4}J_\sigma+J_\tau\big]\vert 2,0\rangle_+\nonumber\\
H\vert 1,0\rangle_-&=\big[-\frac{1}{2}(J_l+J_d)+\frac{1}{4}J_\sigma-J_\tau\big]\vert 1,0\rangle_-
\end{align}
For the ``doubly degenerate'' states we obtain
\begin{align}
H\vert 2,0\rangle_-^{(1)}&=\Big[\frac{1}{2}(J_l+J_d)
+\frac{1}{4}J_\sigma-J_\tau\Big]\vert 2,0\rangle_-^{(1)}\nonumber\\
&+\frac{1}{\sqrt{2}}(J_l-J_d)\vert 2,0\rangle_-^{(2)}\nonumber\\
H\vert 2,0\rangle_-^{(2)}&=\frac{1}{\sqrt{2}}(J_l-J_d)\vert 2,0\rangle_-^{(1)}
-\big(\frac{3}{4}J_\sigma-J_\tau\big)\vert 2,0\rangle_-^{(2)}
\end{align}
\begin{align}
H\vert 0,0\rangle^{(1)}&=\big(-J_l-J_d+\frac{1}{4}J_\sigma-J_\tau\big)\vert 0,0\rangle^{(1)}\nonumber\\
&+\sqrt{2}(J_d-J_l)\vert 0,0\rangle^{(2)}\nonumber\\
H\vert 0,0\rangle^{(2)}&=\sqrt{2}(J_d-J_l)\vert 0,0\rangle^{(1)}
-\big(\frac{3}{4}J_\sigma+2J_\tau\big)\vert 0,0\rangle^{(2)}
\end{align}
and therefore we have just to diagonalize $2\times2$ matrices. For the singlet states we find
\begin{align}\label{eq:EV_singlet}
E=\frac{1}{2}\big(h_{11}+h_{22}\pm\sqrt{(h_{11}-h_{22})^2+4h_{12}^2}\big)\, ,
\end{align}
where
\begin{align}
h_{11}&=-J_l-J_d+\frac{1}{4}J_\sigma-J_\tau\nonumber\\
h_{22}&=-\frac{3}{4}J_\sigma-2J_\tau\nonumber\\
h_{12}&=\sqrt{2}\big(J_d-J_l\big)
\end{align}

The remaining ``triply degenerate'' states satisfy the eigenvalue equation
\begin{align}
H\vert 1,0\rangle_+^{(\alpha)}=\sum_{\beta=1}^3h_{\alpha\beta}\vert 1,0\rangle_+^{(\beta)}\, ,
\, \alpha=1,2,3
\end{align}
with matrix elements
\begin{align}
h_{11}&=-\big(\frac{3}{4}J_\sigma+J_\tau\big)\nonumber\\
h_{12}&=h_{21}=\frac{1}{2}(J_d-J_l)\nonumber\\
h_{13}&=h_{31}=\frac{\sqrt{5}}{2}(J_d-J_l)\nonumber\\
h_{22}&=\frac{1}{4}[-5(J_l+J_d)+J_\sigma+2J_\tau]\nonumber\\
h_{23}&=h_{32}=\frac{\sqrt{5}}{4}(-J_l-J_d+2J_\tau)\nonumber\\
h_{33}&=\frac{1}{4}(-J_l-J_d+J_\sigma-6J_\tau)
\end{align}

Miraculously, for the particular parameter set $J_l=J_\sigma=J_\tau=1$ all eigenvalues are simple linear functions of $J_d$, as shown in Table \ref{tab:EV} and illustrated in Fig. \ref{fig:EV}.

\begin{table}
\centering
$
\begin{array}{c|c|c|c|c}
S&\mbox{Parity}&E_1&E_2&E_3\\
\hline
3&+&\frac{9}{4}+J_d&&\\
2&+&\frac{3}{4}-\frac{1}{2}J_d&&\\
1&+&\frac{1}{4}-2J_d&-\frac{5}{4}-\frac{1}{2}J_d&-\frac{11}{4}+J_d\\
2&-&\frac{3}{4}-\frac{1}{2}J_d&-\frac{3}{4}+J_d&\\
1&-&-\frac{5}{4}-\frac{1}{2}J_d&&\\
0&-&-\frac{3}{4}-2J_d&-\frac{15}{4}+J_d&
\end{array}
$
\caption{Eigenvalues of the plaquette Hamiltonian for $J_l=J_\sigma=J_\tau=1$.}
\label{tab:EV}
\end{table}

\section{Expansion in powers of $J_l/J_r$ for $J_r>0$}\label{app:expansion}
We use perturbation theory to evaluate the ground state for antiferromagnetic rung coupling, $J_r>0$, and small leg coupling, $J_l\ll J_r$. A rung with $\sigma$-spins has four states, a singlet ($S=0$)
\begin{align}
\vert s_\sigma\rangle=\frac{1}{\sqrt{2}}(\vert\!\uparrow\rangle\vert\!\downarrow\rangle-\vert\!\downarrow\rangle\vert\!\uparrow\rangle)
\end{align}
and three triplet states ($S=1$)
\begin{align}
\vert t^+_\s\rangle&=\vert\!\uparrow\rangle\vert\!\uparrow\rangle,\quad \vert t^-_\s\rangle=\vert\!\downarrow\rangle\vert\!\downarrow\rangle,\nonumber\\
\vert t^0_\s\rangle&=\frac{1}{\sqrt{2}}(\vert\!\uparrow\rangle\vert\!\downarrow\rangle+\vert\!\downarrow\rangle\vert\!\uparrow\rangle),
\end{align}
where we have used the convention that in the product states the first ket is on chain 1, the second on chain 2.

A rung with $\tau$-spins has nine states, a quintet ($S=2$)
\begin{align}
\vert q^{\pm 2}_\t\rangle&=\vert\pm\rangle\vert\pm\rangle,\nonumber\\
\vert q^{\pm 1}_\t\rangle&=\frac{1}{\sqrt{2}}(\vert\pm\rangle\vert 0\rangle+\vert 0\rangle\vert\pm\rangle),\nonumber\\
\vert q^0_\t\rangle&=\frac{1}{\sqrt{6}}(\vert +\rangle\vert -\rangle+2\vert 0\rangle\vert 0\rangle+\vert -\rangle\vert +\rangle),
\end{align}
a triplet ($S=1$)
\begin{align}
\vert t^{\pm}_\t\rangle&=\pm\frac{1}{\sqrt{2}}(\vert\pm\rangle\vert 0\rangle-\vert 0\rangle\vert\pm\rangle),\nonumber\\
\vert t^0_\t\rangle&=\frac{1}{\sqrt{2}}(\vert +\rangle\vert -\rangle-\vert -\rangle\vert +\rangle),
\end{align}
and a singlet ($S=0$)
\begin{align}
\vert s_\t\rangle&=\frac{1}{\sqrt{3}}(\vert 0\rangle\vert 0\rangle-\vert +\rangle\vert -\rangle-\vert -\rangle\vert +\rangle),
\end{align}
where $\vert\pm\rangle$ and $\vert 0\rangle$ are the eigenstates of $\tau_z$ with eigenvalues $\pm 1$ and $0$, respectively.

The ground state for $J_l=J_d=0$ and $J_r>0$ is a product of rung singlets $\bigotimes_i\vert s_\s\rangle_i\vert s_\t\rangle_i$. To first order in $J_l$ we have to apply the operator
$\s^{(1)}\cdot\t^{(1)}+\s^{(2)}\cdot\t^{(2)}$ to each pair of adjacent rungs. We find
\begin{align}\label{eq:triplet}
\big(\s^{(1)}\cdot\t^{(1)}+\s^{(2)}\cdot\t^{(2)}\big)\vert s_\s\rangle\vert s_\t\rangle=\nonumber\\
\sqrt{\frac{2}{3}}\big(\vert t^+_\s\rangle\vert t^-_\t\rangle+\vert t^-_\s\rangle\vert t^+_\t\rangle-\vert t^0_\s\rangle\vert t^0_\t\rangle\big).
\end{align}
Triplets appear in first-order perturbation theory for the ground state, but there are no quintets.

To second order in $J_l$ three different terms are generated. If the operator $\s^{(1)}\cdot\t^{(1)}+\s^{(2)}\cdot\t^{(2)}$ acts on four distinct rungs, the expression
(\ref{eq:triplet}) simply appears twice and no quintets are produced. If the same operator is applied twice to a single pair of rungs, quintets neither appear. This is easily understood by applying $\s^{(1)}\cdot\t^{(1)}+\s^{(2)}\cdot\t^{(2)}$ to Eq. (\ref{eq:triplet}). In the third case, where three neighboring rungs are involved, say, at sites $i-1,i,i+1$, quintet states do appear.

\bibliography{Ferriladder}

\begin{thebibliography}{86}%
\makeatletter
\providecommand \@ifxundefined [1]{%
 \@ifx{#1\undefined}
}%
\providecommand \@ifnum [1]{%
 \ifnum #1\expandafter \@firstoftwo
 \else \expandafter \@secondoftwo
 \fi
}%
\providecommand \@ifx [1]{%
 \ifx #1\expandafter \@firstoftwo
 \else \expandafter \@secondoftwo
 \fi
}%
\providecommand \natexlab [1]{#1}%
\providecommand \enquote  [1]{``#1''}%
\providecommand \bibnamefont  [1]{#1}%
\providecommand \bibfnamefont [1]{#1}%
\providecommand \citenamefont [1]{#1}%
\providecommand \href@noop [0]{\@secondoftwo}%
\providecommand \href [0]{\begingroup \@sanitize@url \@href}%
\providecommand \@href[1]{\@@startlink{#1}\@@href}%
\providecommand \@@href[1]{\endgroup#1\@@endlink}%
\providecommand \@sanitize@url [0]{\catcode `\\12\catcode `\$12\catcode
  `\&12\catcode `\#12\catcode `\^12\catcode `\_12\catcode `\%12\relax}%
\providecommand \@@startlink[1]{}%
\providecommand \@@endlink[0]{}%
\providecommand \url  [0]{\begingroup\@sanitize@url \@url }%
\providecommand \@url [1]{\endgroup\@href {#1}{\urlprefix }}%
\providecommand \urlprefix  [0]{URL }%
\providecommand \Eprint [0]{\href }%
\providecommand \doibase [0]{http://dx.doi.org/}%
\providecommand \selectlanguage [0]{\@gobble}%
\providecommand \bibinfo  [0]{\@secondoftwo}%
\providecommand \bibfield  [0]{\@secondoftwo}%
\providecommand \translation [1]{[#1]}%
\providecommand \BibitemOpen [0]{}%
\providecommand \bibitemStop [0]{}%
\providecommand \bibitemNoStop [0]{.\EOS\space}%
\providecommand \EOS [0]{\spacefactor3000\relax}%
\providecommand \BibitemShut  [1]{\csname bibitem#1\endcsname}%
\let\auto@bib@innerbib\@empty
\bibitem [{\citenamefont {Diep}(2005)}]{doi:10.1142/5697}%
  \BibitemOpen
  \bibfield  {author} {\bibinfo {author} {\bibfnamefont {H.~T.}\ \bibnamefont
  {Diep}},\ }\href {\doibase 10.1142/5697} {\emph {\bibinfo {title} {Frustrated
  Spin Systems}}}\ (\bibinfo  {publisher} {WORLD SCIENTIFIC},\ \bibinfo {year}
  {2005})\BibitemShut {NoStop}%
\bibitem [{\citenamefont {Derzhko}\ \emph {et~al.}(2015)\citenamefont
  {Derzhko}, \citenamefont {Richter},\ and\ \citenamefont
  {Maksymenko}}]{Derzhko}%
  \BibitemOpen
  \bibfield  {author} {\bibinfo {author} {\bibfnamefont {O.}~\bibnamefont
  {Derzhko}}, \bibinfo {author} {\bibfnamefont {J.}~\bibnamefont {Richter}}, \
  and\ \bibinfo {author} {\bibfnamefont {M.}~\bibnamefont {Maksymenko}},\
  }\href@noop {} {\bibfield  {journal} {\bibinfo  {journal} {International
  Journal of Modern Physics B}\ }\textbf {\bibinfo {volume} {29}},\ \bibinfo
  {pages} {1530007} (\bibinfo {year} {2015})}\BibitemShut {NoStop}%
\bibitem [{\citenamefont {Schmidt}(2013)}]{Schmidt-toric}%
  \BibitemOpen
  \bibfield  {author} {\bibinfo {author} {\bibfnamefont {K.~P.}\ \bibnamefont
  {Schmidt}},\ }\href {\doibase 10.1103/PhysRevB.88.035118} {\bibfield
  {journal} {\bibinfo  {journal} {Phys. Rev. B}\ }\textbf {\bibinfo {volume}
  {88}},\ \bibinfo {pages} {035118} (\bibinfo {year} {2013})}\BibitemShut
  {NoStop}%
\bibitem [{\citenamefont {Schulz}\ and\ \citenamefont
  {Burnell}(2016)}]{Schulz-toric}%
  \BibitemOpen
  \bibfield  {author} {\bibinfo {author} {\bibfnamefont {M.~D.}\ \bibnamefont
  {Schulz}}\ and\ \bibinfo {author} {\bibfnamefont {F.~J.}\ \bibnamefont
  {Burnell}},\ }\href {\doibase 10.1103/PhysRevB.94.165110} {\bibfield
  {journal} {\bibinfo  {journal} {Phys. Rev. B}\ }\textbf {\bibinfo {volume}
  {94}},\ \bibinfo {pages} {165110} (\bibinfo {year} {2016})}\BibitemShut
  {NoStop}%
\bibitem [{\citenamefont {Zarei}\ and\ \citenamefont {Abouie}(2021)}]{Zarei}%
  \BibitemOpen
  \bibfield  {author} {\bibinfo {author} {\bibfnamefont {M.~H.}\ \bibnamefont
  {Zarei}}\ and\ \bibinfo {author} {\bibfnamefont {J.}~\bibnamefont {Abouie}},\
  }\href {\doibase 10.1103/PhysRevB.104.115141} {\bibfield  {journal} {\bibinfo
   {journal} {Phys. Rev. B}\ }\textbf {\bibinfo {volume} {104}},\ \bibinfo
  {pages} {115141} (\bibinfo {year} {2021})}\BibitemShut {NoStop}%
\bibitem [{\citenamefont {Zhou}\ \emph {et~al.}(2017)\citenamefont {Zhou},
  \citenamefont {Kanoda},\ and\ \citenamefont {Ng}}]{RevModPhys.89.025003}%
  \BibitemOpen
  \bibfield  {author} {\bibinfo {author} {\bibfnamefont {Y.}~\bibnamefont
  {Zhou}}, \bibinfo {author} {\bibfnamefont {K.}~\bibnamefont {Kanoda}}, \ and\
  \bibinfo {author} {\bibfnamefont {T.-K.}\ \bibnamefont {Ng}},\ }\href
  {\doibase 10.1103/RevModPhys.89.025003} {\bibfield  {journal} {\bibinfo
  {journal} {Rev. Mod. Phys.}\ }\textbf {\bibinfo {volume} {89}},\ \bibinfo
  {pages} {025003} (\bibinfo {year} {2017})}\BibitemShut {NoStop}%
\bibitem [{\citenamefont {Nisoli}\ \emph {et~al.}(2013)\citenamefont {Nisoli},
  \citenamefont {Moessner},\ and\ \citenamefont
  {Schiffer}}]{RevModPhys.85.1473}%
  \BibitemOpen
  \bibfield  {author} {\bibinfo {author} {\bibfnamefont {C.}~\bibnamefont
  {Nisoli}}, \bibinfo {author} {\bibfnamefont {R.}~\bibnamefont {Moessner}}, \
  and\ \bibinfo {author} {\bibfnamefont {P.}~\bibnamefont {Schiffer}},\ }\href
  {\doibase 10.1103/RevModPhys.85.1473} {\bibfield  {journal} {\bibinfo
  {journal} {Rev. Mod. Phys.}\ }\textbf {\bibinfo {volume} {85}},\ \bibinfo
  {pages} {1473} (\bibinfo {year} {2013})}\BibitemShut {NoStop}%
\bibitem [{\citenamefont {Ortiz-Ambriz}\ \emph {et~al.}(2019)\citenamefont
  {Ortiz-Ambriz}, \citenamefont {Nisoli}, \citenamefont {Reichhardt},
  \citenamefont {Reichhardt},\ and\ \citenamefont
  {Tierno}}]{RevModPhys.91.041003}%
  \BibitemOpen
  \bibfield  {author} {\bibinfo {author} {\bibfnamefont {A.}~\bibnamefont
  {Ortiz-Ambriz}}, \bibinfo {author} {\bibfnamefont {C.}~\bibnamefont
  {Nisoli}}, \bibinfo {author} {\bibfnamefont {C.}~\bibnamefont {Reichhardt}},
  \bibinfo {author} {\bibfnamefont {C.~J.~O.}\ \bibnamefont {Reichhardt}}, \
  and\ \bibinfo {author} {\bibfnamefont {P.}~\bibnamefont {Tierno}},\ }\href
  {\doibase 10.1103/RevModPhys.91.041003} {\bibfield  {journal} {\bibinfo
  {journal} {Rev. Mod. Phys.}\ }\textbf {\bibinfo {volume} {91}},\ \bibinfo
  {pages} {041003} (\bibinfo {year} {2019})}\BibitemShut {NoStop}%
\bibitem [{\citenamefont {Zapf}\ \emph
  {et~al.}(2014{\natexlab{a}})\citenamefont {Zapf}, \citenamefont {Jaime},\
  and\ \citenamefont {Batista}}]{RevModPhys.86.563}%
  \BibitemOpen
  \bibfield  {author} {\bibinfo {author} {\bibfnamefont {V.}~\bibnamefont
  {Zapf}}, \bibinfo {author} {\bibfnamefont {M.}~\bibnamefont {Jaime}}, \ and\
  \bibinfo {author} {\bibfnamefont {C.~D.}\ \bibnamefont {Batista}},\ }\href
  {\doibase 10.1103/RevModPhys.86.563} {\bibfield  {journal} {\bibinfo
  {journal} {Rev. Mod. Phys.}\ }\textbf {\bibinfo {volume} {86}},\ \bibinfo
  {pages} {563} (\bibinfo {year} {2014}{\natexlab{a}})}\BibitemShut {NoStop}%
\bibitem [{\citenamefont {Zapf}\ \emph
  {et~al.}(2014{\natexlab{b}})\citenamefont {Zapf}, \citenamefont {Jaime},\
  and\ \citenamefont {Batista}}]{RevModPhys.86.1453}%
  \BibitemOpen
  \bibfield  {author} {\bibinfo {author} {\bibfnamefont {V.}~\bibnamefont
  {Zapf}}, \bibinfo {author} {\bibfnamefont {M.}~\bibnamefont {Jaime}}, \ and\
  \bibinfo {author} {\bibfnamefont {C.~D.}\ \bibnamefont {Batista}},\ }\href
  {\doibase 10.1103/RevModPhys.86.1453} {\bibfield  {journal} {\bibinfo
  {journal} {Rev. Mod. Phys.}\ }\textbf {\bibinfo {volume} {86}},\ \bibinfo
  {pages} {1453} (\bibinfo {year} {2014}{\natexlab{b}})}\BibitemShut {NoStop}%
\bibitem [{\citenamefont {F\'ath}\ \emph {et~al.}(2001)\citenamefont {F\'ath},
  \citenamefont {Legeza},\ and\ \citenamefont {S\'olyom}}]{PhysRevB.63.134403}%
  \BibitemOpen
  \bibfield  {author} {\bibinfo {author} {\bibfnamefont {G.}~\bibnamefont
  {F\'ath}}, \bibinfo {author} {\bibfnamefont {O.}~\bibnamefont {Legeza}}, \
  and\ \bibinfo {author} {\bibfnamefont {J.}~\bibnamefont {S\'olyom}},\ }\href
  {\doibase 10.1103/PhysRevB.63.134403} {\bibfield  {journal} {\bibinfo
  {journal} {Phys. Rev. B}\ }\textbf {\bibinfo {volume} {63}},\ \bibinfo
  {pages} {134403} (\bibinfo {year} {2001})}\BibitemShut {NoStop}%
\bibitem [{\citenamefont {Kim}\ \emph {et~al.}(2000)\citenamefont {Kim},
  \citenamefont {F\'ath}, \citenamefont {S\'olyom},\ and\ \citenamefont
  {Scalapino}}]{PhysRevB.62.14965}%
  \BibitemOpen
  \bibfield  {author} {\bibinfo {author} {\bibfnamefont {E.~H.}\ \bibnamefont
  {Kim}}, \bibinfo {author} {\bibfnamefont {G.}~\bibnamefont {F\'ath}},
  \bibinfo {author} {\bibfnamefont {J.}~\bibnamefont {S\'olyom}}, \ and\
  \bibinfo {author} {\bibfnamefont {D.~J.}\ \bibnamefont {Scalapino}},\ }\href
  {\doibase 10.1103/PhysRevB.62.14965} {\bibfield  {journal} {\bibinfo
  {journal} {Phys. Rev. B}\ }\textbf {\bibinfo {volume} {62}},\ \bibinfo
  {pages} {14965} (\bibinfo {year} {2000})}\BibitemShut {NoStop}%
\bibitem [{\citenamefont {Hijii}\ \emph {et~al.}(2005)\citenamefont {Hijii},
  \citenamefont {Kitazawa},\ and\ \citenamefont {Nomura}}]{PhysRevB.72.014449}%
  \BibitemOpen
  \bibfield  {author} {\bibinfo {author} {\bibfnamefont {K.}~\bibnamefont
  {Hijii}}, \bibinfo {author} {\bibfnamefont {A.}~\bibnamefont {Kitazawa}}, \
  and\ \bibinfo {author} {\bibfnamefont {K.}~\bibnamefont {Nomura}},\ }\href
  {\doibase 10.1103/PhysRevB.72.014449} {\bibfield  {journal} {\bibinfo
  {journal} {Phys. Rev. B}\ }\textbf {\bibinfo {volume} {72}},\ \bibinfo
  {pages} {014449} (\bibinfo {year} {2005})}\BibitemShut {NoStop}%
\bibitem [{\citenamefont {Karrasch}\ \emph {et~al.}(2015)\citenamefont
  {Karrasch}, \citenamefont {Kennes},\ and\ \citenamefont
  {Heidrich-Meisner}}]{PhysRevB.91.115130}%
  \BibitemOpen
  \bibfield  {author} {\bibinfo {author} {\bibfnamefont {C.}~\bibnamefont
  {Karrasch}}, \bibinfo {author} {\bibfnamefont {D.~M.}\ \bibnamefont
  {Kennes}}, \ and\ \bibinfo {author} {\bibfnamefont {F.}~\bibnamefont
  {Heidrich-Meisner}},\ }\href {\doibase 10.1103/PhysRevB.91.115130} {\bibfield
   {journal} {\bibinfo  {journal} {Phys. Rev. B}\ }\textbf {\bibinfo {volume}
  {91}},\ \bibinfo {pages} {115130} (\bibinfo {year} {2015})}\BibitemShut
  {NoStop}%
\bibitem [{\citenamefont {Ramos}\ and\ \citenamefont
  {Xavier}(2014)}]{PhysRevB.89.094424}%
  \BibitemOpen
  \bibfield  {author} {\bibinfo {author} {\bibfnamefont {F.~B.}\ \bibnamefont
  {Ramos}}\ and\ \bibinfo {author} {\bibfnamefont {J.~C.}\ \bibnamefont
  {Xavier}},\ }\href {\doibase 10.1103/PhysRevB.89.094424} {\bibfield
  {journal} {\bibinfo  {journal} {Phys. Rev. B}\ }\textbf {\bibinfo {volume}
  {89}},\ \bibinfo {pages} {094424} (\bibinfo {year} {2014})}\BibitemShut
  {NoStop}%
\bibitem [{\citenamefont {Aristov}\ \emph {et~al.}(2010)\citenamefont
  {Aristov}, \citenamefont {Br\"unger}, \citenamefont {Assaad}, \citenamefont
  {Kiselev}, \citenamefont {Weichselbaum}, \citenamefont {Capponi},\ and\
  \citenamefont {Alet}}]{PhysRevB.82.174410}%
  \BibitemOpen
  \bibfield  {author} {\bibinfo {author} {\bibfnamefont {D.~N.}\ \bibnamefont
  {Aristov}}, \bibinfo {author} {\bibfnamefont {C.}~\bibnamefont {Br\"unger}},
  \bibinfo {author} {\bibfnamefont {F.~F.}\ \bibnamefont {Assaad}}, \bibinfo
  {author} {\bibfnamefont {M.~N.}\ \bibnamefont {Kiselev}}, \bibinfo {author}
  {\bibfnamefont {A.}~\bibnamefont {Weichselbaum}}, \bibinfo {author}
  {\bibfnamefont {S.}~\bibnamefont {Capponi}}, \ and\ \bibinfo {author}
  {\bibfnamefont {F.}~\bibnamefont {Alet}},\ }\href {\doibase
  10.1103/PhysRevB.82.174410} {\bibfield  {journal} {\bibinfo  {journal} {Phys.
  Rev. B}\ }\textbf {\bibinfo {volume} {82}},\ \bibinfo {pages} {174410}
  (\bibinfo {year} {2010})}\BibitemShut {NoStop}%
\bibitem [{\citenamefont {Batchelor}\ \emph {et~al.}(2007)\citenamefont
  {Batchelor}, \citenamefont {Guan}, \citenamefont {Oelkers},\ and\
  \citenamefont {Tsuboi}}]{doi:10.1080/00018730701265383}%
  \BibitemOpen
  \bibfield  {author} {\bibinfo {author} {\bibfnamefont {M.~T.}\ \bibnamefont
  {Batchelor}}, \bibinfo {author} {\bibfnamefont {X.~W.}\ \bibnamefont {Guan}},
  \bibinfo {author} {\bibfnamefont {N.}~\bibnamefont {Oelkers}}, \ and\
  \bibinfo {author} {\bibfnamefont {Z.}~\bibnamefont {Tsuboi}},\ }\href
  {\doibase 10.1080/00018730701265383} {\bibfield  {journal} {\bibinfo
  {journal} {Advances in Physics}\ }\textbf {\bibinfo {volume} {56}},\ \bibinfo
  {pages} {465} (\bibinfo {year} {2007})},\ \Eprint
  {http://arxiv.org/abs/https://doi.org/10.1080/00018730701265383}
  {https://doi.org/10.1080/00018730701265383} \BibitemShut {NoStop}%
\bibitem [{\citenamefont {Hikihara}\ and\ \citenamefont
  {Starykh}(2010)}]{PhysRevB.81.064432}%
  \BibitemOpen
  \bibfield  {author} {\bibinfo {author} {\bibfnamefont {T.}~\bibnamefont
  {Hikihara}}\ and\ \bibinfo {author} {\bibfnamefont {O.~A.}\ \bibnamefont
  {Starykh}},\ }\href {\doibase 10.1103/PhysRevB.81.064432} {\bibfield
  {journal} {\bibinfo  {journal} {Phys. Rev. B}\ }\textbf {\bibinfo {volume}
  {81}},\ \bibinfo {pages} {064432} (\bibinfo {year} {2010})}\BibitemShut
  {NoStop}%
\bibitem [{\citenamefont {Mishra}\ \emph {et~al.}(2015)\citenamefont {Mishra},
  \citenamefont {Greschner},\ and\ \citenamefont
  {Santos}}]{PhysRevA.91.043614}%
  \BibitemOpen
  \bibfield  {author} {\bibinfo {author} {\bibfnamefont {T.}~\bibnamefont
  {Mishra}}, \bibinfo {author} {\bibfnamefont {S.}~\bibnamefont {Greschner}}, \
  and\ \bibinfo {author} {\bibfnamefont {L.}~\bibnamefont {Santos}},\ }\href
  {\doibase 10.1103/PhysRevA.91.043614} {\bibfield  {journal} {\bibinfo
  {journal} {Phys. Rev. A}\ }\textbf {\bibinfo {volume} {91}},\ \bibinfo
  {pages} {043614} (\bibinfo {year} {2015})}\BibitemShut {NoStop}%
\bibitem [{\citenamefont {Bhaseen}\ and\ \citenamefont
  {Tsvelik}(2003)}]{PhysRevB.68.094405}%
  \BibitemOpen
  \bibfield  {author} {\bibinfo {author} {\bibfnamefont {M.~J.}\ \bibnamefont
  {Bhaseen}}\ and\ \bibinfo {author} {\bibfnamefont {A.~M.}\ \bibnamefont
  {Tsvelik}},\ }\href {\doibase 10.1103/PhysRevB.68.094405} {\bibfield
  {journal} {\bibinfo  {journal} {Phys. Rev. B}\ }\textbf {\bibinfo {volume}
  {68}},\ \bibinfo {pages} {094405} (\bibinfo {year} {2003})}\BibitemShut
  {NoStop}%
\bibitem [{\citenamefont {Sugimoto}\ \emph {et~al.}(2018)\citenamefont
  {Sugimoto}, \citenamefont {Mori}, \citenamefont {Tohyama},\ and\
  \citenamefont {Maekawa}}]{PhysRevB.97.144424}%
  \BibitemOpen
  \bibfield  {author} {\bibinfo {author} {\bibfnamefont {T.}~\bibnamefont
  {Sugimoto}}, \bibinfo {author} {\bibfnamefont {M.}~\bibnamefont {Mori}},
  \bibinfo {author} {\bibfnamefont {T.}~\bibnamefont {Tohyama}}, \ and\
  \bibinfo {author} {\bibfnamefont {S.}~\bibnamefont {Maekawa}},\ }\href
  {\doibase 10.1103/PhysRevB.97.144424} {\bibfield  {journal} {\bibinfo
  {journal} {Phys. Rev. B}\ }\textbf {\bibinfo {volume} {97}},\ \bibinfo
  {pages} {144424} (\bibinfo {year} {2018})}\BibitemShut {NoStop}%
\bibitem [{\citenamefont {Sasaki}\ \emph {et~al.}(2020)\citenamefont {Sasaki},
  \citenamefont {Sugimoto}, \citenamefont {Tohyama},\ and\ \citenamefont
  {Sota}}]{PhysRevB.101.144407}%
  \BibitemOpen
  \bibfield  {author} {\bibinfo {author} {\bibfnamefont {K.}~\bibnamefont
  {Sasaki}}, \bibinfo {author} {\bibfnamefont {T.}~\bibnamefont {Sugimoto}},
  \bibinfo {author} {\bibfnamefont {T.}~\bibnamefont {Tohyama}}, \ and\
  \bibinfo {author} {\bibfnamefont {S.}~\bibnamefont {Sota}},\ }\href {\doibase
  10.1103/PhysRevB.101.144407} {\bibfield  {journal} {\bibinfo  {journal}
  {Phys. Rev. B}\ }\textbf {\bibinfo {volume} {101}},\ \bibinfo {pages}
  {144407} (\bibinfo {year} {2020})}\BibitemShut {NoStop}%
\bibitem [{\citenamefont {Hung}\ \emph {et~al.}(2006)\citenamefont {Hung},
  \citenamefont {Gong}, \citenamefont {Chen},\ and\ \citenamefont
  {Yang}}]{PhysRevB.73.224433}%
  \BibitemOpen
  \bibfield  {author} {\bibinfo {author} {\bibfnamefont {H.-H.}\ \bibnamefont
  {Hung}}, \bibinfo {author} {\bibfnamefont {C.-D.}\ \bibnamefont {Gong}},
  \bibinfo {author} {\bibfnamefont {Y.-C.}\ \bibnamefont {Chen}}, \ and\
  \bibinfo {author} {\bibfnamefont {M.-F.}\ \bibnamefont {Yang}},\ }\href
  {\doibase 10.1103/PhysRevB.73.224433} {\bibfield  {journal} {\bibinfo
  {journal} {Phys. Rev. B}\ }\textbf {\bibinfo {volume} {73}},\ \bibinfo
  {pages} {224433} (\bibinfo {year} {2006})}\BibitemShut {NoStop}%
\bibitem [{\citenamefont {Barbarino}\ \emph {et~al.}(2018)\citenamefont
  {Barbarino}, \citenamefont {Dalmonte}, \citenamefont {Fazio},\ and\
  \citenamefont {Santoro}}]{PhysRevA.97.013634}%
  \BibitemOpen
  \bibfield  {author} {\bibinfo {author} {\bibfnamefont {S.}~\bibnamefont
  {Barbarino}}, \bibinfo {author} {\bibfnamefont {M.}~\bibnamefont {Dalmonte}},
  \bibinfo {author} {\bibfnamefont {R.}~\bibnamefont {Fazio}}, \ and\ \bibinfo
  {author} {\bibfnamefont {G.~E.}\ \bibnamefont {Santoro}},\ }\href {\doibase
  10.1103/PhysRevA.97.013634} {\bibfield  {journal} {\bibinfo  {journal} {Phys.
  Rev. A}\ }\textbf {\bibinfo {volume} {97}},\ \bibinfo {pages} {013634}
  (\bibinfo {year} {2018})}\BibitemShut {NoStop}%
\bibitem [{\citenamefont {Vekua}\ and\ \citenamefont
  {Honecker}(2006)}]{PhysRevB.73.214427}%
  \BibitemOpen
  \bibfield  {author} {\bibinfo {author} {\bibfnamefont {T.}~\bibnamefont
  {Vekua}}\ and\ \bibinfo {author} {\bibfnamefont {A.}~\bibnamefont
  {Honecker}},\ }\href {\doibase 10.1103/PhysRevB.73.214427} {\bibfield
  {journal} {\bibinfo  {journal} {Phys. Rev. B}\ }\textbf {\bibinfo {volume}
  {73}},\ \bibinfo {pages} {214427} (\bibinfo {year} {2006})}\BibitemShut
  {NoStop}%
\bibitem [{\citenamefont {Liu}\ \emph {et~al.}(2008)\citenamefont {Liu},
  \citenamefont {Wang},\ and\ \citenamefont {Tian}}]{PhysRevB.77.214418}%
  \BibitemOpen
  \bibfield  {author} {\bibinfo {author} {\bibfnamefont {G.-H.}\ \bibnamefont
  {Liu}}, \bibinfo {author} {\bibfnamefont {H.-L.}\ \bibnamefont {Wang}}, \
  and\ \bibinfo {author} {\bibfnamefont {G.-S.}\ \bibnamefont {Tian}},\ }\href
  {\doibase 10.1103/PhysRevB.77.214418} {\bibfield  {journal} {\bibinfo
  {journal} {Phys. Rev. B}\ }\textbf {\bibinfo {volume} {77}},\ \bibinfo
  {pages} {214418} (\bibinfo {year} {2008})}\BibitemShut {NoStop}%
\bibitem [{\citenamefont {Dagotto}\ and\ \citenamefont
  {Rice}(1996)}]{Dagotto618}%
  \BibitemOpen
  \bibfield  {author} {\bibinfo {author} {\bibfnamefont {E.}~\bibnamefont
  {Dagotto}}\ and\ \bibinfo {author} {\bibfnamefont {T.~M.}\ \bibnamefont
  {Rice}},\ }\href {\doibase 10.1126/science.271.5249.618} {\bibfield
  {journal} {\bibinfo  {journal} {Science}\ }\textbf {\bibinfo {volume}
  {271}},\ \bibinfo {pages} {618} (\bibinfo {year} {1996})}\BibitemShut
  {NoStop}%
\bibitem [{\citenamefont {Allen}\ \emph {et~al.}(2000)\citenamefont {Allen},
  \citenamefont {Essler},\ and\ \citenamefont {Nersesyan}}]{PhysRevB.61.8871}%
  \BibitemOpen
  \bibfield  {author} {\bibinfo {author} {\bibfnamefont {D.}~\bibnamefont
  {Allen}}, \bibinfo {author} {\bibfnamefont {F.~H.~L.}\ \bibnamefont
  {Essler}}, \ and\ \bibinfo {author} {\bibfnamefont {A.~A.}\ \bibnamefont
  {Nersesyan}},\ }\href {\doibase 10.1103/PhysRevB.61.8871} {\bibfield
  {journal} {\bibinfo  {journal} {Phys. Rev. B}\ }\textbf {\bibinfo {volume}
  {61}},\ \bibinfo {pages} {8871} (\bibinfo {year} {2000})}\BibitemShut
  {NoStop}%
\bibitem [{\citenamefont {Fouet}\ \emph {et~al.}(2006)\citenamefont {Fouet},
  \citenamefont {Mila}, \citenamefont {Clarke}, \citenamefont {Youk},
  \citenamefont {Tchernyshyov}, \citenamefont {Fendley},\ and\ \citenamefont
  {Noack}}]{PhysRevB.73.214405}%
  \BibitemOpen
  \bibfield  {author} {\bibinfo {author} {\bibfnamefont {J.-B.}\ \bibnamefont
  {Fouet}}, \bibinfo {author} {\bibfnamefont {F.}~\bibnamefont {Mila}},
  \bibinfo {author} {\bibfnamefont {D.}~\bibnamefont {Clarke}}, \bibinfo
  {author} {\bibfnamefont {H.}~\bibnamefont {Youk}}, \bibinfo {author}
  {\bibfnamefont {O.}~\bibnamefont {Tchernyshyov}}, \bibinfo {author}
  {\bibfnamefont {P.}~\bibnamefont {Fendley}}, \ and\ \bibinfo {author}
  {\bibfnamefont {R.~M.}\ \bibnamefont {Noack}},\ }\href {\doibase
  10.1103/PhysRevB.73.214405} {\bibfield  {journal} {\bibinfo  {journal} {Phys.
  Rev. B}\ }\textbf {\bibinfo {volume} {73}},\ \bibinfo {pages} {214405}
  (\bibinfo {year} {2006})}\BibitemShut {NoStop}%
\bibitem [{\citenamefont {Abouie}\ and\ \citenamefont
  {Langari}(2004)}]{PhysRevB.70.184416}%
  \BibitemOpen
  \bibfield  {author} {\bibinfo {author} {\bibfnamefont {J.}~\bibnamefont
  {Abouie}}\ and\ \bibinfo {author} {\bibfnamefont {A.}~\bibnamefont
  {Langari}},\ }\href {\doibase 10.1103/PhysRevB.70.184416} {\bibfield
  {journal} {\bibinfo  {journal} {Phys. Rev. B}\ }\textbf {\bibinfo {volume}
  {70}},\ \bibinfo {pages} {184416} (\bibinfo {year} {2004})}\BibitemShut
  {NoStop}%
\bibitem [{\citenamefont {Abouie}\ \emph {et~al.}(2006)\citenamefont {Abouie},
  \citenamefont {Ghasemi},\ and\ \citenamefont {Langari}}]{PhysRevB.73.014411}%
  \BibitemOpen
  \bibfield  {author} {\bibinfo {author} {\bibfnamefont {J.}~\bibnamefont
  {Abouie}}, \bibinfo {author} {\bibfnamefont {S.~A.}\ \bibnamefont {Ghasemi}},
  \ and\ \bibinfo {author} {\bibfnamefont {A.}~\bibnamefont {Langari}},\ }\href
  {\doibase 10.1103/PhysRevB.73.014411} {\bibfield  {journal} {\bibinfo
  {journal} {Phys. Rev. B}\ }\textbf {\bibinfo {volume} {73}},\ \bibinfo
  {pages} {014411} (\bibinfo {year} {2006})}\BibitemShut {NoStop}%
\bibitem [{\citenamefont {Heydarinasab}\ and\ \citenamefont
  {Abouie}(2017)}]{PhysRevB.96.104406}%
  \BibitemOpen
  \bibfield  {author} {\bibinfo {author} {\bibfnamefont {F.}~\bibnamefont
  {Heydarinasab}}\ and\ \bibinfo {author} {\bibfnamefont {J.}~\bibnamefont
  {Abouie}},\ }\href {\doibase 10.1103/PhysRevB.96.104406} {\bibfield
  {journal} {\bibinfo  {journal} {Phys. Rev. B}\ }\textbf {\bibinfo {volume}
  {96}},\ \bibinfo {pages} {104406} (\bibinfo {year} {2017})}\BibitemShut
  {NoStop}%
\bibitem [{\citenamefont {Rezai}\ \emph {et~al.}(2010)\citenamefont {Rezai},
  \citenamefont {Langari},\ and\ \citenamefont {Abouie}}]{PhysRevB.81.060401}%
  \BibitemOpen
  \bibfield  {author} {\bibinfo {author} {\bibfnamefont {M.}~\bibnamefont
  {Rezai}}, \bibinfo {author} {\bibfnamefont {A.}~\bibnamefont {Langari}}, \
  and\ \bibinfo {author} {\bibfnamefont {J.}~\bibnamefont {Abouie}},\ }\href
  {\doibase 10.1103/PhysRevB.81.060401} {\bibfield  {journal} {\bibinfo
  {journal} {Phys. Rev. B}\ }\textbf {\bibinfo {volume} {81}},\ \bibinfo
  {pages} {060401} (\bibinfo {year} {2010})}\BibitemShut {NoStop}%
\bibitem [{\citenamefont {Heydarinasab}\ and\ \citenamefont
  {Abouie}(2020)}]{Heydarinasab_2020}%
  \BibitemOpen
  \bibfield  {author} {\bibinfo {author} {\bibfnamefont {F.}~\bibnamefont
  {Heydarinasab}}\ and\ \bibinfo {author} {\bibfnamefont {J.}~\bibnamefont
  {Abouie}},\ }\href {\doibase 10.1088/1361-648x/ab61ca} {\bibfield  {journal}
  {\bibinfo  {journal} {Journal of Physics: Condensed Matter}\ }\textbf
  {\bibinfo {volume} {32}},\ \bibinfo {pages} {165804} (\bibinfo {year}
  {2020})}\BibitemShut {NoStop}%
\bibitem [{\citenamefont {Langari}\ \emph {et~al.}(2011)\citenamefont
  {Langari}, \citenamefont {Abouie}, \citenamefont {Asadzadeh},\ and\
  \citenamefont {Rezai}}]{Langari_2011}%
  \BibitemOpen
  \bibfield  {author} {\bibinfo {author} {\bibfnamefont {A.}~\bibnamefont
  {Langari}}, \bibinfo {author} {\bibfnamefont {J.}~\bibnamefont {Abouie}},
  \bibinfo {author} {\bibfnamefont {M.~Z.}\ \bibnamefont {Asadzadeh}}, \ and\
  \bibinfo {author} {\bibfnamefont {M.}~\bibnamefont {Rezai}},\ }\href
  {\doibase 10.1088/1742-5468/2011/08/p08001} {\bibfield  {journal} {\bibinfo
  {journal} {Journal of Statistical Mechanics: Theory and Experiment}\ }\textbf
  {\bibinfo {volume} {2011}},\ \bibinfo {pages} {P08001} (\bibinfo {year}
  {2011})}\BibitemShut {NoStop}%
\bibitem [{\citenamefont {Lieb}\ and\ \citenamefont {Mattis}(1962)}]{Lieb_62}%
  \BibitemOpen
  \bibfield  {author} {\bibinfo {author} {\bibfnamefont {E.}~\bibnamefont
  {Lieb}}\ and\ \bibinfo {author} {\bibfnamefont {D.}~\bibnamefont {Mattis}},\
  }\href@noop {} {\bibfield  {journal} {\bibinfo  {journal} {J. Math. Phys.}\
  }\textbf {\bibinfo {volume} {3}},\ \bibinfo {pages} {749} (\bibinfo {year}
  {1962})}\BibitemShut {NoStop}%
\bibitem [{\citenamefont {Chandra}\ \emph {et~al.}(2010)\citenamefont
  {Chandra}, \citenamefont {Ivanov},\ and\ \citenamefont
  {Richter}}]{PhysRevB.81.024409}%
  \BibitemOpen
  \bibfield  {author} {\bibinfo {author} {\bibfnamefont {V.~R.}\ \bibnamefont
  {Chandra}}, \bibinfo {author} {\bibfnamefont {N.~B.}\ \bibnamefont {Ivanov}},
  \ and\ \bibinfo {author} {\bibfnamefont {J.}~\bibnamefont {Richter}},\ }\href
  {\doibase 10.1103/PhysRevB.81.024409} {\bibfield  {journal} {\bibinfo
  {journal} {Phys. Rev. B}\ }\textbf {\bibinfo {volume} {81}},\ \bibinfo
  {pages} {024409} (\bibinfo {year} {2010})}\BibitemShut {NoStop}%
\bibitem [{\citenamefont {Starykh}\ and\ \citenamefont
  {Balents}(2004)}]{PhysRevLett.93.127202}%
  \BibitemOpen
  \bibfield  {author} {\bibinfo {author} {\bibfnamefont {O.~A.}\ \bibnamefont
  {Starykh}}\ and\ \bibinfo {author} {\bibfnamefont {L.}~\bibnamefont
  {Balents}},\ }\href {\doibase 10.1103/PhysRevLett.93.127202} {\bibfield
  {journal} {\bibinfo  {journal} {Phys. Rev. Lett.}\ }\textbf {\bibinfo
  {volume} {93}},\ \bibinfo {pages} {127202} (\bibinfo {year}
  {2004})}\BibitemShut {NoStop}%
\bibitem [{\citenamefont {Kim}\ \emph {et~al.}(2008)\citenamefont {Kim},
  \citenamefont {Legeza},\ and\ \citenamefont {S\'olyom}}]{PhysRevB.77.205121}%
  \BibitemOpen
  \bibfield  {author} {\bibinfo {author} {\bibfnamefont {E.~H.}\ \bibnamefont
  {Kim}}, \bibinfo {author} {\bibfnamefont {O.}~\bibnamefont {Legeza}}, \ and\
  \bibinfo {author} {\bibfnamefont {J.}~\bibnamefont {S\'olyom}},\ }\href
  {\doibase 10.1103/PhysRevB.77.205121} {\bibfield  {journal} {\bibinfo
  {journal} {Phys. Rev. B}\ }\textbf {\bibinfo {volume} {77}},\ \bibinfo
  {pages} {205121} (\bibinfo {year} {2008})}\BibitemShut {NoStop}%
\bibitem [{\citenamefont {Weihong}\ \emph {et~al.}(1998)\citenamefont
  {Weihong}, \citenamefont {Kotov},\ and\ \citenamefont
  {Oitmaa}}]{PhysRevB.57.11439}%
  \BibitemOpen
  \bibfield  {author} {\bibinfo {author} {\bibfnamefont {Z.}~\bibnamefont
  {Weihong}}, \bibinfo {author} {\bibfnamefont {V.}~\bibnamefont {Kotov}}, \
  and\ \bibinfo {author} {\bibfnamefont {J.}~\bibnamefont {Oitmaa}},\ }\href
  {\doibase 10.1103/PhysRevB.57.11439} {\bibfield  {journal} {\bibinfo
  {journal} {Phys. Rev. B}\ }\textbf {\bibinfo {volume} {57}},\ \bibinfo
  {pages} {11439} (\bibinfo {year} {1998})}\BibitemShut {NoStop}%
\bibitem [{\citenamefont {Barcza}\ \emph {et~al.}(2012)\citenamefont {Barcza},
  \citenamefont {Legeza}, \citenamefont {Noack},\ and\ \citenamefont
  {S\'olyom}}]{PhysRevB.86.075133}%
  \BibitemOpen
  \bibfield  {author} {\bibinfo {author} {\bibfnamefont {G.}~\bibnamefont
  {Barcza}}, \bibinfo {author} {\bibfnamefont {O.}~\bibnamefont {Legeza}},
  \bibinfo {author} {\bibfnamefont {R.~M.}\ \bibnamefont {Noack}}, \ and\
  \bibinfo {author} {\bibfnamefont {J.}~\bibnamefont {S\'olyom}},\ }\href
  {\doibase 10.1103/PhysRevB.86.075133} {\bibfield  {journal} {\bibinfo
  {journal} {Phys. Rev. B}\ }\textbf {\bibinfo {volume} {86}},\ \bibinfo
  {pages} {075133} (\bibinfo {year} {2012})}\BibitemShut {NoStop}%
\bibitem [{\citenamefont {Chen}\ \emph {et~al.}(2017)\citenamefont {Chen},
  \citenamefont {Cho}, \citenamefont {Batchelor},\ and\ \citenamefont
  {Zhou}}]{Chen_16}%
  \BibitemOpen
  \bibfield  {author} {\bibinfo {author} {\bibfnamefont {X.~H.}\ \bibnamefont
  {Chen}}, \bibinfo {author} {\bibfnamefont {S.~Y.}\ \bibnamefont {Cho}},
  \bibinfo {author} {\bibfnamefont {M.~T.}\ \bibnamefont {Batchelor}}, \ and\
  \bibinfo {author} {\bibfnamefont {H.~Q.}\ \bibnamefont {Zhou}},\ }\href
  {\doibase 10.21468/SciPostPhys.3.1.005} {\bibfield  {journal} {\bibinfo
  {journal} {SciPost Phys.}\ }\textbf {\bibinfo {volume} {3}},\ \bibinfo
  {pages} {005} (\bibinfo {year} {2017})}\BibitemShut {NoStop}%
\bibitem [{\citenamefont {Wessel}\ \emph {et~al.}(2016)\citenamefont {Wessel},
  \citenamefont {Normand}, \citenamefont {Mila},\ and\ \citenamefont
  {Honecker}}]{SciPostPhys.3.1.005}%
  \BibitemOpen
  \bibfield  {author} {\bibinfo {author} {\bibfnamefont {S.}~\bibnamefont
  {Wessel}}, \bibinfo {author} {\bibfnamefont {B.}~\bibnamefont {Normand}},
  \bibinfo {author} {\bibfnamefont {F.}~\bibnamefont {Mila}}, \ and\ \bibinfo
  {author} {\bibfnamefont {A.}~\bibnamefont {Honecker}},\ }\href@noop {}
  {\bibfield  {journal} {\bibinfo  {journal} {J.. Korean Phys. Soc.}\ }\textbf
  {\bibinfo {volume} {68}},\ \bibinfo {pages} {1114} (\bibinfo {year}
  {2016})}\BibitemShut {NoStop}%
\bibitem [{\citenamefont {Poilblanc}(2010)}]{PhysRevLett.105.077202}%
  \BibitemOpen
  \bibfield  {author} {\bibinfo {author} {\bibfnamefont {D.}~\bibnamefont
  {Poilblanc}},\ }\href {\doibase 10.1103/PhysRevLett.105.077202} {\bibfield
  {journal} {\bibinfo  {journal} {Phys. Rev. Lett.}\ }\textbf {\bibinfo
  {volume} {105}},\ \bibinfo {pages} {077202} (\bibinfo {year}
  {2010})}\BibitemShut {NoStop}%
\bibitem [{\citenamefont {White}(1992)}]{PhysRevLett.69.2863}%
  \BibitemOpen
  \bibfield  {author} {\bibinfo {author} {\bibfnamefont {S.~R.}\ \bibnamefont
  {White}},\ }\href {\doibase 10.1103/PhysRevLett.69.2863} {\bibfield
  {journal} {\bibinfo  {journal} {Phys. Rev. Lett.}\ }\textbf {\bibinfo
  {volume} {69}},\ \bibinfo {pages} {2863} (\bibinfo {year}
  {1992})}\BibitemShut {NoStop}%
\bibitem [{\citenamefont {Vidal}(2003)}]{PhysRevLett.91.147902}%
  \BibitemOpen
  \bibfield  {author} {\bibinfo {author} {\bibfnamefont {G.}~\bibnamefont
  {Vidal}},\ }\href {\doibase 10.1103/PhysRevLett.91.147902} {\bibfield
  {journal} {\bibinfo  {journal} {Phys. Rev. Lett.}\ }\textbf {\bibinfo
  {volume} {91}},\ \bibinfo {pages} {147902} (\bibinfo {year}
  {2003})}\BibitemShut {NoStop}%
\bibitem [{\citenamefont {Schollw\"ock}(2005)}]{RevModPhys.77.259}%
  \BibitemOpen
  \bibfield  {author} {\bibinfo {author} {\bibfnamefont {U.}~\bibnamefont
  {Schollw\"ock}},\ }\href {\doibase 10.1103/RevModPhys.77.259} {\bibfield
  {journal} {\bibinfo  {journal} {Rev. Mod. Phys.}\ }\textbf {\bibinfo {volume}
  {77}},\ \bibinfo {pages} {259} (\bibinfo {year} {2005})}\BibitemShut
  {NoStop}%
\bibitem [{\citenamefont {Vidal}(2004)}]{PhysRevLett.93.040502}%
  \BibitemOpen
  \bibfield  {author} {\bibinfo {author} {\bibfnamefont {G.}~\bibnamefont
  {Vidal}},\ }\href {\doibase 10.1103/PhysRevLett.93.040502} {\bibfield
  {journal} {\bibinfo  {journal} {Phys. Rev. Lett.}\ }\textbf {\bibinfo
  {volume} {93}},\ \bibinfo {pages} {040502} (\bibinfo {year}
  {2004})}\BibitemShut {NoStop}%
\bibitem [{\citenamefont {Jordan}\ \emph {et~al.}(2008)\citenamefont {Jordan},
  \citenamefont {Or\'us}, \citenamefont {Vidal}, \citenamefont {Verstraete},\
  and\ \citenamefont {Cirac}}]{PhysRevLett.101.250602}%
  \BibitemOpen
  \bibfield  {author} {\bibinfo {author} {\bibfnamefont {J.}~\bibnamefont
  {Jordan}}, \bibinfo {author} {\bibfnamefont {R.}~\bibnamefont {Or\'us}},
  \bibinfo {author} {\bibfnamefont {G.}~\bibnamefont {Vidal}}, \bibinfo
  {author} {\bibfnamefont {F.}~\bibnamefont {Verstraete}}, \ and\ \bibinfo
  {author} {\bibfnamefont {J.~I.}\ \bibnamefont {Cirac}},\ }\href {\doibase
  10.1103/PhysRevLett.101.250602} {\bibfield  {journal} {\bibinfo  {journal}
  {Phys. Rev. Lett.}\ }\textbf {\bibinfo {volume} {101}},\ \bibinfo {pages}
  {250602} (\bibinfo {year} {2008})}\BibitemShut {NoStop}%
\bibitem [{\citenamefont {Haghshenas}\ and\ \citenamefont
  {Sheng}(2018)}]{PhysRevB.97.174408}%
  \BibitemOpen
  \bibfield  {author} {\bibinfo {author} {\bibfnamefont {R.}~\bibnamefont
  {Haghshenas}}\ and\ \bibinfo {author} {\bibfnamefont {D.~N.}\ \bibnamefont
  {Sheng}},\ }\href {\doibase 10.1103/PhysRevB.97.174408} {\bibfield  {journal}
  {\bibinfo  {journal} {Phys. Rev. B}\ }\textbf {\bibinfo {volume} {97}},\
  \bibinfo {pages} {174408} (\bibinfo {year} {2018})}\BibitemShut {NoStop}%
\bibitem [{\citenamefont {Haghshenas}\ \emph
  {et~al.}(2019{\natexlab{a}})\citenamefont {Haghshenas}, \citenamefont
  {Gong},\ and\ \citenamefont {Sheng}}]{Haghshenas:2019}%
  \BibitemOpen
  \bibfield  {author} {\bibinfo {author} {\bibfnamefont {R.}~\bibnamefont
  {Haghshenas}}, \bibinfo {author} {\bibfnamefont {S.-S.}\ \bibnamefont
  {Gong}}, \ and\ \bibinfo {author} {\bibfnamefont {D.~N.}\ \bibnamefont
  {Sheng}},\ }\href {\doibase 10.1103/PhysRevB.99.174423} {\bibfield  {journal}
  {\bibinfo  {journal} {Phys. Rev. B}\ }\textbf {\bibinfo {volume} {99}},\
  \bibinfo {pages} {174423} (\bibinfo {year} {2019}{\natexlab{a}})}\BibitemShut
  {NoStop}%
\bibitem [{\citenamefont {Haghshenas}\ \emph
  {et~al.}(2019{\natexlab{b}})\citenamefont {Haghshenas}, \citenamefont
  {O'Rourke},\ and\ \citenamefont {Chan}}]{PhysRevB.100.054404}%
  \BibitemOpen
  \bibfield  {author} {\bibinfo {author} {\bibfnamefont {R.}~\bibnamefont
  {Haghshenas}}, \bibinfo {author} {\bibfnamefont {M.~J.}\ \bibnamefont
  {O'Rourke}}, \ and\ \bibinfo {author} {\bibfnamefont {G.~K.-L.}\ \bibnamefont
  {Chan}},\ }\href {\doibase 10.1103/PhysRevB.100.054404} {\bibfield  {journal}
  {\bibinfo  {journal} {Phys. Rev. B}\ }\textbf {\bibinfo {volume} {100}},\
  \bibinfo {pages} {054404} (\bibinfo {year} {2019}{\natexlab{b}})}\BibitemShut
  {NoStop}%
\bibitem [{\citenamefont {Osterloh}\ \emph {et~al.}(2002)\citenamefont
  {Osterloh}, \citenamefont {Amico}, \citenamefont {Falci},\ and\ \citenamefont
  {Fazio}}]{osterlo}%
  \BibitemOpen
  \bibfield  {author} {\bibinfo {author} {\bibfnamefont {A.}~\bibnamefont
  {Osterloh}}, \bibinfo {author} {\bibfnamefont {L.}~\bibnamefont {Amico}},
  \bibinfo {author} {\bibfnamefont {G.}~\bibnamefont {Falci}}, \ and\ \bibinfo
  {author} {\bibfnamefont {R.}~\bibnamefont {Fazio}},\ }\href
  {http://dx.doi.org/10.1038/416608a} {\bibfield  {journal} {\bibinfo
  {journal} {Nature}\ }\textbf {\bibinfo {volume} {416}},\ \bibinfo {pages}
  {608} (\bibinfo {year} {2002})}\BibitemShut {NoStop}%
\bibitem [{\citenamefont {Levin}\ and\ \citenamefont
  {Wen}(2006)}]{PhysRevLett.96.110405}%
  \BibitemOpen
  \bibfield  {author} {\bibinfo {author} {\bibfnamefont {M.}~\bibnamefont
  {Levin}}\ and\ \bibinfo {author} {\bibfnamefont {X.-G.}\ \bibnamefont
  {Wen}},\ }\href {\doibase 10.1103/PhysRevLett.96.110405} {\bibfield
  {journal} {\bibinfo  {journal} {Phys. Rev. Lett.}\ }\textbf {\bibinfo
  {volume} {96}},\ \bibinfo {pages} {110405} (\bibinfo {year}
  {2006})}\BibitemShut {NoStop}%
\bibitem [{\citenamefont {Kitaev}\ and\ \citenamefont
  {Preskill}(2006)}]{PhysRevLett.96.110404}%
  \BibitemOpen
  \bibfield  {author} {\bibinfo {author} {\bibfnamefont {A.}~\bibnamefont
  {Kitaev}}\ and\ \bibinfo {author} {\bibfnamefont {J.}~\bibnamefont
  {Preskill}},\ }\href {\doibase 10.1103/PhysRevLett.96.110404} {\bibfield
  {journal} {\bibinfo  {journal} {Phys. Rev. Lett.}\ }\textbf {\bibinfo
  {volume} {96}},\ \bibinfo {pages} {110404} (\bibinfo {year}
  {2006})}\BibitemShut {NoStop}%
\bibitem [{\citenamefont {Li}\ and\ \citenamefont
  {Haldane}(2008)}]{PhysRevLett.101.010504}%
  \BibitemOpen
  \bibfield  {author} {\bibinfo {author} {\bibfnamefont {H.}~\bibnamefont
  {Li}}\ and\ \bibinfo {author} {\bibfnamefont {F.~D.~M.}\ \bibnamefont
  {Haldane}},\ }\href {\doibase 10.1103/PhysRevLett.101.010504} {\bibfield
  {journal} {\bibinfo  {journal} {Phys. Rev. Lett.}\ }\textbf {\bibinfo
  {volume} {101}},\ \bibinfo {pages} {010504} (\bibinfo {year}
  {2008})}\BibitemShut {NoStop}%
\bibitem [{\citenamefont {Pollmann}\ \emph {et~al.}(2010)\citenamefont
  {Pollmann}, \citenamefont {Turner}, \citenamefont {Berg},\ and\ \citenamefont
  {Oshikawa}}]{PhysRevB.81.064439}%
  \BibitemOpen
  \bibfield  {author} {\bibinfo {author} {\bibfnamefont {F.}~\bibnamefont
  {Pollmann}}, \bibinfo {author} {\bibfnamefont {A.~M.}\ \bibnamefont
  {Turner}}, \bibinfo {author} {\bibfnamefont {E.}~\bibnamefont {Berg}}, \ and\
  \bibinfo {author} {\bibfnamefont {M.}~\bibnamefont {Oshikawa}},\ }\href
  {\doibase 10.1103/PhysRevB.81.064439} {\bibfield  {journal} {\bibinfo
  {journal} {Phys. Rev. B}\ }\textbf {\bibinfo {volume} {81}},\ \bibinfo
  {pages} {064439} (\bibinfo {year} {2010})}\BibitemShut {NoStop}%
\bibitem [{\citenamefont {Haghshenas}\ \emph {et~al.}(2014)\citenamefont
  {Haghshenas}, \citenamefont {Langari},\ and\ \citenamefont
  {Rezakhani}}]{Haghshenas:2014}%
  \BibitemOpen
  \bibfield  {author} {\bibinfo {author} {\bibfnamefont {R.}~\bibnamefont
  {Haghshenas}}, \bibinfo {author} {\bibfnamefont {A.}~\bibnamefont {Langari}},
  \ and\ \bibinfo {author} {\bibfnamefont {A.~T.}\ \bibnamefont {Rezakhani}},\
  }\href {http://stacks.iop.org/0953-8984/26/i=45/a=456001} {\bibfield
  {journal} {\bibinfo  {journal} {Journal of Physics: Condensed Matter}\
  }\textbf {\bibinfo {volume} {26}},\ \bibinfo {pages} {456001} (\bibinfo
  {year} {2014})}\BibitemShut {NoStop}%
\bibitem [{\citenamefont {Ahmadi}\ \emph {et~al.}(2020)\citenamefont {Ahmadi},
  \citenamefont {Abouie},\ and\ \citenamefont {Baeriswyl}}]{Ahmadi2020}%
  \BibitemOpen
  \bibfield  {author} {\bibinfo {author} {\bibfnamefont {N.}~\bibnamefont
  {Ahmadi}}, \bibinfo {author} {\bibfnamefont {J.}~\bibnamefont {Abouie}}, \
  and\ \bibinfo {author} {\bibfnamefont {D.}~\bibnamefont {Baeriswyl}},\ }\href
  {\doibase 10.1103/PhysRevB.101.195117} {\bibfield  {journal} {\bibinfo
  {journal} {Phys. Rev. B}\ }\textbf {\bibinfo {volume} {101}},\ \bibinfo
  {pages} {195117} (\bibinfo {year} {2020})}\BibitemShut {NoStop}%
\bibitem [{\citenamefont {Nakamura}(1999)}]{Nakamura-JPSJ}%
  \BibitemOpen
  \bibfield  {author} {\bibinfo {author} {\bibfnamefont {M.}~\bibnamefont
  {Nakamura}},\ }\href@noop {} {\bibfield  {journal} {\bibinfo  {journal}
  {Journal of the Physical Society of Japan}\ }\textbf {\bibinfo {volume}
  {68}},\ \bibinfo {pages} {3123} (\bibinfo {year} {1999})}\BibitemShut
  {NoStop}%
\bibitem [{\citenamefont {Nakamura}(2000)}]{Nakamura-PRB}%
  \BibitemOpen
  \bibfield  {author} {\bibinfo {author} {\bibfnamefont {M.}~\bibnamefont
  {Nakamura}},\ }\href {\doibase 10.1103/PhysRevB.61.16377} {\bibfield
  {journal} {\bibinfo  {journal} {Phys. Rev. B}\ }\textbf {\bibinfo {volume}
  {61}},\ \bibinfo {pages} {16377} (\bibinfo {year} {2000})}\BibitemShut
  {NoStop}%
\bibitem [{\citenamefont {Sengupta}\ \emph {et~al.}(2002)\citenamefont
  {Sengupta}, \citenamefont {Sandvik},\ and\ \citenamefont
  {Campbell}}]{Sangupta}%
  \BibitemOpen
  \bibfield  {author} {\bibinfo {author} {\bibfnamefont {P.}~\bibnamefont
  {Sengupta}}, \bibinfo {author} {\bibfnamefont {A.~W.}\ \bibnamefont
  {Sandvik}}, \ and\ \bibinfo {author} {\bibfnamefont {D.~K.}\ \bibnamefont
  {Campbell}},\ }\href {\doibase 10.1103/PhysRevB.65.155113} {\bibfield
  {journal} {\bibinfo  {journal} {Phys. Rev. B}\ }\textbf {\bibinfo {volume}
  {65}},\ \bibinfo {pages} {155113} (\bibinfo {year} {2002})}\BibitemShut
  {NoStop}%
\bibitem [{\citenamefont {White}\ and\ \citenamefont
  {Affleck}(1996)}]{White-Affleck}%
  \BibitemOpen
  \bibfield  {author} {\bibinfo {author} {\bibfnamefont {S.~R.}\ \bibnamefont
  {White}}\ and\ \bibinfo {author} {\bibfnamefont {I.}~\bibnamefont
  {Affleck}},\ }\href {\doibase 10.1103/PhysRevB.54.9862} {\bibfield  {journal}
  {\bibinfo  {journal} {Phys. Rev. B}\ }\textbf {\bibinfo {volume} {54}},\
  \bibinfo {pages} {9862} (\bibinfo {year} {1996})}\BibitemShut {NoStop}%
\bibitem [{\citenamefont {Itoi}\ and\ \citenamefont {Qin}(2001)}]{J1J2-chigak}%
  \BibitemOpen
  \bibfield  {author} {\bibinfo {author} {\bibfnamefont {C.}~\bibnamefont
  {Itoi}}\ and\ \bibinfo {author} {\bibfnamefont {S.}~\bibnamefont {Qin}},\
  }\href {\doibase 10.1103/PhysRevB.63.224423} {\bibfield  {journal} {\bibinfo
  {journal} {Phys. Rev. B}\ }\textbf {\bibinfo {volume} {63}},\ \bibinfo
  {pages} {224423} (\bibinfo {year} {2001})}\BibitemShut {NoStop}%
\bibitem [{\citenamefont {Kumar}\ \emph
  {et~al.}(2010{\natexlab{a}})\citenamefont {Kumar}, \citenamefont
  {Ramasesha},\ and\ \citenamefont {Soos}}]{Kumar1}%
  \BibitemOpen
  \bibfield  {author} {\bibinfo {author} {\bibfnamefont {M.}~\bibnamefont
  {Kumar}}, \bibinfo {author} {\bibfnamefont {S.}~\bibnamefont {Ramasesha}}, \
  and\ \bibinfo {author} {\bibfnamefont {Z.~G.}\ \bibnamefont {Soos}},\ }\href
  {\doibase 10.1103/PhysRevB.81.054413} {\bibfield  {journal} {\bibinfo
  {journal} {Phys. Rev. B}\ }\textbf {\bibinfo {volume} {81}},\ \bibinfo
  {pages} {054413} (\bibinfo {year} {2010}{\natexlab{a}})}\BibitemShut
  {NoStop}%
\bibitem [{\citenamefont {Kumar}\ \emph
  {et~al.}(2010{\natexlab{b}})\citenamefont {Kumar}, \citenamefont {Soos},
  \citenamefont {Sen},\ and\ \citenamefont {Ramasesha}}]{Kumar2}%
  \BibitemOpen
  \bibfield  {author} {\bibinfo {author} {\bibfnamefont {M.}~\bibnamefont
  {Kumar}}, \bibinfo {author} {\bibfnamefont {Z.~G.}\ \bibnamefont {Soos}},
  \bibinfo {author} {\bibfnamefont {D.}~\bibnamefont {Sen}}, \ and\ \bibinfo
  {author} {\bibfnamefont {S.}~\bibnamefont {Ramasesha}},\ }\href {\doibase
  10.1103/PhysRevB.81.104406} {\bibfield  {journal} {\bibinfo  {journal} {Phys.
  Rev. B}\ }\textbf {\bibinfo {volume} {81}},\ \bibinfo {pages} {104406}
  (\bibinfo {year} {2010}{\natexlab{b}})}\BibitemShut {NoStop}%
\bibitem [{\citenamefont {Brehmer}\ \emph {et~al.}(1997)\citenamefont
  {Brehmer}, \citenamefont {Mikeska},\ and\ \citenamefont
  {Yamamoto}}]{Brehmer_1997}%
  \BibitemOpen
  \bibfield  {author} {\bibinfo {author} {\bibfnamefont {S.}~\bibnamefont
  {Brehmer}}, \bibinfo {author} {\bibfnamefont {H.-J.}\ \bibnamefont
  {Mikeska}}, \ and\ \bibinfo {author} {\bibfnamefont {S.}~\bibnamefont
  {Yamamoto}},\ }\href {\doibase 10.1088/0953-8984/9/19/012} {\bibfield
  {journal} {\bibinfo  {journal} {Journal of Physics: Condensed Matter}\
  }\textbf {\bibinfo {volume} {9}},\ \bibinfo {pages} {3921} (\bibinfo {year}
  {1997})}\BibitemShut {NoStop}%
\bibitem [{\citenamefont {Yamamoto}\ \emph {et~al.}(1998)\citenamefont
  {Yamamoto}, \citenamefont {Brehmer},\ and\ \citenamefont
  {Mikeska}}]{PhysRevB.57.13610}%
  \BibitemOpen
  \bibfield  {author} {\bibinfo {author} {\bibfnamefont {S.}~\bibnamefont
  {Yamamoto}}, \bibinfo {author} {\bibfnamefont {S.}~\bibnamefont {Brehmer}}, \
  and\ \bibinfo {author} {\bibfnamefont {H.-J.}\ \bibnamefont {Mikeska}},\
  }\href {\doibase 10.1103/PhysRevB.57.13610} {\bibfield  {journal} {\bibinfo
  {journal} {Phys. Rev. B}\ }\textbf {\bibinfo {volume} {57}},\ \bibinfo
  {pages} {13610} (\bibinfo {year} {1998})}\BibitemShut {NoStop}%
\bibitem [{\citenamefont {Yamamoto}\ \emph {et~al.}(2000)\citenamefont
  {Yamamoto}, \citenamefont {Fukui},\ and\ \citenamefont
  {Sakai}}]{Yamamoto2000}%
  \BibitemOpen
  \bibfield  {author} {\bibinfo {author} {\bibfnamefont {S.}~\bibnamefont
  {Yamamoto}}, \bibinfo {author} {\bibfnamefont {T.}~\bibnamefont {Fukui}}, \
  and\ \bibinfo {author} {\bibfnamefont {T.}~\bibnamefont {Sakai}},\ }\href
  {\doibase 10.1007/s100510051118} {\bibfield  {journal} {\bibinfo  {journal}
  {The European Physical Journal B - Condensed Matter and Complex Systems}\
  }\textbf {\bibinfo {volume} {15}},\ \bibinfo {pages} {211} (\bibinfo {year}
  {2000})}\BibitemShut {NoStop}%
\bibitem [{\citenamefont {Trumper}\ and\ \citenamefont
  {Gazza}(2001)}]{PhysRevB.64.134408}%
  \BibitemOpen
  \bibfield  {author} {\bibinfo {author} {\bibfnamefont {A.~E.}\ \bibnamefont
  {Trumper}}\ and\ \bibinfo {author} {\bibfnamefont {C.}~\bibnamefont
  {Gazza}},\ }\href {\doibase 10.1103/PhysRevB.64.134408} {\bibfield  {journal}
  {\bibinfo  {journal} {Phys. Rev. B}\ }\textbf {\bibinfo {volume} {64}},\
  \bibinfo {pages} {134408} (\bibinfo {year} {2001})}\BibitemShut {NoStop}%
\bibitem [{\citenamefont {\"Ostlund}\ and\ \citenamefont
  {Rommer}(1995)}]{PhysRevLett.75.3537}%
  \BibitemOpen
  \bibfield  {author} {\bibinfo {author} {\bibfnamefont {S.}~\bibnamefont
  {\"Ostlund}}\ and\ \bibinfo {author} {\bibfnamefont {S.}~\bibnamefont
  {Rommer}},\ }\href {\doibase 10.1103/PhysRevLett.75.3537} {\bibfield
  {journal} {\bibinfo  {journal} {Phys. Rev. Lett.}\ }\textbf {\bibinfo
  {volume} {75}},\ \bibinfo {pages} {3537} (\bibinfo {year}
  {1995})}\BibitemShut {NoStop}%
\bibitem [{\citenamefont {Cirac}\ \emph {et~al.}(2021)\citenamefont {Cirac},
  \citenamefont {P\'erez-Garc\'{\i}a}, \citenamefont {Schuch},\ and\
  \citenamefont {Verstraete}}]{RevModPhys.93.045003}%
  \BibitemOpen
  \bibfield  {author} {\bibinfo {author} {\bibfnamefont {J.~I.}\ \bibnamefont
  {Cirac}}, \bibinfo {author} {\bibfnamefont {D.}~\bibnamefont
  {P\'erez-Garc\'{\i}a}}, \bibinfo {author} {\bibfnamefont {N.}~\bibnamefont
  {Schuch}}, \ and\ \bibinfo {author} {\bibfnamefont {F.}~\bibnamefont
  {Verstraete}},\ }\href {\doibase 10.1103/RevModPhys.93.045003} {\bibfield
  {journal} {\bibinfo  {journal} {Rev. Mod. Phys.}\ }\textbf {\bibinfo {volume}
  {93}},\ \bibinfo {pages} {045003} (\bibinfo {year} {2021})}\BibitemShut
  {NoStop}%
\bibitem [{\citenamefont {Eisert}\ \emph {et~al.}(2010)\citenamefont {Eisert},
  \citenamefont {Cramer},\ and\ \citenamefont {Plenio}}]{RevModPhys.82.277}%
  \BibitemOpen
  \bibfield  {author} {\bibinfo {author} {\bibfnamefont {J.}~\bibnamefont
  {Eisert}}, \bibinfo {author} {\bibfnamefont {M.}~\bibnamefont {Cramer}}, \
  and\ \bibinfo {author} {\bibfnamefont {M.~B.}\ \bibnamefont {Plenio}},\
  }\href {\doibase 10.1103/RevModPhys.82.277} {\bibfield  {journal} {\bibinfo
  {journal} {Rev. Mod. Phys.}\ }\textbf {\bibinfo {volume} {82}},\ \bibinfo
  {pages} {277} (\bibinfo {year} {2010})}\BibitemShut {NoStop}%
\bibitem [{\citenamefont {Franco-Rubio}\ and\ \citenamefont
  {Cirac}(2022)}]{Rubio_22}%
  \BibitemOpen
  \bibfield  {author} {\bibinfo {author} {\bibfnamefont {A.}~\bibnamefont
  {Franco-Rubio}}\ and\ \bibinfo {author} {\bibfnamefont {J.~I.}\ \bibnamefont
  {Cirac}},\ }\href {\doibase 10.48550/ARXIV.2204.02478} {\enquote {\bibinfo
  {title} {Gaussian matrix product states cannot efficiently describe critical
  systems},}\ } (\bibinfo {year} {2022})\BibitemShut {NoStop}%
\bibitem [{\citenamefont {White}\ and\ \citenamefont
  {Feiguin}(2004)}]{PhysRevLett.93.076401}%
  \BibitemOpen
  \bibfield  {author} {\bibinfo {author} {\bibfnamefont {S.~R.}\ \bibnamefont
  {White}}\ and\ \bibinfo {author} {\bibfnamefont {A.~E.}\ \bibnamefont
  {Feiguin}},\ }\href {\doibase 10.1103/PhysRevLett.93.076401} {\bibfield
  {journal} {\bibinfo  {journal} {Phys. Rev. Lett.}\ }\textbf {\bibinfo
  {volume} {93}},\ \bibinfo {pages} {076401} (\bibinfo {year}
  {2004})}\BibitemShut {NoStop}%
\bibitem [{\citenamefont {Daley}\ \emph {et~al.}(2004)\citenamefont {Daley},
  \citenamefont {Kollath}, \citenamefont {Schollwöck},\ and\ \citenamefont
  {Vidal}}]{1742-5468-2004-04-P04005}%
  \BibitemOpen
  \bibfield  {author} {\bibinfo {author} {\bibfnamefont {A.~J.}\ \bibnamefont
  {Daley}}, \bibinfo {author} {\bibfnamefont {C.}~\bibnamefont {Kollath}},
  \bibinfo {author} {\bibfnamefont {U.}~\bibnamefont {Schollwöck}}, \ and\
  \bibinfo {author} {\bibfnamefont {G.}~\bibnamefont {Vidal}},\ }\href
  {http://stacks.iop.org/1742-5468/2004/i=04/a=P04005} {\bibfield  {journal}
  {\bibinfo  {journal} {Journal of Statistical Mechanics: Theory and
  Experiment}\ }\textbf {\bibinfo {volume} {2004}},\ \bibinfo {pages} {P04005}
  (\bibinfo {year} {2004})}\BibitemShut {NoStop}%
\bibitem [{\citenamefont {Or\'us}\ and\ \citenamefont
  {Vidal}(2008)}]{PhysRevB.78.155117}%
  \BibitemOpen
  \bibfield  {author} {\bibinfo {author} {\bibfnamefont {R.}~\bibnamefont
  {Or\'us}}\ and\ \bibinfo {author} {\bibfnamefont {G.}~\bibnamefont {Vidal}},\
  }\href {\doibase 10.1103/PhysRevB.78.155117} {\bibfield  {journal} {\bibinfo
  {journal} {Phys. Rev. B}\ }\textbf {\bibinfo {volume} {78}},\ \bibinfo
  {pages} {155117} (\bibinfo {year} {2008})}\BibitemShut {NoStop}%
\bibitem [{\citenamefont {Pollmann}\ \emph {et~al.}(2012)\citenamefont
  {Pollmann}, \citenamefont {Berg}, \citenamefont {Turner},\ and\ \citenamefont
  {Oshikawa}}]{PhysRevB.85.075125}%
  \BibitemOpen
  \bibfield  {author} {\bibinfo {author} {\bibfnamefont {F.}~\bibnamefont
  {Pollmann}}, \bibinfo {author} {\bibfnamefont {E.}~\bibnamefont {Berg}},
  \bibinfo {author} {\bibfnamefont {A.~M.}\ \bibnamefont {Turner}}, \ and\
  \bibinfo {author} {\bibfnamefont {M.}~\bibnamefont {Oshikawa}},\ }\href
  {\doibase 10.1103/PhysRevB.85.075125} {\bibfield  {journal} {\bibinfo
  {journal} {Phys. Rev. B}\ }\textbf {\bibinfo {volume} {85}},\ \bibinfo
  {pages} {075125} (\bibinfo {year} {2012})}\BibitemShut {NoStop}%
\bibitem [{\citenamefont {Huang}\ and\ \citenamefont
  {Lin}(2011)}]{PhysRevB.84.125110}%
  \BibitemOpen
  \bibfield  {author} {\bibinfo {author} {\bibfnamefont {C.-Y.}\ \bibnamefont
  {Huang}}\ and\ \bibinfo {author} {\bibfnamefont {F.-L.}\ \bibnamefont
  {Lin}},\ }\href {\doibase 10.1103/PhysRevB.84.125110} {\bibfield  {journal}
  {\bibinfo  {journal} {Phys. Rev. B}\ }\textbf {\bibinfo {volume} {84}},\
  \bibinfo {pages} {125110} (\bibinfo {year} {2011})}\BibitemShut {NoStop}%
\bibitem [{\citenamefont {Ye}\ \emph {et~al.}(2016)\citenamefont {Ye},
  \citenamefont {Mu},\ and\ \citenamefont {Fan}}]{PhysRevB.94.165167}%
  \BibitemOpen
  \bibfield  {author} {\bibinfo {author} {\bibfnamefont {B.-T.}\ \bibnamefont
  {Ye}}, \bibinfo {author} {\bibfnamefont {L.-Z.}\ \bibnamefont {Mu}}, \ and\
  \bibinfo {author} {\bibfnamefont {H.}~\bibnamefont {Fan}},\ }\href {\doibase
  10.1103/PhysRevB.94.165167} {\bibfield  {journal} {\bibinfo  {journal} {Phys.
  Rev. B}\ }\textbf {\bibinfo {volume} {94}},\ \bibinfo {pages} {165167}
  (\bibinfo {year} {2016})}\BibitemShut {NoStop}%
\bibitem [{\citenamefont {De~Chiara}\ \emph {et~al.}(2012)\citenamefont
  {De~Chiara}, \citenamefont {Lepori}, \citenamefont {Lewenstein},\ and\
  \citenamefont {Sanpera}}]{PhysRevLett.109.237208}%
  \BibitemOpen
  \bibfield  {author} {\bibinfo {author} {\bibfnamefont {G.}~\bibnamefont
  {De~Chiara}}, \bibinfo {author} {\bibfnamefont {L.}~\bibnamefont {Lepori}},
  \bibinfo {author} {\bibfnamefont {M.}~\bibnamefont {Lewenstein}}, \ and\
  \bibinfo {author} {\bibfnamefont {A.}~\bibnamefont {Sanpera}},\ }\href
  {\doibase 10.1103/PhysRevLett.109.237208} {\bibfield  {journal} {\bibinfo
  {journal} {Phys. Rev. Lett.}\ }\textbf {\bibinfo {volume} {109}},\ \bibinfo
  {pages} {237208} (\bibinfo {year} {2012})}\BibitemShut {NoStop}%
\bibitem [{\citenamefont {Weichselbaum}\ \emph {et~al.}(2018)\citenamefont
  {Weichselbaum}, \citenamefont {Capponi}, \citenamefont {Lecheminant},
  \citenamefont {Tsvelik},\ and\ \citenamefont {L\"auchli}}]{SUN-ladder}%
  \BibitemOpen
  \bibfield  {author} {\bibinfo {author} {\bibfnamefont {A.}~\bibnamefont
  {Weichselbaum}}, \bibinfo {author} {\bibfnamefont {S.}~\bibnamefont
  {Capponi}}, \bibinfo {author} {\bibfnamefont {P.}~\bibnamefont
  {Lecheminant}}, \bibinfo {author} {\bibfnamefont {A.~M.}\ \bibnamefont
  {Tsvelik}}, \ and\ \bibinfo {author} {\bibfnamefont {A.~M.}\ \bibnamefont
  {L\"auchli}},\ }\href {\doibase 10.1103/PhysRevB.98.085104} {\bibfield
  {journal} {\bibinfo  {journal} {Phys. Rev. B}\ }\textbf {\bibinfo {volume}
  {98}},\ \bibinfo {pages} {085104} (\bibinfo {year} {2018})}\BibitemShut
  {NoStop}%
\bibitem [{\citenamefont {Meng}\ \emph {et~al.}(2010)\citenamefont {Meng},
  \citenamefont {Lang}, \citenamefont {Wessel}, \citenamefont {Assaad},\ and\
  \citenamefont {Muramatsu}}]{Meng_2010}%
  \BibitemOpen
  \bibfield  {author} {\bibinfo {author} {\bibfnamefont {Z.~Y.}\ \bibnamefont
  {Meng}}, \bibinfo {author} {\bibfnamefont {T.~C.}\ \bibnamefont {Lang}},
  \bibinfo {author} {\bibfnamefont {S.}~\bibnamefont {Wessel}}, \bibinfo
  {author} {\bibfnamefont {F.~F.}\ \bibnamefont {Assaad}}, \ and\ \bibinfo
  {author} {\bibfnamefont {A.}~\bibnamefont {Muramatsu}},\ }\href
  {https://doi.org/10.1038/nature08942} {\bibfield  {journal} {\bibinfo
  {journal} {Nature}\ }\textbf {\bibinfo {volume} {464}},\ \bibinfo {pages}
  {847} (\bibinfo {year} {2010})}\BibitemShut {NoStop}%
\bibitem [{\citenamefont {Sorella}\ \emph {et~al.}(2012)\citenamefont
  {Sorella}, \citenamefont {Otsuka},\ and\ \citenamefont
  {Yunoki}}]{Sorella_2012}%
  \BibitemOpen
  \bibfield  {author} {\bibinfo {author} {\bibfnamefont {S.}~\bibnamefont
  {Sorella}}, \bibinfo {author} {\bibfnamefont {Y.}~\bibnamefont {Otsuka}}, \
  and\ \bibinfo {author} {\bibfnamefont {S.}~\bibnamefont {Yunoki}},\ }\href
  {\doibase 10.1038/srep00992} {\bibfield  {journal} {\bibinfo  {journal} {Sci.
  Rep.}\ }\textbf {\bibinfo {volume} {2}},\ \bibinfo {pages} {992} (\bibinfo
  {year} {2012})}\BibitemShut {NoStop}%
\bibitem [{\citenamefont {Kao}\ \emph {et~al.}(2015)\citenamefont {Kao},
  \citenamefont {Hsieh},\ and\ \citenamefont {Chen}}]{Kao2015Uni10AO}%
  \BibitemOpen
  \bibfield  {author} {\bibinfo {author} {\bibfnamefont {Y.}~\bibnamefont
  {Kao}}, \bibinfo {author} {\bibfnamefont {Y.-D.}\ \bibnamefont {Hsieh}}, \
  and\ \bibinfo {author} {\bibfnamefont {P.}~\bibnamefont {Chen}}\ }(\bibinfo
  {year} {2015})\BibitemShut {NoStop}%
\bibitem [{\citenamefont {Fishman}\ \emph {et~al.}(2020)\citenamefont
  {Fishman}, \citenamefont {White},\ and\ \citenamefont
  {Stoudenmire}}]{itensor}%
  \BibitemOpen
  \bibfield  {author} {\bibinfo {author} {\bibfnamefont {M.}~\bibnamefont
  {Fishman}}, \bibinfo {author} {\bibfnamefont {S.~R.}\ \bibnamefont {White}},
  \ and\ \bibinfo {author} {\bibfnamefont {E.~M.}\ \bibnamefont
  {Stoudenmire}},\ }\href@noop {} {\enquote {\bibinfo {title} {The
  \mbox{ITensor} software library for tensor network calculations},}\ }
  (\bibinfo {year} {2020}),\ \Eprint {http://arxiv.org/abs/2007.14822}
  {arXiv:2007.14822} \BibitemShut {NoStop}%
\end{thebibliography}%
\end{document}